\documentclass[reprint,superscriptaddress]{revtex4-1}

\usepackage{amsmath}
\usepackage{amsfonts}
\usepackage{amssymb}

\usepackage{graphicx}

\usepackage{tikz}
\usetikzlibrary{shapes}
\usetikzlibrary{decorations.pathreplacing}

\usepackage{color}

\usepackage{lineno}

\newcommand{\result}[1]{#1}

\DeclareMathOperator*{\sometext}{\bigcup}

\begin{document}

\title{
iDQ: Statistical Inference of Non-Gaussian Noise with Auxiliary Degrees of Freedom in Gravitational-Wave Detectors
}

\author{Reed Essick}
\affiliation{Kavli Institute for Cosmological Physics, The University of Chicago, 5640 South Ellis Avenue, Chicago, Illinois, 60637, USA}

\author{Patrick Godwin}
\affiliation{Department of Physics, The Pennsylvania State University, University Park, PA 16802, USA}
\affiliation{Institute for Gravitation and the Cosmos, The Pennsylvania State University, University Park, PA 16802, USA}

\author{Chad Hanna}
\affiliation{Department of Physics, The Pennsylvania State University, University Park, PA 16802, USA}
\affiliation{Institute for Gravitation and the Cosmos, The Pennsylvania State University, University Park, PA 16802, USA}
\affiliation{Department of Astronomy and Astrophysics, The Pennsylvania State University, University Park, PA 16802, USA}
\affiliation{Institute for CyberScience, The Pennsylvania State University, University Park, PA 16802, USA}

\author{Lindy Blackburn}
\affiliation{Center for Astrophysics, Harvard \& Smithsonian, 60 Garden Street, Cambridge, MA 02138, USA}
\affiliation{LIGO, Massachusetts Institute of Technology, Cambridge, MA 02139, USA}

\author{Erik Katsavounidis}
\affiliation{LIGO, Massachusetts Institute of Technology, Cambridge, MA 02139, USA}

\date{\today}
\begin{abstract}
Gravitational-wave detectors are exquisitely sensitive instruments and routinely enable ground-breaking observations of novel astronomical phenomena.
However, they also witness non-stationary, non-Gaussian noise that can be mistaken for astrophysical sources, lower detection confidence, or simply complicate the extraction of signal parameters from noisy data.
To address this, we present iDQ, a supervised learning framework to autonomously detect noise artifacts in gravitational-wave detectors based only on auxiliary degrees of freedom insensitive to gravitational waves.
iDQ has operated in low latency throughout the advanced detector era at each of the two LIGO interferometers, providing invaluable data quality information about each detection to date in real-time.
We document the algorithm, describing the statistical framework and possible applications within gravitational-wave searches.
In particular, we construct a likelihood-ratio test that simultaneously accounts for the presence of non-Gaussian noise artifacts and utilizes information from both the observed gravitational-wave strain signal and thousands of auxiliary degrees of freedom.
We also present several examples of iDQ's performance with modern interferometers, showing iDQ's ability to autonomously reproduce known data quality monitors and identify noise artifacts not flagged by other analyses.
\end{abstract}

\maketitle

\section{Introduction}\label{sec:introduction}

Gravitational-wave (GW) detectors, like the advanced LIGO~\cite{LIGO} and Virgo~\cite{Virgo} interferometers (IFOs), are exquisitely sensitive machines.
This sensitivity requires complex control schemes to isolate the instruments from their surroundings ~\cite{Matichard:2015eva,Rollins:2016hlk} and detailed calibration to infer the correct astrophysical strain incident on the detectors \cite{Viets:2017yvy}.
Their success, including the first direct detection of GWs~\cite{GW150914}, the now routine detection of binary black hole coalescences~\cite{catalog}, and the detection of coalescing neutron stars~\cite{GW170817, GW190425}, which enabled ground-breaking multi-messenger observations~\cite{GW170817MMA, GW170817GRB, Coulter1556, Goldstein_2017}, is due to a combination of the detectors' sensitivity and advanced signal processing techniques.

However, several sources of noise still limit the detectors' sensitivity.
The most fundamental is stationary Gaussian noise \cite{Martynov:2016fzi}, which can be completely characterized by a power spectral density (PSD) and describes the detectors' behavior reasonably well most of the time.
Another common noise source, referred to as \emph{non-Gaussian noise transients} (colloquially termed \emph{glitches}~\cite{LIGOScientific:2019hgc}), manifests as bursts of excess power in the detectors above and beyond what would be expected from stationary Gaussian noise alone.
Because this is also the hallmark of a GW signal, non-Gaussian noise transients can be mistaken for real GW signals if they occur simultaneously in multiple detectors and currently limit searches' sensitivity to many expected astrophysical signals (e.g., \cite{catalog, PhysRevD.100.024017, Abbott_2018}).
Throughout the advanced detector era, an extensive zoology has been developed to categorize and mitigate the impact of non-Gaussian noise transients.
This includes examining the morphology of the noise transients themselves in GW strain data as well as searching for correlations between the noise transients and other degrees of freedom that are not sensitive to GWs.
Many other works have explored the former ~\cite{Powell:2015ona,Powell:2016rkl,Zevin2017}.
We focus on the latter.

Information from the detectors is recorded in a set of discretely sampled timeseries, referred to as \emph{channels}.
These channels observe many different degrees of freedom within the interferometers, including control signals used to stabilize the device~\cite{Mueller:2016hex, Matichard:2015eva, Staley:2015nie, Rollins:2016hlk} and monitors of the physical environment~\cite{Effler:2014zpa}.
In total, there are more than $2\times10^5$ channels recorded at each LIGO detector, although only $\mathcal{O}(10^4)$ are sampled at frequencies high enough to be within the detectors' sensitive band.
Typically the auxiliary features used within iDQ are derived from this subset of channels, which are sampled at $\gtrsim 256\, \mathrm{Hz}$.
Each channel may witness a variety of possible noise sources, some of which may couple to the measurement of GWs and some of which may not.
Because of the large number of channels, it is impractical to measure the couplings between all of them directly via targeted injection campaigns.
Instead, we rely on statistical correlations between channels to infer the noise's source.
If channels insensitive to GWs routinely glitch in close proximity to non-Gaussian transients within the GW channel, we infer that the transients in the GW channel are mostly likely due to terrestrial noise rather than being of astrophysical origin.
Fig.~\ref{fig:probabilistic graph} depicts a probabilistic graphical model representing this inference.

iDQ~\cite{idq-repo,idq-docs}, a statistical framework for this inference, has operated throughout the advanced detector era and continues to provide robust, real-time measures of correlations between thousands of degrees of freedom within each detector and non-Gaussian noise in the GW channel.
The speed and reliability of this information has proven invaluable for several GW detections (e.g., GW170817~\cite{GW170817} and examples in Section~\ref{sec:examples}).
With the expected increases in detector sensitivity and corresponding elevated detection rates over the next few years~\cite{ObservingScenarios}, real-time data quality information will only become more important.

iDQ, first described in Ref.~\cite{Essick2017}, was developed as an extension of Ref.~\cite{Biswas2013} in preparation for the first observing run, which began in September 2015.
Although there is a long history of algorithmic development in the field, including hierarchical veto application schemes based on approximations of the likelihood ratio~\cite{Essick2013}, the Poisson significance of coincident noise~\cite{Smith2011}, the percentage of time witnesses remove noise~\cite{Isogai2010}, and applications of more general machine learning algorithms~\cite{Biswas2013, Cavaglia2018, Colgan:2019lyo}, many of these algorithms fail to produce probabilistic statements about their predictions.
Furthermore, such algorithms did not operate in real-time, a key component of searches for multi-messenger astrophysical events~\cite{GBM:2017lvd, LIGOScientific:2019gag}.
Additionally, GW interferometers are not stationary over long periods of time; the characteristics of the noise change.
This means that correlations measured by any particular algorithm at any particular time may not generalize well to data recorded later, complicating the inference process.
We note that there is a distinction between non-Gaussian noise and non-stationarity.
For example, non-Gaussian noise may be described by stationary Poisson processes.
In fact, many algorithms make this assumption~\cite{Essick2013, Smith2011}, although it is not guaranteed to be the case.

iDQ provides a framework in which any supervised learning algorithm can be run in real-time.
It calibrates their output into statistical statements about the confidence that non-Gaussian noise is present in the GW channel.
This is accomplished via two-class classification, amenable to many machine learning algorithms.
Additionally, iDQ automatically re-trains and re-calibrates the algorithms to capture non-stationarity within the detectors.
This means that iDQ autonomously adapts to new sources of non-Gaussian noise within the detectors, identifying witnesses of previously unseen noise sources and flagging data as problematic without human intervention.

We describe iDQ's formalism in Section~\ref{sec:formalism}, including our use of supervised learning in Section~\ref{sec:supervised learning}.
Section~\ref{sec:decomposition} describes how we structure the inference, including how we construct vectorized representations of a detector's state (Section~\ref{sec:features}), how we train machine learning algorithms, including ways to extract feature importance from the trained models (Section~\ref{sec:training}), as well as how iDQ manages cross-validation (Section~\ref{sec:evaluation}) and calibrates its predictions into probabilistic statements (Section~\ref{sec:calibration}).
A few examples of iDQ's performance are shown in Section~\ref{sec:examples} and possible applications within searches are discussed in Section~\ref{sec:applications}, including a likelihood ratio test based on first-principles noise models which account for our imperfect knowledge of the presence of non-Gaussian noise in our detectors.
We conclude in Section~\ref{sec:discussion}.

\section{Formalism}\label{sec:formalism}

We couch our statistical inference as two-class classification, which we approach within a supervised learning framework.
This produces predictions for the presence or absence of non-Gaussian noise.
It is worth noting that this is not the only approach, and one could instead attempt to regress the full waveform of the non-Gaussian noise based on auxiliary degrees of freedom (e.g.,~\cite{PhysRevD.101.042003, Ormiston:2020ele}).
However, classification is more tractable at this time, and we construct our inference in that framework using a vectorized representation of the detector's auxiliary state.

\begin{figure*}
    \tikzstyle{target} = [circle, draw, fill=black!20, text centered, rounded corners, minimum width=4em, minimum height=4em]
    \tikzstyle{unsafe} = [circle, draw, fill=red!20, text centered, rounded corners, minimum width=4em, minimum height=4em]
    \tikzstyle{safe} = [circle, draw, fill=blue!20, text centered, rounded corners, minimum width=4em, minimum height=4em]

    \tikzstyle{gw} = [circle, draw, text centered, rounded corners, minimum width=4em, minimum height=4em]
    \tikzstyle{latent} = [circle, draw, text centered, rounded corners, minimum width=4em, minimum height=4em]

    \tikzstyle{blank} = [rectangle, text centered, rounded corners]

    \tikzstyle{arrow} = [draw, ->, line width=1.0pt, rounded corners, shorten >=2pt, black!50]
    \tikzstyle{thick-arrow} = [draw, ->, line width=1.5pt, rounded corners, shorten >=2pt, black!100]
    \tikzstyle{dash-arrow} = [draw, ->, line width=1.0pt, rounded corners, dashed, shorten >=2pt]

    \begin{center}
    \begin{tikzpicture}

        \node [gw] (signal) {\textcolor{black}{$s$}};

        \node [target, node distance=2.0cm, below of=signal, yshift=-2.0cm] (hoft) {$h$};

        \node [unsafe, node distance=2.0cm, right of=hoft] (unsafe1) {$a_1$};
        \node [unsafe, node distance=2.0cm, right of=unsafe1] (unsafe2) {$a_2$};
        \node [blank, node distance=1.25cm, right of=unsafe2] (unsafeDots) {\large{$\cdots$}};
        \node [unsafe, node distance=1.25cm, right of=unsafeDots] (unsafeN) {$a_N$};

        \node [blank, draw=red!50, line width=1.5pt, minimum width=19em, minimum height=6em, right of=unsafe1, xshift=+1.2cm] (unsafe) {};

        \node [safe, node distance=2.0cm, right of=unsafeN] (safe1) {$a_{N+1}$};
        \node [safe, node distance=2.0cm, right of=safe1] (safe2) {$a_{N+2}$};
        \node [blank, node distance=1.25cm, right of=safe2] (safeDots) {\large{$\cdots$}};
        \node [safe, node distance=1.25cm, right of=safeDots] (safeN) {$a_{N+M}$};

        \node [blank, draw=blue!50, line width=1.5pt, minimum width=19em, minimum height=6em, right of=safe1, xshift=+1.2cm] (unsafe) {};

        \node [latent, node distance=5.0cm, right of=signal] (noise1) {$G_1$};
        \node [latent, node distance=2.0cm, right of=noise1] (noise2) {$G_2$};
        \node [blank, node distance=1.25cm, right of=noise2] (noiseDots) {\large{$\cdots$}};
        \node [latent, node distance=1.25cm, right of=noiseDots] (noiseN) {$G_\mathcal{N}$};

        \node [latent, node distance=7.0cm, right of=signal, yshift=-8.0cm] (indep-noise1) {$G_{\mathcal{N}+1}$};
        \node [latent, node distance=2.0cm, right of=indep-noise1] (indep-noise2) {$G_{\mathcal{N}+2}$};
        \node [blank, node distance=1.25cm, right of=indep-noise2] (indep-noiseDots) {\large{$\cdots$}};
        \node [latent, node distance=1.25cm, right of=indep-noiseDots] (indep-noiseN) {$G_{\mathcal{N}+\mathcal{M}}$};


        \path [arrow] (hoft) to [out=315, in=225] (unsafe1);
        \path [arrow] (hoft) to [out=315, in=225] (unsafe2);
        \path [arrow] (hoft) to [out=315, in=225] (unsafeN);

        \path [thick-arrow] (signal.south) -- (hoft.north);
        \path [arrow] (signal.south) -- (unsafe1.north);
        \path [arrow] (signal.south) -- (unsafe2.north);
        \path [arrow] (signal.south) -- (unsafeN.north);

        \path [thick-arrow] (noise1.south) -- (hoft.north);
        \path [arrow] (noise1.south) -- (unsafe1.north);
        \path [arrow] (noise1.south) -- (unsafe2.north);
        \path [arrow] (noise1.south) -- (unsafeN.north);
        \path [arrow] (noise1.south) -- (safe1.north);
        \path [arrow] (noise1.south) -- (safe2.north);
        \path [arrow] (noise1.south) -- (safeN.north);

        \path [thick-arrow] (noise2.south) -- (hoft.north);
        \path [arrow] (noise2.south) -- (unsafe1.north);
        \path [arrow] (noise2.south) -- (unsafe2.north);
        \path [arrow] (noise2.south) -- (unsafeN.north);
        \path [arrow] (noise2.south) -- (safe1.north);
        \path [arrow] (noise2.south) -- (safe2.north);
        \path [arrow] (noise2.south) -- (safeN.north);

        \path [thick-arrow] (noiseN.south) -- (hoft.north);
        \path [arrow] (noiseN.south) -- (unsafe1.north);
        \path [arrow] (noiseN.south) -- (unsafe2.north);
        \path [arrow] (noiseN.south) -- (unsafeN.north);
        \path [arrow] (noiseN.south) -- (safe1.north);
        \path [arrow] (noiseN.south) -- (safe2.north);
        \path [arrow] (noiseN.south) -- (safeN.north);

        \path [arrow] (indep-noise1.north) -- (unsafe1.south);
        \path [arrow] (indep-noise1.north) -- (unsafe2.south);
        \path [arrow] (indep-noise1.north) -- (unsafeN.south);
        \path [arrow] (indep-noise1.north) -- (safe1.south);
        \path [arrow] (indep-noise1.north) -- (safe2.south);
        \path [arrow] (indep-noise1.north) -- (safeN.south);

        \path [arrow] (indep-noise2.north) -- (unsafe1.south);
        \path [arrow] (indep-noise2.north) -- (unsafe2.south);
        \path [arrow] (indep-noise2.north) -- (unsafeN.south);
        \path [arrow] (indep-noise2.north) -- (safe1.south);
        \path [arrow] (indep-noise2.north) -- (safe2.south);
        \path [arrow] (indep-noise2.north) -- (safeN.south);

        \path [arrow] (indep-noiseN.north) -- (unsafe1.south);
        \path [arrow] (indep-noiseN.north) -- (unsafe2.south);
        \path [arrow] (indep-noiseN.north) -- (unsafeN.south);
        \path [arrow] (indep-noiseN.north) -- (safe1.south);
        \path [arrow] (indep-noiseN.north) -- (safe2.south);
        \path [arrow] (indep-noiseN.north) -- (safeN.south);


        \node [blank, right of=safeN, xshift=+1.4cm] {observed processes};

        \node [blank, right of=noiseN, xshift=+4.7cm, yshift=+0.17cm] {latent processes};
        \node [blank, right of=noiseN, xshift=+4.7cm, yshift=-0.18cm] {that influence $h$};

        \node [blank, right of=indep-noiseN, xshift=+2.8cm, yshift=+0.17cm] {latent processes};
        \node [blank, right of=indep-noiseN, xshift=+2.4cm, yshift=-0.18cm] {that do not influence $h$};

        \node [blank, above right of=signal, xshift=+1.0cm, yshift=+0.3cm] {\textcolor{black}{astrophysical signals}};
        \node [blank, above of=noiseN] {\textcolor{black}{non-Gaussian noise sources}};
        \node [blank, left of=indep-noise1, xshift=-2.0cm] {\textcolor{black}{non-Gaussian noise sources}};

        \node [blank, fill=black!40, below of=hoft, yshift=-0.5cm] (target) {\textcolor{white}{target channel}};
        \node [blank, fill=red!40, below of=unsafe2, xshift=+0.2cm, yshift=-0.5cm] (unsafe) {\textcolor{white}{unsafe auxiliary channels}};
        \node [blank, fill=blue!40, below of=safe2, xshift=+0.2cm, yshift=-0.5cm] (safe) {\textcolor{white}{safe auxiliary channels}};

    \end{tikzpicture}
    \end{center}
    \caption{
        Probabilistic graphical model representing different sources of noise within GW interferometers.
        Each circle represents a different process, some of which are observed and some of which are not, and conditional dependencies are represented by directed arrows.
        We assume that each separate process additionally includes independent stationary additive Gaussian noise, which we omit for clarity.
        Latent processes, which we cannot observe directly, are shown as unshaded circles.
        The top row corresponds to processes that can influence the target channel ($h$), including astrophysical signals ($s$) and a subset of possible sources of non-Gaussian noise ($G_1$ -- $G_\mathcal{N}$).
        Conditional dependencies from these channels to $h$ (thick arrows) are how non-Gaussian noise couples into the target channel.
        The bottom row ($G_{\mathcal{N}+1}$ -- $G_{\mathcal{N}+\mathcal{M}}$) corresponds to possible sources of non-Gaussian noise that do not affect $h$ but may affect auxiliary witnesses.
        Observed processes are shown in the middle row, which include $h$ (\emph{grey shaded circle}) as well as $N$ \emph{unsafe} auxiliary channels (\emph{red shaded circles}), which may be influenced by $s$ or $h$ (or both), and $M$ \emph{safe} auxiliary channels (\emph{blue shaded circles}), which only witness sources of non-Gaussian noise.
        If we marginalize over the latent processes in this graph, we will introduce conditional dependencies between $h$ and the auxiliary channels as well as between different sets of auxiliary channels.
        iDQ only uses information from safe auxiliary channels to infer these induced correlations, thereby predicting the presence or absence of non-Gaussian noise in the target channel without observing the target channel itself.
    }
    \label{fig:probabilistic graph}
\end{figure*}
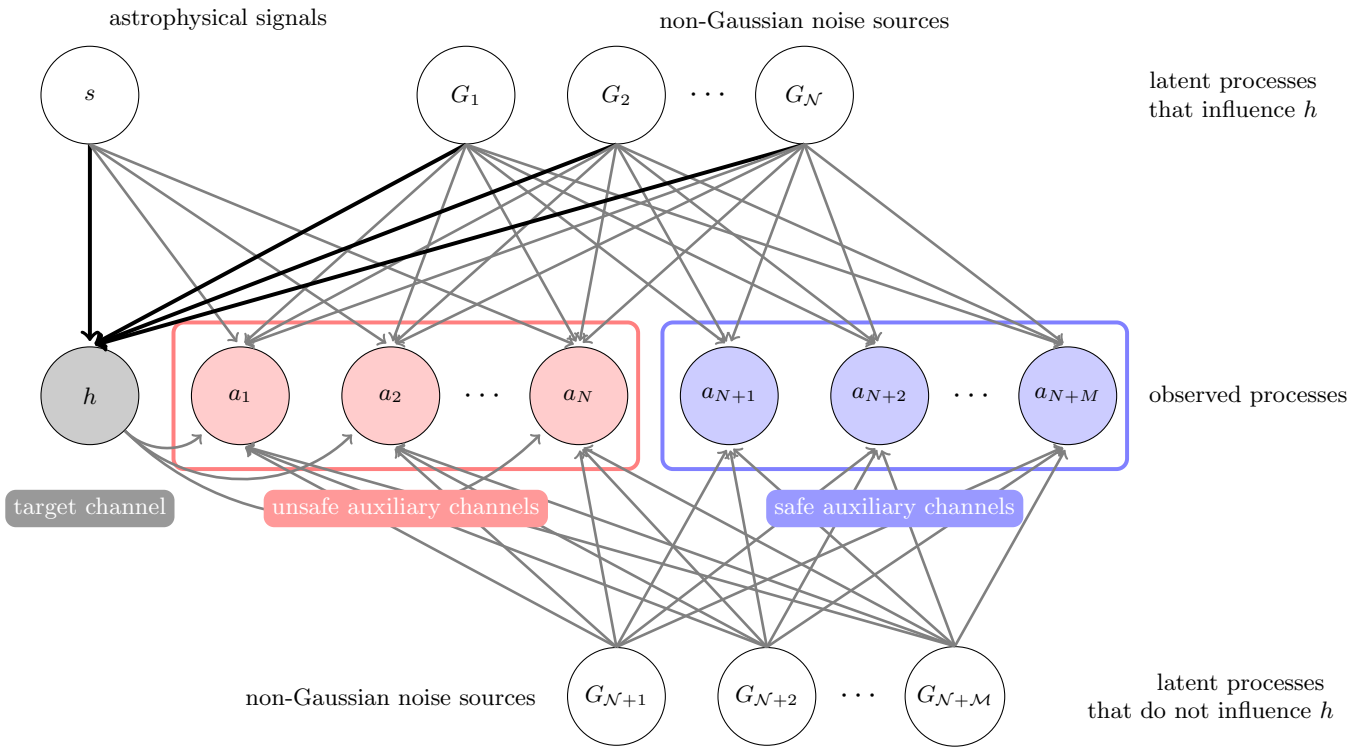

First, a bit of nomenclature.
The \emph{target channel} $h(t)$ refers to the degree of freedom containing non-Gaussian noise we would like to identify, typically a proxy for the GW channel.
This is shown in grey in Fig.~\ref{fig:probabilistic graph}.
\emph{Auxiliary channels} refer to all other channels.
Because GW detectors are complicated devices that require active control, several auxiliary channels may be nearly identical to the target channel (i.e., contain signals derived from $h$ used to control the interferometer).
These are referred to as \emph{unsafe auxiliary channels}, as they are likely to witness real GW signals, and it would be \emph{unsafe} to use them to construct data quality flags as they could systematically veto real GW signals~\cite{Isogai2010,TheLIGOScientific:2016zmo}.
These are shown in red in Fig.~\ref{fig:probabilistic graph}.
The remaining auxiliary channels, which are not sensitive to GWs, are referred to as \emph{safe auxiliary channels} $\vec{a}(t)$, shown in blue in Fig.~\ref{fig:probabilistic graph}.
Safety is typically determined through a series of hardware injections in which excitations are injected into the interferometer to mimic the effect of a real GW.
Auxiliary channels which correlate strongly with the hardware injections may similarly correlate with real GW signals and are deemed unsafe.
\emph{iDQ only uses information from safe auxiliary channels}, although labels for supervised learning are derived from $h$.

To put this more formally, we assume the target channel is composed of three independent components
\begin{equation}
    h(t) = n(t) + s(t) + g(t)
\end{equation}
where $n$ is stationary Gaussian noise, $s$ is the astrophysical strain induced in the detector, and $g$ represents non-Gaussian noise artifacts.
We can only observe $h$ and therefore model $n$, $s$, and $g$ as latent processes which may or may not be correlated with other degrees of freedom.
Furthermore, we assume $n$ is stationary over timescales much longer than either $s$ or $g$ and can therefore be completely described by a single PSD.

Because we adopt a two-class classification scheme, we must define our classes.
We take $G$ to be the union of all possible types of non-Gaussian noise ($G_i$) without explicitly enumerating each class (as opposed to, e.g., GravitySpy's explicit multi-class classification~\cite{Zevin2017}).

\begin{equation}
    G = \sometext\limits_{i \in \substack{\text{glitch} \\ \text{classes}}} G_i
\end{equation}
Typically, this is defined by a set of thresholds on $h$, such as signal-to-noise ratio ($\rho$) and a frequency range.
For instance, one may target only loud, low-frequency noise relevant for high-mass binary black hole searches.
$G$ then consists of all time samples that correspond to $h$ within these thresholds.
The complement of $G$, referred to as $C$, corresponds to \emph{clean} times when the detector does not display non-Gaussian noise such that
\begin{equation}
    p(G)+p(C)=1\ \forall\ t
\end{equation}
and
\begin{equation}
    p(G \cap C) = 0\ \forall\ t
\end{equation}
Furthermore, clean states axiomatically imply $g(t)=0\ \forall\ t\in C$, in that there are no non-Gaussian noise transients within clean times.

Because we derive labels based on $h$ instead of $g$, true signals may also fall within the thresholds defining $G$.
However, the true signal rate is expected to be orders of magnitude less than the rate of non-Gaussian noise artifacts ($\lesssim 1/\mathrm{day}$ as opposed to $\sim1/\mathrm{minute}$), and we do not expect GWs to significantly pollute our training set~\footnote{We note that elevated detection rates expected with advanced detectors at design sensitivity and other planned detectors may cause this assumption to break down, necessitating further curation of training sets within our supervised learning framework.}.
Furthermore, because we only use safe auxiliary channels, defined by their insensitivity to GW signals, we expect $s$ to be independent of $\vec{a}$, and $h$ containing GW signals is indistinguishable from $h$ without GW signals based on $\vec{a}$ alone.
Nonetheless, one could remove almost all true signals by removing any element of $G$ that is coincident between multiple detectors~\footnote{Note that a few key detections were essentially made with data from a single interferometer (e.g., GW170817 was initially detected as a a single-interferometer trigger at LIGO Hanford) and therefore requiring coincidences between detectors may not remove all astrophysical signals.}.
This would also accidentally remove elements of $G$ due to $g$ that just happened to be coincident, but we expect the processes producing $g$ to be independent in each detector~\footnote{The assumption of independent noise in each detector may break down in certain cases, like correlated magnetic noise due to Schumann resonances~\cite{Schumann1, Schumann2}.}.
Removing such a random subset of $G$ would only decrease our sample size without biasing the training sets.
However, because of the additional complexity associated with synchronizing processes running at geographically disparate locations, and the fact that the impact on our training sets is negligible, iDQ does not currently remove elements coincident between detectors from its training set.

Again, this is not the only way to construct the inference.
Instead of classification, one could use the fact that $\rho$ measures the probability that Gaussian noise alone could have produced $h$, and can therefore be used to estimate the probability that a non-Gaussian transient is present.
Instead of classifying samples separated by hard thresholds, one could regress $\rho$ directly or use $\rho$ to define weighed training sets.
We leave such extensions to future work, but note that similar model comparisons are implicit within our marginal-maximized likelihood ratio test (Section~\ref{sec:optimal applications}).

iDQ infers the probability of the presence of non-Gaussian noise artifacts ($g\neq0$) within $h$ based on safe auxiliary channels.
Specifically, iDQ estimates
\begin{align}\label{eqn:p_G}
    p_G(t) & = p(G|\vec{a}(t)) \nonumber \\
           & = \frac{p(\vec{a}|G)p(G)}{p(\vec{a}|G)p(G) + p(\vec{a}|C)p(C)}.
\end{align}
using supervised learning to estimate the likelihoods $p(\vec{a}|G)$ and $p(\vec{a}|C)$.

\subsection{Supervised Learning as Dimensional Reduction}\label{sec:supervised learning}

As described in Section~\ref{sec:introduction}, GW interferometers monitor a large number of auxiliary degrees of freedom as discretely sampled timeseries recorded at different rates.
iDQ represents the information in these auxiliary channels as a set of \emph{features} extracted from each channel separately and compiled into an array of fixed dimension.

Several example feature extractors are described in Ref.~\cite{Chatterji2004, godwin-thesis}, and these generally rely on a wavelet decomposition to identify excess power beyond what is expected from stationary Gaussian noise alone.
These feature extractors map discretely sampled timeseries into tabular data, such as the frequency, amplitude, and duration of non-Gaussian transients.
iDQ constructs high-dimensional representations of the detector's auxiliary state based on this tabular data, typically recording $\mathcal{O}(5)$ features for each of $\mathcal{O}(10^3)$ auxiliary channels.
Although efforts to extract better feature sets are on-going~\cite{pointypoisson}, iDQ implicitly assumes that features extracted in this way from each channel are sufficient statistics.
Indeed, the wavelet decompositions at the core of many feature extractors form overcomplete bases and contain all information available in the original channel.
There is also evidence that the precise algorithmic details of the feature extractor may not significantly impact the overall inference (see Section 6.2 of Ref.~\cite{GW150914DetChar}).

The Neyman-Pearson lemma \cite{Neyman:1933wgr} states that an optimal classification scheme orders samples by their likelihood ratio
\begin{equation}\label{eqn:likelihood}
    \Lambda^G_C(\vec{a}) = \frac{p(\vec{a}|G)}{p(\vec{a}|C)} = \frac{p(C)}{p(G)}\left(\frac{p_G}{1-p_G}\right)
\end{equation}
However, we do not know the functional form of the likelihoods \textit{a priori} and must estimate them from observed samples.
Compounding this, the dimensionality of $\vec{a}$ can be very large, typically $\mathcal{O}(10^4)$ or more.
This drives us to supervised machine learning as a way to approximate $\Lambda^G_C$ and reduce the dimensionality to something tractable.

We therefore consider the main product of supervised learning algorithms, referred to as \emph{classifiers}, to be a map from the high dimensional input space to the unit interval $\mathcal{M}(\vec{a}): \mathcal{R}^{N\gg1} \rightarrow \mathcal{R} \in [0,1]$.
The precise functional form of the map is determined by the details of the algorithm and is unimportant for the rest of the inference, but we expect elements of $G$ to be mapped to values near 1 and elements of $C$ to be mapped to values near 0.
We then construct a likelihood ratio in this lower dimensional space for each classifier separately, \textit{de facto} estimating
\begin{equation}
    p_G = \frac{p(\mathcal{M}(\vec{a})|G)p(G)}{p(\mathcal{M}(\vec{a})|G)p(G) + p(\mathcal{M}(\vec{a})|C)p(C)}
\end{equation}
Estimates of $p(\mathcal{M}(\vec{a})|G)$ and $p(\mathcal{M}(\vec{a})|C)$ are obtained by evaluating labeled samples with trained classifiers and directly modeling the resulting distributions (Section~\ref{sec:calibration} and Appendix~\ref{sec:kde}).

Optimal classifiers will order samples according to $\Lambda^G_C$, so we expect $\Lambda^G_C$ to be monotonic in the classifier's output $\mathcal{M}(\vec{a})$.
Therefore iDQ also calculates the \emph{efficiency} and \emph{false alarm probability} as cumulative conditioned likelihoods integrated over classifier predictions
\begin{align}
    \text{efficiency}              & = P(\mathcal{M}(\vec{a})\geq r|G) \nonumber \\
                                   & = \int\limits_r^1 dx\, p(\mathcal{M}(\vec{a})=x|G) 
\end{align}
and
\begin{align}
    \text{false alarm probability} & = P(\mathcal{M}(\vec{a})\geq r|C) \nonumber \\
                                   & = \int\limits_r^1 dx\, p(\mathcal{M}(\vec{a})=x|C) 
\end{align}
These cumulative statistics define the receiver operating characteristic (ROC) curve for a classifier, the standard metric for classification performance.
Likelihood ratio tests optimize the efficiency at all false alarm probabilities.

We also note that iDQ can run multiple classifiers in parallel over the same data.
Because each classifier produces a different map, and therefore may be able to better identify different subsets of glitches, we should be able to extract more information by combining classifiers.
\textit{De facto}, this would amount to using supervised learning to map a very high-dimensional space into a more manageable size, which may be amenable to direct modeling of the likelihood.
This is the case for one-dimensional output from a single classifier.
Combining a handful of classifiers should not pose a more complicated conceptual issue, although running enough classifiers in parallel may require us to use supervised learning again to model the joint likelihood over their output.
This type of \textit{boosted classifier}, previously explored within iDQ~\cite{Essick2017}, would again reduce the problem to likelihoods defined over a one-dimensional space.

\begin{figure*}
    \tikzstyle{block} = [rectangle, draw, fill=blue!20, text centered, rounded corners, minimum width=3em, minimum height=3em]
    \tikzstyle{blank} = [rectangle, text centered, rounded corners]
    \tikzstyle{mds} = [cloud, draw, cloud puffs=9, cloud puff arc=120, aspect=2, minimum width=2.0cm, minimum height=1.5cm]

    \tikzstyle{box} = [rectangle, draw, rounded corners, solid]

    \tikzstyle{arrow} = [draw, ->, line width=1.5pt, rounded corners, shorten >=2pt]
    \tikzstyle{dasharrow} = [draw, ->, line width=1.0pt, rounded corners, dashed, shorten >=2pt]

    \begin{center}
    \begin{tikzpicture}
        \node [blank] (chan1) {channel};
        \node [blank, above of=chan1] (chan2) {channel};
        \node [blank, below of=chan1] (chan3) {channel};
        \node [blank, below of=chan3] (chan4) {$\vdots$};
        \node [blank, below of=chan4] (chan5) {channel};

        \node [block, left of=chan3, node distance=2.0cm] (ifo) {IFO};

        \node [blank, right of=chan1, node distance=4.0cm] (features1) {features};
        \node [blank, above of=features1] (features2) {features};
        \node [blank, below of=features1] (features3) {features};
        \node [blank, below of=features3] (features4) {$\vdots$};
        \node [blank, below of=features4] (features5) {features};

        \node [blank, right of=features3, node distance=1.0cm] (point) {};

        \node [block, right of=features2, node distance=4.0cm] (train) {train};
        \node [mds, right of=train, node distance=3.0cm] (models) {models};
        \node [blank, right of=models, node distance=3.0cm] (models explanation) {};
        \node [blank, above of=models explanation, node distance=0.15cm] (models explanation1) {dimensional reduction and};
        \node [blank, below of=models explanation, node distance=0.15cm] (models explanation2) {feature importance};

        \node [block, below of=train, node distance=2.0cm] (evaluate) {evaluate};
        \node [mds, right of=evaluate, node distance=3.0cm] (quivers) {quivers};
        \node [blank, right of=quivers, node distance=3.0cm] (quivers explanation) {};
        \node [blank, above of=quivers explanation, node distance=0.15cm] (quivers explanation1) {cross-validation and};
        \node [blank, below of=quivers explanation, node distance=0.15cm] (quivers explanation2) {generalization error};

        \node [block, below of=evaluate, node distance=2.0cm] (calibrate) {calibrate};
        \node [mds, right of=calibrate, node distance=3.0cm] (calibmaps) {maps};
        \node [blank, right of=calibmaps, node distance=3.0cm] (calibmaps explanation) {};
        \node [blank, above of=calibmaps explanation, node distance=0.15cm] (calibmaps explanation1) {probabilistic statements};
        \node [blank, below of=calibmaps explanation, node distance=0.15cm] (calibmaps explanation2) {and uncertainty};

        \node [block, below of=calibrate, node distance=2.0cm] (timeseries) {timeseries};
        \node [mds, right of=timeseries, node distance=3.0cm] (gwf) {predictions};
        \node [blank, right of=gwf, node distance=3.0cm] (gwf explanation) {};
        \node [blank, above of=gwf explanation, node distance=0.15cm] (gwf explanation1) {input for searches and};
        \node [blank, below of=gwf explanation, node distance=0.15cm] (gwf explanation2) {candidate follow-up};

        \node [box, xshift=11cm, yshift=-2cm, minimum width=10.5cm, minimum height=9cm] (outline) {};
        \node [blank, above of=outline, yshift=+4cm] (label) {\textbf{\large{iDQ}}};


        \draw [decorate, decoration={brace, amplitude=10pt}, xshift=-1.25cm, yshift=-3.75cm] (0.5, 0.5) -- (0.5, 5.0) node [black, midway] {};
        \draw [decorate, decoration={brace, amplitude=10pt, mirror}, xshift=+4.25cm, yshift=-3.75cm] (0.5, 0.5) -- (0.5, 5.0) node [black, midway] {};


        \path [arrow] (chan1) -- (features1);
        \path [arrow] (chan2) -- (features2);
        \path [arrow] (chan3) -- (features3);
        \path [arrow] (chan5) -- (features5);

        \node [block, right of=chan3, rotate=270, minimum width=15em, node distance=2.0cm] (etg) {Feature Extractor};
        \node [mds, below of=etg, xshift=+1.0cm, node distance=4.5cm] (segdb) {segments};

        \path [arrow] (train) -- (models);
        \path [arrow] (evaluate) -- (quivers);
        \path [arrow] (calibrate) -- (calibmaps);
        \path [arrow] (timeseries) -- (gwf);

        \draw[arrow] (models) to [out=230, in=90] (evaluate.north);
        \draw[arrow] (models) to [out=230, in=90] ++(-1.75cm, -1.75cm) to ++(0.0cm, -2.25cm) to [out=270, in=90] (timeseries.north);
        \draw[arrow] (quivers) to [out=230, in=90] (calibrate.north);
        \draw[arrow] (calibmaps) to [out=230, in=90] (timeseries.north);

        \draw[arrow] (point) to [out=0, in=180] (train.west);
        \draw[arrow] (point) to [out=0, in=180] (evaluate.west);
        \draw[arrow] (point) to [out=0, in=180] (calibrate.west);
        \draw[arrow] (point) to [out=0, in=180] (timeseries.west);

        \draw[dasharrow] (segdb.east) to [out=45, in=225] (train);
        \draw[dasharrow] (segdb.east) to [out=0, in=215] (evaluate);
        \draw[dasharrow] (segdb.east) to [out=0, in=215] (calibrate);
        \draw[dasharrow] (segdb.east) to [out=-45, in=210] (timeseries);

    \end{tikzpicture}
    \end{center}
    \caption{
        iDQ's workflow.
        Channels are processed into feature sets $\vec{a}(t)$, which are then used in several asynchronous processes to train classifiers, calibrate their output, and produce real-time probabilistic predictions for the presence of non-Gaussian noise $G$ in the target channel $h(t)$.
        The large block on the right identifies the steps within the iDQ pipeline, while the source of features and segments on the left identify input used at various points in the workflow.
    }
    \label{fig:flowchart}
\end{figure*}
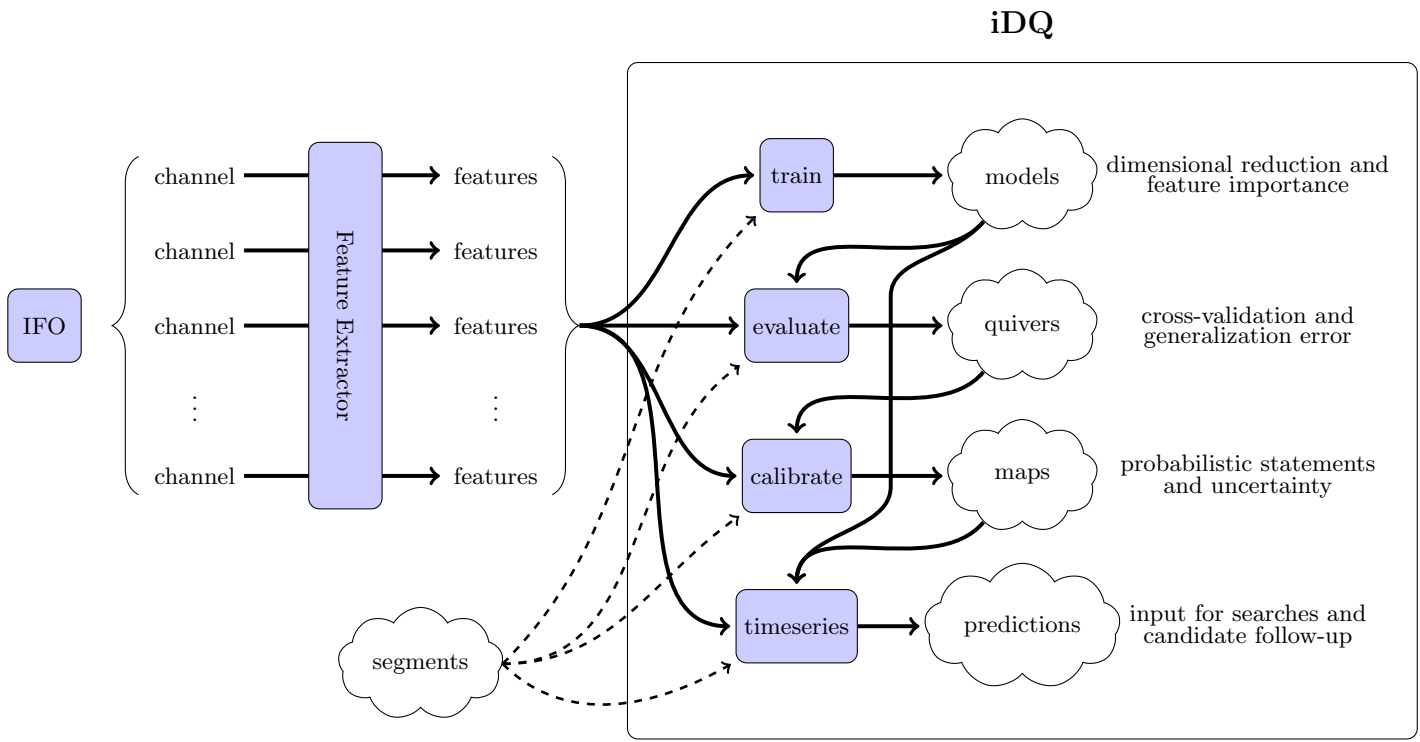

\section{Decomposition of the Statistical Inference}\label{sec:decomposition}

iDQ divides the workflow into several asynchronous processes which communicate to share updated models and calibration.
As Figure~\ref{fig:flowchart} shows, features are generated for each IFO and retrieved by various processes (Section~\ref{sec:data discovery}).
Vectorized representations of the detector's auxiliary state are constructed as needed in each process (Section~\ref{sec:features}).
Training (Section~\ref{sec:training}) produces models for each classifier, which are then used in both evaluation (Section~\ref{sec:evaluation}) and timeseries production (Section~\ref{sec:timeseries}).
The evaluated output is used to calibrate the model (Section~\ref{sec:calibration}), and the resulting map transforms the low-latency predictions made during timeseries production into probabilistic measures with associated uncertainties.
These measures, such as $p_G$, can then be ingested by GW searches in real-time.

\subsection{Data Discovery}\label{sec:data discovery}

iDQ relies upon an external source of features, typically taken to be tabular data denoting the location and properties of non-Gaussian transients in a set of channels.
Although not strictly necessary (e.g., iDQ could ingest the raw timeseries directly from the detectors), we find that this preprocessing efficiently extracts the features relevant for our classification problem.
Figure~\ref{fig:flowchart} represents these feature streams as directed arrows exiting the feature extractor and entering the iDQ workflow.
Pragmatically, these are implemented as abstractions that manage data discovery and produce a consistent tabular output format regardless of the features' source, thereby simplifying any client interactions throughout the pipeline.

Most feature extractors operate in one of several wavelet domains, extracting excess power as collections of time-frequency pixels.
Common choices are the Haar wavelet transform and the $Q$-transform~\cite{Chatterji2004}.
However, the precise form of the feature extractor is unimportant beyond the fact that different wavelet transforms are able to better resolve different aspects of non-Gaussian noise transients.
What's more, not all feature extractors produce the same set of features, although all provide some measure of the transient's central time, duration, frequency content, and amplitude or significance (typically measuring how rare the transient would be in stationary Gaussian noise).
Ref.~\cite{Biswas2013} explored the relative importance of features in the Haar domain, finding the relative time offset and significance to be most important.

Samples from $G$ and $C$ are identified based on the features present in $h$.
Specifically, any transient which meets the criteria for $G$ (see Section~\ref{sec:formalism}) is recorded as a \emph{target time}.
Samples from $C$, called \emph{random times}, are drawn according to a Poisson process in stretches of data sufficiently far away from target times.
These clean segments are defined by another set of thresholds on $h$, which are typically chosen to be slightly looser than the bounds defining target times.
This creates an effective buffer between samples in $G$ and $C$, which is believed to avoid threshold effects and improve classifiers' ability to distinguish the sets.
At the end of the day, any time segment not declared clean is considered \emph{dirty} (i.e., $t\in G$), and this dirty time is accounted for within the calibration's prior odds (Section~\ref{sec:calibration}).

Additionally, iDQ gathers IFO-state information from a remote database ~\cite{dqsegdb}.
Segments produced outside of iDQ record high-level state information about the detectors, such as when the IFOs record science-quality data, but without fine enough temporal resolution to flag subsecond non-Gaussian noise transients.
iDQ polls the database for such segments, filtering the training and evaluation samples to retain only times within science-quality data.
This may not be strictly necessary in all cases and can introduce additional latency before segments are available.
Segment information is, therefore, optional (directed arrows in Figure~\ref{fig:flowchart} are dashed instead of solid), but is almost always used in practice except during low-latency timeseries production.
Timeseries production only applies existing models and calibration to IFO data.
Therefore, it does not care whether the detectors are currently recording science-quality data.

\subsection{Feature Vector Construction}\label{sec:features}

Given a stream of features, we must still determine which features to use.
Most, if not all, supervised classification schemes require consistent dimensionality in the input feature space ($\vec{a}\in\mathcal{R}^N$ with fixed $N$).
Therefore iDQ must downselect features if too many auxiliary transients are nearby or fill in default values if no auxiliary transients are available.
Mapping the streams of features from many auxiliary channels into an array of fixed dimension is referred to as \emph{vectorization}, and is carried out on-the-fly as needed within the pipeline.

Although other types of features have been investigated in the literature, such as averaging over small neighboring time windows~\cite{Colgan:2019lyo}, iDQ implements the following vectorization scheme following Ref.~\cite{Biswas2013}, which was also employed in~\cite{Cavaglia2018}.
For each auxiliary channel, iDQ queries all transients in that channel with central times within some window surrounding the time of interest.
If no auxiliary transients are available, default values are returned for all requested features, thereby denoting auxiliary channels that were inactive.
Otherwise, iDQ will extract features from the loudest auxiliary transient (largest $\rho$) within the window.
We note that this is not the only choice, and quiet transients in closer coincidence with the time of interest may be more relevant than louder transients further away~\cite{pointypoisson}.
Nonetheless, this \emph{select-loudest} algorithm works well in practice~\footnote{There is some evidence that our feature vectors are relatively sparse (i.e., it is rare to have multiple coincident auxiliary transients in a single channel) and the relevant information encoded in the feature sets is simply whether or not there were any non-Gaussian transients in the auxiliary channel within the specified window. In that case, it is perhaps not surprising that the \emph{select-loudest} algorithm retains all the relevant information.} with coincidence windows $\sim100\,\mathrm{ms}$.

The resulting \emph{feature vectors} are collected into sets for training, evaluation, and timeseries generation.
Each vector is labeled according to whether the time is associated with an element of either $G$ or $C$ based on $h$, thereby constituting a supervised learning training set.
These labels are completely ignored during evaluation and timeseries generation.
As implemented, each \emph{feature vector} retains a reference to the data discovery abstraction used to retrieve the features (see Section~\ref{sec:data discovery}).
In this way, classifiers can access the full feature set if desired, although most rely on the vectorization scheme articulated above.
A notable exception is OVL~\cite{Essick2013}, which directly ingests the data discovery abstraction during training.
This is done to avoid additional overhead associated with vectorization and is peculiar to the OVL algorithm, although this type of behavior is more broadly supported within iDQ.

\subsection{Training}\label{sec:training}

Given vectorized representations of the auxiliary features for each time of interest, iDQ then trains classifiers to separate the labeled samples.
Again, classifiers are only given features extracted from \emph{safe auxiliary channels} and cannot construct decision surfaces based on the vectors' $G$ or $C$ labels.
The details of each classifier's training algorithm are specific to each classifier, and iDQ generally relies on external libraries for their implementation (e.g., scikit-learn ~\cite{scikit-learn}, keras ~\cite{chollet2015keras}, and xgboost ~\cite{Chen:2016btl}).
However, a few algorithms are implemented directly within iDQ, such as OVL~\cite{Essick2013}.

In this way, iDQ uses supervised learning on labeled auxiliary feature vectors to generate maps from high-dimensional input spaces to a single scalar \emph{rank}.
We note that vectorization itself also introduces dimensional reduction, as we extract features from at most a single auxiliary transient per channel, but beyond that it is the classifiers themselves that determine which features are relevant and which are not.
This establishes a \emph{model} for each classifier ($\alpha$):
\begin{equation}
    \mathcal{M}_\alpha: \vec{a} \in \mathcal{R}^N \rightarrow r_\alpha \in [0, 1].
\end{equation}
Each classifier generates a separate model, and different models may be able to separate the training sets to different degrees.

iDQ requires ranks to be within the unit interval, although this choice is arbitrary.
The important aspect of the model is the relative ordering of samples, not the precise value of the rank.
Therefore, any monotonic mapping from the unit interval to another range will preserve all relevant information.

Each classifier additionally manages internal cross-validation or provides techniques to prevent over-fitting (see Appendix~\ref{sec:OVL} for an example).
Again, the details may be specific to each classifier, but iDQ also provides a cross-validation scheme independent of the classifiers themselves.
Section~\ref{sec:evaluation} describes the various techniques used to evaluate a classifier's performance fairly, making sure to account for any generalization error associated with the derived models.

Each trained model additionally records the range of data used during training as a unique \emph{hash}.
These hashes are used in evaluation (Section~\ref{sec:evaluation}) and timeseries generation (Section~\ref{sec:timeseries}) to track how data was manipulated as it progressed through the pipeline.

Some algorithms also support measures of \emph{feature importance} within trained models.
These are often related to the directional derivative of the model with respect to each auxiliary feature: $\partial \mathcal{M}/\partial a_i|_{\vec{a}}$.
Features with larger directional derivatives tend to be more important, although each classifier's measure of feature importance may adopt different specific details (Appendix~\ref{sec:OVL} describes how OVL extracts the feature importance measures shown in Figures~\ref{fig:example gw170817} and~\ref{fig:example whistle}).
Not all classifiers provide this information.
Many that do only provide global estimates averaged over all samples instead of the local estimates at a particular auxiliary vector.
However, we leave further investigations of standardized measures of feature importance, such as Ref.~\cite{2016arXiv160204938T}, to future work.

\subsection{Evaluation}\label{sec:evaluation}

Supervised learning relies on cross-validation to evaluate generalization errors.
This typically consists of subdividing the data into distinct sets, one of which is used for training and the other to evaluate performance.
iDQ supports two main ways to subdivide the data into different \emph{bins} for cross-validation.

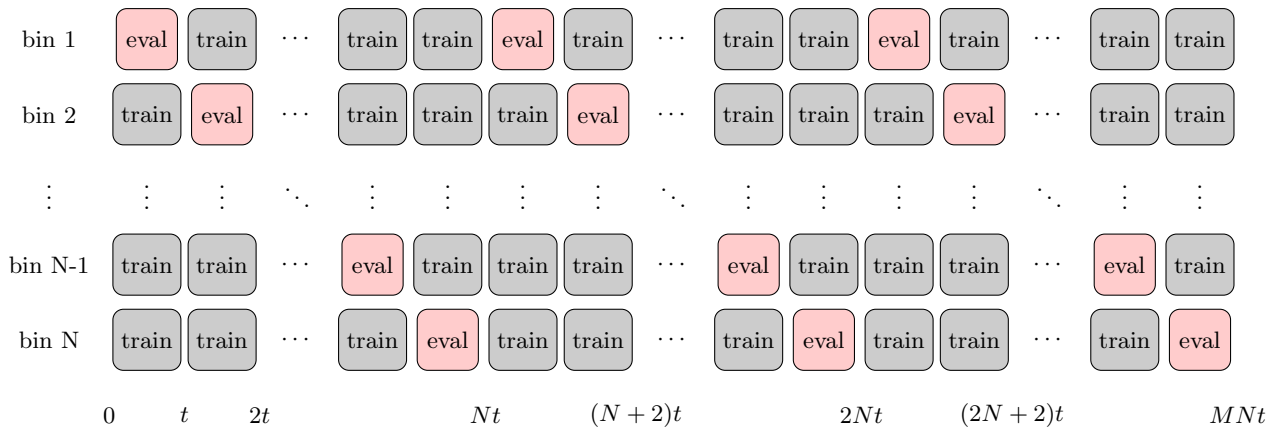
\begin{figure*}

    \begin{center}
    \begin{tikzpicture}

    \tikzstyle{grey-block} = [rectangle, draw, fill=black!20, text centered, rounded corners, minimum width=2.5em, minimum height=2.5em]
    \tikzstyle{red-block} = [rectangle, draw, fill=red!20, text centered, rounded corners, minimum width=2.5em, minimum height=2.5em]
    \tikzstyle{blank} = [rectangle, text centered, rounded corners]


    \node [red-block] (bin 1 1) {eval};
    \node [grey-block, right of=bin 1 1] (bin 1 2) {train};
    \node [blank, right of=bin 1 2] (bin 1 3) {$\cdots$};
    \node [grey-block, right of=bin 1 3] (bin 1 4) {train};
    \node [grey-block, right of=bin 1 4] (bin 1 5) {train};

    \node [red-block, right of=bin 1 5] (bin 1 6) {eval};
    \node [grey-block, right of=bin 1 6] (bin 1 7) {train};
    \node [blank, right of=bin 1 7] (bin 1 8) {$\cdots$};
    \node [grey-block, right of=bin 1 8] (bin 1 9) {train};
    \node [grey-block, right of=bin 1 9] (bin 1 10) {train};

    \node [red-block, right of=bin 1 10] (bin 1 11) {eval};
    \node [grey-block, right of=bin 1 11] (bin 1 12) {train};
    \node [blank, right of=bin 1 12] (bin 1 13) {$\cdots$};
    \node [grey-block, right of=bin 1 13] (bin 1 14) {train};
    \node [grey-block, right of=bin 1 14] (bin 1 15) {train};


    \node [grey-block, below of=bin 1 1] (bin 2 1) {train};
    \node [red-block, right of=bin 2 1] (bin 2 2) {eval};
    \node [blank, right of=bin 2 2] (bin 2 3) {$\cdots$};
    \node [grey-block, right of=bin 2 3] (bin 2 4) {train};
    \node [grey-block, right of=bin 2 4] (bin 2 5) {train};

    \node [grey-block, right of=bin 2 5] (bin 2 6) {train};
    \node [red-block, right of=bin 2 6] (bin 2 7) {eval};
    \node [blank, right of=bin 2 7] (bin 2 8) {$\cdots$};
    \node [grey-block, right of=bin 2 8] (bin 2 9) {train};
    \node [grey-block, right of=bin 2 9] (bin 2 10) {train};

    \node [grey-block, right of=bin 2 10] (bin 2 11) {train};
    \node [red-block, right of=bin 2 11] (bin 2 12) {eval};
    \node [blank, right of=bin 2 12] (bin 2 13) {$\cdots$};
    \node [grey-block, right of=bin 2 13] (bin 2 14) {train};
    \node [grey-block, right of=bin 2 14] (bin 2 15) {train};


    \node [blank, below of=bin 2 1] (bin vdots 1) {$\vdots$};
    \node [blank, below of=bin 2 2] (bin vdots 2) {$\vdots$};
    \node [blank, right of=bin vdots 2] (bin vdots 3) {$\ddots$};
    \node [blank, right of=bin vdots 3] (bin vdots 4) {$\vdots$};
    \node [blank, right of=bin vdots 4] (bin vdots 5) {$\vdots$};
    \node [blank, right of=bin vdots 5] (bin vdots 6) {$\vdots$};
    \node [blank, right of=bin vdots 6] (bin vdots 7) {$\vdots$};
    \node [blank, right of=bin vdots 7] (bin vdots 8) {$\ddots$};
    \node [blank, right of=bin vdots 8] (bin vdots 9) {$\vdots$};
    \node [blank, right of=bin vdots 9] (bin vdots 10) {$\vdots$};
    \node [blank, right of=bin vdots 10] (bin vdots 11) {$\vdots$};
    \node [blank, right of=bin vdots 11] (bin vdots 12) {$\vdots$};
    \node [blank, right of=bin vdots 12] (bin vdots 13) {$\ddots$};
    \node [blank, right of=bin vdots 13] (bin vdots 14) {$\vdots$};
    \node [blank, right of=bin vdots 14] (bin vdots 15) {$\vdots$};


    \node [grey-block, below of=bin vdots 1] (bin N-1 1) {train};
    \node [grey-block, right of=bin N-1 1] (bin N-1 2) {train};
    \node [blank, right of=bin N-1 2] (bin N-1 3) {$\cdots$};
    \node [red-block, right of=bin N-1 3] (bin N-1 4) {eval};
    \node [grey-block, right of=bin N-1 4] (bin N-1 5) {train};

    \node [grey-block, right of=bin N-1 5] (bin N-1 6) {train};
    \node [grey-block, right of=bin N-1 6] (bin N-1 7) {train};
    \node [blank, right of=bin N-1 7] (bin N-1 8) {$\cdots$};
    \node [red-block, right of=bin N-1 8] (bin N-1 9) {eval};
    \node [grey-block, right of=bin N-1 9] (bin N-1 10) {train};

    \node [grey-block, right of=bin N-1 10] (bin N-1 11) {train};
    \node [grey-block, right of=bin N-1 11] (bin N-1 12) {train};
    \node [blank, right of=bin N-1 12] (bin N-1 13) {$\cdots$};
    \node [red-block, right of=bin N-1 13] (bin N-1 14) {eval};
    \node [grey-block, right of=bin N-1 14] (bin N-1 15) {train};


    \node [grey-block, below of=bin N-1 1] (bin N 1) {train};
    \node [grey-block, right of=bin N 1] (bin N 2) {train};
    \node [blank, right of=bin N 2] (bin N 3) {$\cdots$};
    \node [grey-block, right of=bin N 3] (bin N 4) {train};
    \node [red-block, right of=bin N 4] (bin N 5) {eval};

    \node [grey-block, right of=bin N 5] (bin N 6) {train};
    \node [grey-block, right of=bin N 6] (bin N 7) {train};
    \node [blank, right of=bin N 7] (bin N 8) {$\cdots$};
    \node [grey-block, right of=bin N 8] (bin N 9) {train};
    \node [red-block, right of=bin N 9] (bin N 10) {eval};

    \node [grey-block, right of=bin N 10] (bin N 11) {train};
    \node [grey-block, right of=bin N 11] (bin N 12) {train};
    \node [blank, right of=bin N 12] (bin N 13) {$\cdots$};
    \node [grey-block, right of=bin N 13] (bin N 14) {train};
    \node [red-block, right of=bin N 14] (bin N 15) {eval};


    \node [blank, left of=bin 1 1, xshift=-0.3cm] (bin 1 name) {bin 1};
    \node [blank, left of=bin 2 1, xshift=-0.3cm] (bin 2 name) {bin 2};
    \node [blank, left of=bin vdots 1, xshift=-0.3cm] (bin vdots name) {$\vdots$};
    \node [blank, left of=bin N-1 1, xshift=-0.3cm] (bin 3 name) {bin N-1};
    \node [blank, left of=bin N 1, xshift=-0.3cm] (bin 4 name) {bin N};

    \node [blank, below of=bin N 1, xshift=-0.5cm] {$0$};
    \node [blank, below of=bin N 1, xshift=+0.5cm] {$t$};
    \node [blank, below of=bin N 2, xshift=+0.5cm] {$2t$};
    \node [blank, below of=bin N 3, xshift=+0.5cm] {};
    \node [blank, below of=bin N 4, xshift=+0.5cm] {};
    \node [blank, below of=bin N 5, xshift=+0.5cm] {$Nt$};
    \node [blank, below of=bin N 6, xshift=+0.5cm] {};
    \node [blank, below of=bin N 7, xshift=+0.5cm] {$(N+2)t$};
    \node [blank, below of=bin N 8, xshift=+0.5cm] {};
    \node [blank, below of=bin N 9, xshift=+0.5cm] {};
    \node [blank, below of=bin N 10, xshift=+0.5cm] {$2Nt$};
    \node [blank, below of=bin N 11, xshift=+0.5cm] {};
    \node [blank, below of=bin N 12, xshift=+0.5cm] {$(2N+2)t$};
    \node [blank, below of=bin N 13, xshift=+0.5cm] {};
    \node [blank, below of=bin N 14, xshift=+0.5cm] {};
    \node [blank, below of=bin N 15, xshift=+0.5cm] {$MNt$};

    \end{tikzpicture}
    \end{center}
    \caption{
        Schematic diagram showing data segmentation for acausal batch operation.
        Each row corresponds to one of $N$ bins, and each column corresponds to one of $N\times M$ segments.
        Data from sequential segments are assigned in order to different bins, effectively giving each bin access to samples from throughout the entire analysis period during training.
        Evaluation sets from each bin remain disjoint from training sets, though, allowing for meaningful cross-validation of classifier performance.
    }
    \label{figure:acausal segmentation}
\end{figure*}

\emph{Acausal} or \emph{round-robin} binning divides the data into a sequence of small segments, mixing samples independently of their time ordering.
Fig.~\ref{figure:acausal segmentation} demonstrates how segments are assigned to different data sets.
Briefly, for $N$ different bins with $M$ segments per bin, iDQ generates $N_\mathrm{segs}=N \times M$ segments of equal duration.
Data from the first segment is assigned to the first bin, the second segment to the second bin, and so on.
Data from the $(M+1)^\mathrm{th}$ segment is assigned to the first bin, the $(M+2)^\mathrm{th}$ segment to the second bin, and this repeats until all data has been assigned.
We then train over all data outside of the $i^\mathrm{th}$ bin to generate a model used to evaluate data inside the $i^\mathrm{th}$ bin, repeating the procedure for all bins.
This approach provides features drawn from consistent distributions in both training and evaluation, even if the instantaneous feature distributions change over time.
With features drawn from consistent distributions in both training and evaluation, acausal binning measures the best performance that should ever be expected from a classifier.

\begin{figure}

    \begin{center}
    \begin{tikzpicture}

    \tikzstyle{grey-block} = [rectangle, draw, fill=black!20, text centered, rounded corners, minimum width=2.5em, minimum height=2.5em]
    \tikzstyle{red-block} = [rectangle, draw, fill=red!20, text centered, rounded corners, minimum width=2.5em, minimum height=2.5em]
    \tikzstyle{empty-block} = [rectangle, draw=black!40, text centered, rounded corners, minimum width=2.5em, minimum height=2.5em]
    \tikzstyle{blank} = [rectangle, text centered, rounded corners]


    \node [grey-block, minimum width=5em] (bin 1 1) {train};
    \node [red-block, right of=bin 1 1, xshift=+0.4cm] (bin 1 2) {eval};
    \node [empty-block, right of=bin 1 2] (bin 1 3) {};
    \node [empty-block, right of=bin 1 3] (bin 1 4) {};
    \node [blank, right of=bin 1 4] (bin 1 cdots) {$\cdots$};
    \node [empty-block, right of=bin 1 cdots] (bin 1 N) {};

    \node [grey-block, minimum width=5em, below of=bin 1 1] (bin 2 1) {train};
    \node [grey-block, right of=bin 2 1, xshift=+0.4cm] (bin 2 2) {train};
    \node [red-block, right of=bin 2 2] (bin 2 3) {eval};
    \node [empty-block, right of=bin 2 3] (bin 2 4) {};
    \node [blank, right of=bin 2 4] (bin 2 cdots) {$\cdots$};
    \node [empty-block, right of=bin 2 cdots] (bin 2 N) {};

    \node [grey-block, minimum width=5em, below of=bin 2 1] (bin 3 1) {train};
    \node [grey-block, right of=bin 3 1, xshift=+0.4cm] (bin 3 2) {train};
    \node [grey-block, right of=bin 3 2] (bin 3 3) {train};
    \node [red-block, right of=bin 3 3] (bin 3 4) {eval};
    \node [blank, right of=bin 3 4] (bin 3 cdots) {$\cdots$};
    \node [empty-block, right of=bin 3 cdots] (bin 3 N) {};

    \node [blank, below of=bin 3 1] (bin vdots 1) {$\vdots$};
    \node [blank, below of=bin 3 2] (bin vdots 2) {$\vdots$};
    \node [blank, right of=bin vdots 2] (bin vdots 3) {$\vdots$};
    \node [blank, right of=bin vdots 3] (bin vdots 4) {$\vdots$};
    \node [blank, right of=bin vdots 4] (bin angledots) {$\ddots$};
    \node [blank, right of=bin angledots] (bin vdots N) {$\vdots$};

    \node [grey-block, minimum width=5em, below of=bin vdots 1] (bin N 1) {train};
    \node [grey-block, right of=bin N 1, xshift=+0.4cm] (bin N 2) {train};
    \node [grey-block, right of=bin N 2] (bin N 3) {train};
    \node [grey-block, right of=bin N 3] (bin N 4) {train};
    \node [blank, right of=bin N 4] (bin N cdots) {$\cdots$};
    \node [red-block, right of=bin N cdots] (bin N N) {eval};


    \node [blank, left of=bin 1 1, xshift=-0.3cm] (bin 1 name) {bin 1};
    \node [blank, left of=bin 2 1, xshift=-0.3cm] (bin 2 name) {bin 2};
    \node [blank, left of=bin 3 1, xshift=-0.3cm] (bin 3 name) {bin 3};
    \node [blank, left of=bin vdots 1, xshift=-0.3cm] (bin vdots name) {$\vdots$};
    \node [blank, left of=bin N 1, xshift=-0.3cm] (bin 3 name) {bin N};

    \node [blank, below of=bin N 1, xshift=-0.9cm] {$-T$};
    \node [blank, below of=bin N 1, xshift=+0.9cm] {$0$};
    \node [blank, below of=bin N 2, xshift=+0.5cm] {$t$};
    \node [blank, below of=bin N 3, xshift=+0.5cm] {$2t$};
    \node [blank, below of=bin N 4, xshift=+0.5cm] {$3t$};
    \node [blank, below of=bin N N, xshift=+0.5cm] {$Nt$};

    \node [blank, above of=bin 1 1] {initial};
    \node [blank, above of=bin 1 1, yshift=-0.35cm] {lookback};


    \end{tikzpicture}
    \end{center}
    \caption{
        Schematic of data segmentation for causal batch analyses.
        Each row illustrates the data included in a single bin, and columns represent different segments of data.
        Each bin uses historical data cumulatively when training (grey shaded boxes), including an initial lookback period before the first evaluation segment (red boxes).
        Bins progressively train on more and more time-ordered data, thereby testing algorithmic sensitivity to detector non-stationarity in a similar way to what is experienced during streaming operation.
    }
    \label{figure:causal segmentation}
\end{figure}
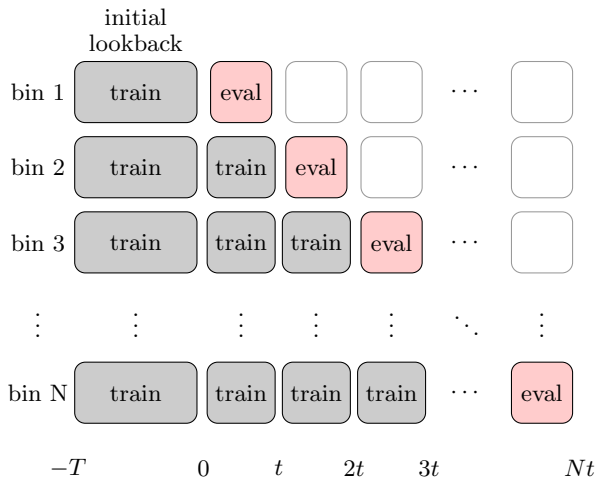

\emph{Causal} binning again divides the data into many small segments.
Fig.~\ref{figure:causal segmentation} shows this schematically.
Each analysis then trains on a cumulative set of historical segments and uses the resulting model to evaluate the next segment, preserving the relative time ordering.
Again, this is repeated for all segments, each using all historical data available during training.
This allows users to investigate how detector non-stationarity affects their algorithm.

Each approach has its uses, and both produce ROC curves that measure classifiers' performance in a fair manner.
Importantly, iDQ can simultaneously run multiple classifiers, guaranteeing they see identical data sets and that comparisons are as fair as possible.
Evaluation also records each model's hash within each feature vector to maintain a record of how each feature vector was classified.

\subsection{Calibration}\label{sec:calibration}

\begin{figure*}
    \begin{minipage}{0.49\textwidth}
        \begin{center}
        \includegraphics[width=1.0\textwidth, clip=True, trim=1.0cm 0.5cm 9.93cm 1.25cm]{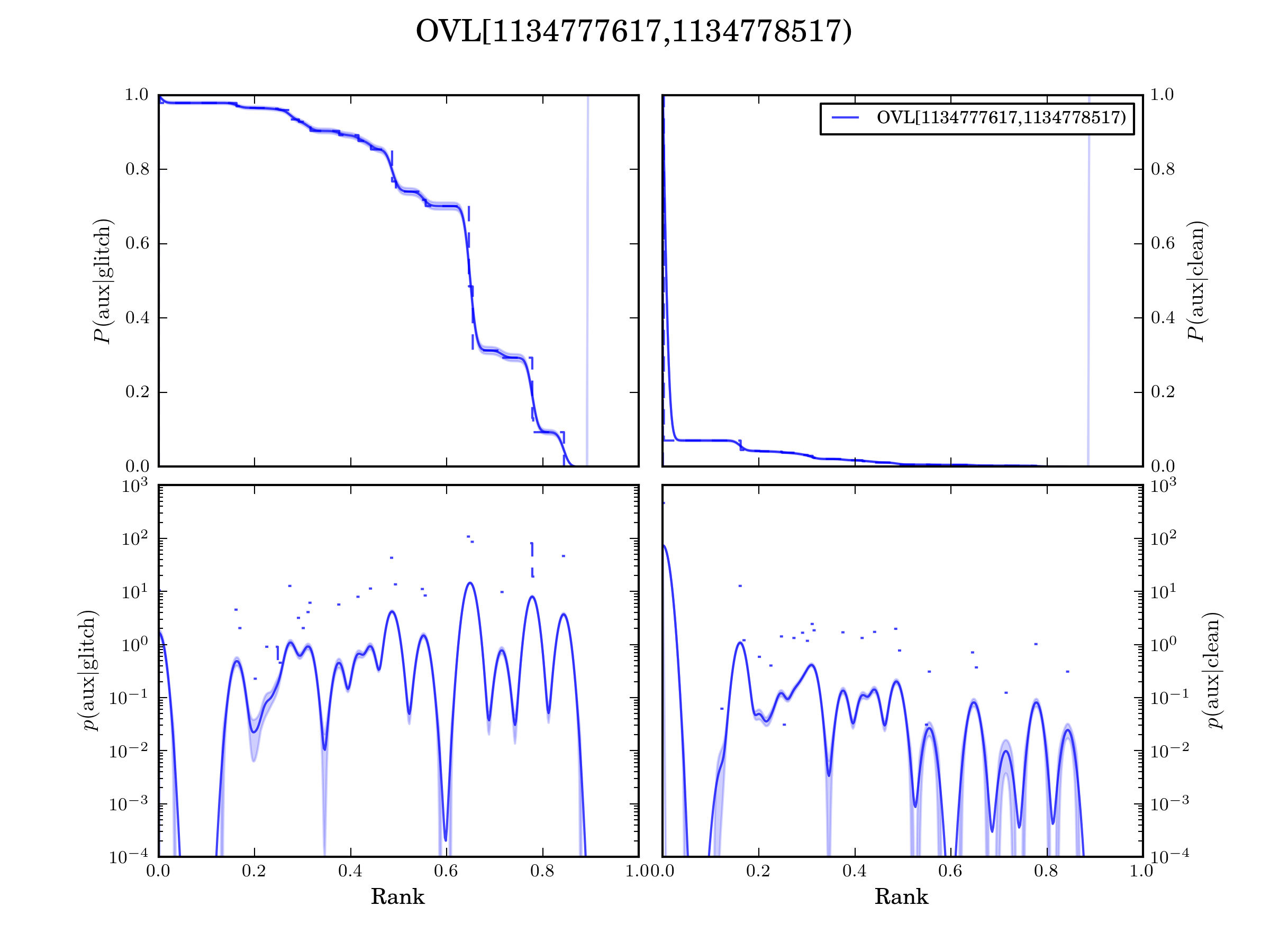}
        \end{center}
    \end{minipage}
    \begin{minipage}{0.49\textwidth}
        \begin{center}
        \includegraphics[width=1.0\textwidth, clip=True, trim=1.0cm 0.5cm 9.85cm 1.25cm]{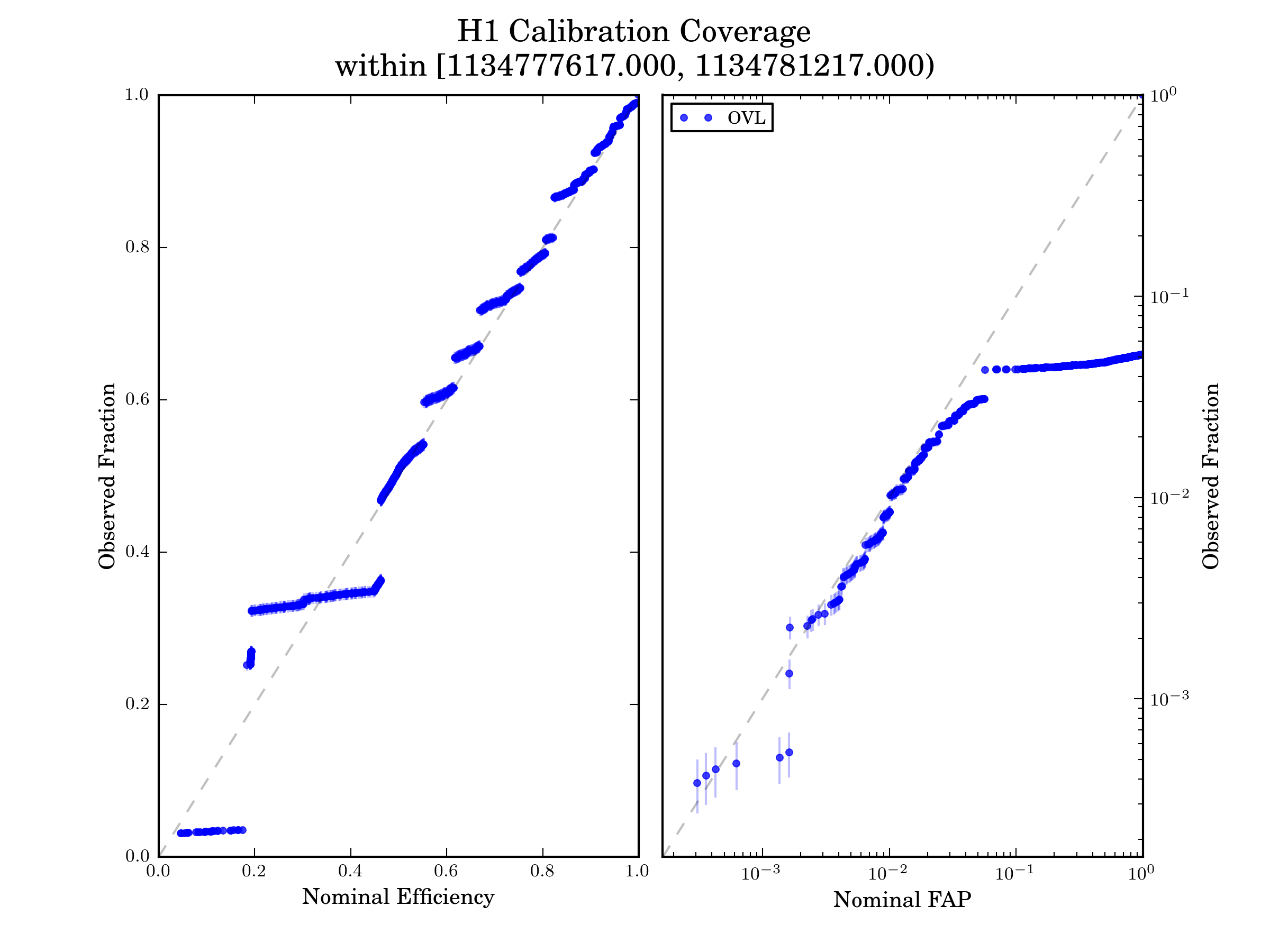}
        \end{center}
    \end{minipage}
    \caption{
        (\emph{left}) Example calibration distribution representing $p(\mathcal{M}(\vec{a})|G)$ from a batch analysis during a period of elevated radio-frequency noise at the LIGO Hanford detector on Monday December 22, 2015 (see Fig.~\ref{fig:example roc}).
        Solid lines denote the optimized Gaussian KDE representing the distributions; shaded regions correspond to the expected error in that fit; and dashed lines show histograms of the observed ranks.
        A cumulative histogram is shown above the differential histogram as it more clearly shows the rounding effects of the finite bandwidth assumed within the KDE model.
        (\emph{right}) Calibration coverage, showing the approximate fraction of glitches assigned a particular nominal efficiency the calibration map.
        Perfect calibration corresponds to a diagonal line.
        The step-like behavior observed here comes from interpolation artifacts when assigning nominal efficiencies to glitches based on the regularly sampled timeseries.
    }
    \label{figure:example calibration distributions}
\end{figure*}

While training and evaluation are the foundation of iDQ's supervised learning approach, calibrating the resulting ranks into probabilistic statements is of equal importance.
iDQ does this by directly modeling the observed conditioned likelihoods for each classifier's rank: $p(r_\alpha|G)$ and $p(r_\alpha|C)$.
Just like each classifier's model retains a hash to track provenance, each pair of conditioned likelihoods, or \emph{calibration map}, records a unique hash to denote the evaluated samples from which it was generated.

iDQ can model the conditioned likelihoods in two ways, related to different assumptions about the nature of ranks produced by a classifier's model.
Both quantify the uncertainty in their estimates, and both work well in practice.
Figure~\ref{figure:example calibration distributions} shows how they correctly model non-trivial conditioned likelihoods.
Calibration maps further estimate prior odds for $G$ and $C$ samples.
We describe these procedures in more detail below.

\subsubsection{Continuous Calibration Maps}
\label{sec:continuous calibration maps}

If classifiers produce ranks that can take any real value within the unit interval, with no discrete lumps of probability so that
\begin{equation}
    \lim\limits_{\varepsilon\rightarrow0}\left(\int\limits_{r-\varepsilon}^{r+\varepsilon} dx\, p(x|X)\right) \propto \varepsilon\quad \forall\ r\in[0,1],
\end{equation}
then it is appropriate to model the underlying distribution with a continuous kernel density estimate (KDE).
iDQ implements a Gaussian kernel and dynamically optimizes the kernel's bandwidth to maximize a cross-validation likelihood quantifying how well the KDE reproduces the observed sample set.
iDQ reflects observed samples around rank=0 and 1 to avoid edge effects within the KDE while numerically enforcing proper normalization.
Uncertainty in the KDE model for the true likelihood at each rank is modeled by a $\beta$-distribution (see Appendix~\ref{sec:kde}).
This procedure accurately predicts the observed variance in KDEs obtained from different realizations of sample sets drawn from the same underlying distributions, regardless of the true distribution.

\subsubsection{Discrete Calibration Maps}
\label{sec:discrete calibration maps}

If the trained model only produces a finite number of possible ranks, the resulting distribution may be better modeled as a weighed sum of $\delta$-functions
\begin{equation}
    p(r|X) = \sum\limits_i w_i \delta(r-r_i)\quad \left|\quad \sum\limits_i w_i = 1 \right.
\end{equation}
equivalent to our continuous Gaussian KDE in the limit of vanishingly small bandwidths.
Weights are estimated as the fraction of observed samples assigned to that rank, and uncertainty in the weights is again modeled as a $\beta$-distribution such that
\begin{equation}
    p(w_i|n_i, N) \propto w_i^{n_i}\left(1-w_i\right)^{N-n_i}
\end{equation}
where $n_i$ out of $N$ total samples were assigned rank $r_i$.

\subsubsection{Prior Odds}
\label{sec:prior odds}

While iDQ's conditioned likelihoods, and therefore the likelihood ratio $\Lambda^G_C$, do not depend on the prior odds between $G$ and $C$, most applications instead rely on $p_G$ (Eqn.~\ref{eqn:p_G}).
While $p_G$ is monotonic in $\Lambda^G_C$, the exact value can be made arbitrarily large or small based on the prior odds assumed.
This forces us to carefully consider our assumptions \textit{a priori} about the relative frequencies of $G$ and $C$.
iDQ implements several choices, either allowing users to specify fixed prior odds, estimating them based on the relative fractions of samples within training sets, or estimating them based on the fraction of time declared clean within a training set.
All these are based on the premise that the prior odds are approximately the ratio of the rates at which each type of sample occurs, adopting different techniques for approximating the rates of $G$ ($R_G$) and $C$ ($R_C$) samples.

As appropriate for inference over tabulated data, one approximation is $R_G/R_C \approx N_G/N_C$.
This asks what the chance is that one would select either a $G$ or $C$ sample when randomly choosing an element of the fixed training set.
Generally, $N_G$ is set by the true $R_G$ in the detector and the amount of time over which we collect samples.
Similarly, $N_C$ depends on the rate at which we generate clean samples, which is an arbitrary choice typically chosen to balance training accuracy and computational expense.
While formally correct for tabulated data, and therefore useful in some applications, this model does not necessarily represent the prior odds relevant for timeseries production.

Alternatively, we can model $R_G/R_C\approx T_G/T_C = (T/T_C) - 1$ where $T_C/T$ is the fraction of analysis time declared clean when constructing the sample set.
We measure $T_C$ directly from the conditions defining $C$ based on $h$ (Section~\ref{sec:data discovery}).
This approach models the relative frequency of times declared glitchy or clean, rather than selecting an element from tabular data, as is more appropriate for timeseries production.

\subsection{Timeseries Production}\label{sec:timeseries}

iDQ produces streaming estimates of statistical quantities as the culmination of training and calibration.
These timeseries should be thought of as the main data product generated within iDQ and are the most applicable to GW searches.
iDQ generates vectorized feature sets on a regular grid in time, typically sampled at $\geq 128\,\mathrm{Hz}$.
The regularly spaced vectors are then evaluated using a trained model, and the resulting array of ranks is calibrated into several statistical quantities using a calibration map.
These quantities are then distributed in real-time to GW searches, and we discuss ways to incorporate them within searches in Section~\ref{sec:applications}.

Section~\ref{sec:batch vs stream} describes the differences between offline (\emph{batch}) and online (\emph{stream}) modes of operation in more detail, but both utilize asynchronous processes to manage training, evaluation, calibration, and timeseries production.
Because some tasks require the output from other tasks, iDQ synchronizes them by polling for specific models and calibration maps from various repositories, referred to as \emph{modular data servers} (MDSs).
In this way, timeseries jobs can obtain the most relevant model and calibration map for any stretch of data.
As with data discovery, iDQ does not depend on the particular implementation of a MDS as long as it allows tasks to \emph{get} and \emph{put} results through a consistent interface.
\textit{De facto}, most MDSs are implemented via local filesystems.

As noted in Section~\ref{sec:calibration}, the prior odds assumed during calibration can have a significant impact on the interpretability of the resulting timeseries.
In particular, we expect $p(G) \ll p(C)$ because the typical duration of non-Gaussian noise transients is usually much less than their separation: $\mathcal{O}(10^{-1}\,\mathrm{sec}) \ll \mathcal{O}(10^2\,\mathrm{sec})$.

Because of the ambiguity associated with the choice of prior odds, iDQ produces timeseries for multiple statistical quantiles besides $p_G$, including $\Lambda^G_C$, the conditioned survival functions (\emph{efficiency} and \emph{false alarm probability}), as well as the raw rank produced by the model.
Access to $\Lambda^G_C$ allows users to estimate $p_G$ with whatever prior odds they choose.
Furthermore, the \emph{false alarm probability} approximates the amount of time discarded by the classifier, and therefore could be used to set a convenient working point.
What's more, the calibration map also provides uncertainty estimates based on the finite calibration sample size.

\begin{figure*}
    \includegraphics[width=\textwidth, clip=True, trim=1.3cm 1.2cm 1.7cm 1.25cm]{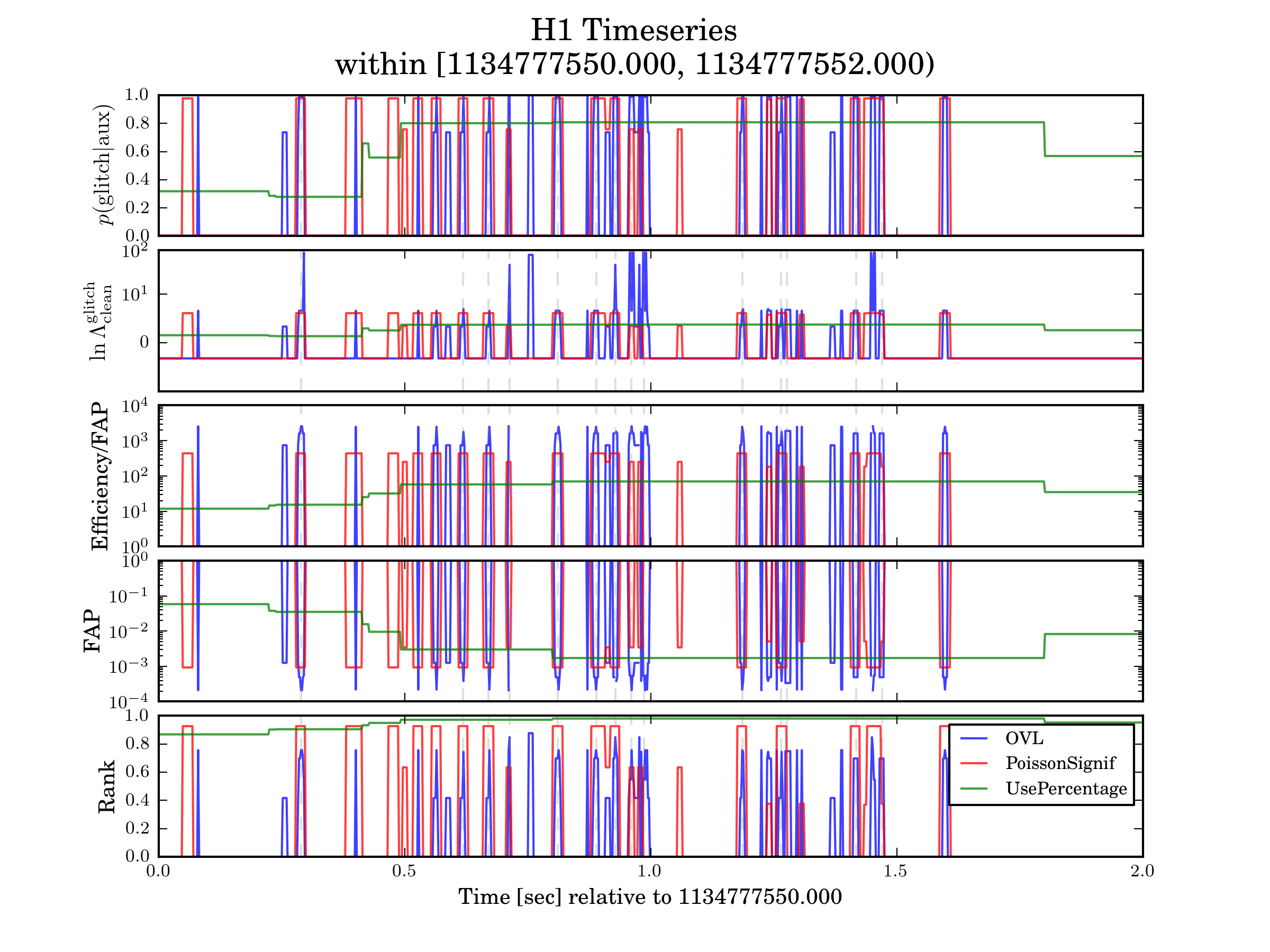}
    \begin{center}
        Time [sec] 
    \end{center}
    \caption{
        Example iDQ timeseries for OVL~\cite{Essick2013} with 3 different ranking metrics (see Appendix~\ref{sec:OVL}) from a batch analysis of $\mathcal{O}(10^2)$ sec during a period of elevated radio-frequency noise at the LIGO Hanford detector on Monday December 21, 2015.
        From the top to the bottom, we show the estimated probability of a glitch given the auxiliary data, the log likelihood-ratio between the glitch and clean models, the ratio of the detection efficiency to the false alarm probability (FAP), the FAP itself, and the raw rank produced by the classifier.
        For reference, central times for glitches derived from $h$ are shown with light grey lines, occuring at a rate of $\sim10\,\mathrm{Hz}$ at this time.
        We see that OVL correctly identifies the times at which glitches are present in $h$ and that our prior ($p(C)\gg p(G)$) correctly drive $p_G$ to small values when the likelihood is uninformative.
    }
    \label{figure:example timeseries}
\end{figure*}

\section{Batch vs. Stream modes}\label{sec:batch vs stream}

In addition to the decomposition described in Section~\ref{sec:decomposition}, iDQ supports two modes of operation related to how it synchronizes processes.

The offline, or \emph{batch}, mode targets specific stretches of data and can support both \emph{causal} and \emph{acausal} binning schemes (Section~\ref{sec:evaluation}).
Additionally, \emph{batch} jobs run tasks synchronously within each bin, which is to say that training must complete before evaluation begins, evaluation must complete before calibration begins, and calibration must complete before timeseries are produced.
However, separate bins are independent and can be processed in parallel.
Batch analyses have loose latency requirements.

The online, or \emph{stream}, mode instead runs in low-latency, typically producing timeseries within $\mathcal{O}(10^{-1}\,\mathrm{sec})$ of receiving features.
The dominant source of latency for iDQ is feature generation and vectorization.
The time required for feature extractors to process $\mathcal{O}(10^3)$ channels has recently been reduced to $\sim5\,\mathrm{sec}$ ~\cite{godwin-thesis}, as opposed to $\sim32\,\mathrm{sec}$ for the implementation of the Haar transform (KleineWelle~\cite{kleine-welle}) used throughout the first two observing runs.
Vectorization can also limit iDQ's latency, but the pipeline can generate vectors consisting of $\mathcal{O}(5)$ features for each of $\mathcal{O}(10^3)$ channels at rates above $128\,\mathrm{Hz}$, which is typical of production configurations as most glitches have durations $\sim 100\,\mathrm{ms}$.

Online jobs process all data causally and manage tasks asynchronously.
However, because some tasks require input from others before they can begin, the streaming iDQ pipeline will run small batch pipelines if models and/or calibration maps are not initially available for all classifiers.
Once initial models and calibration maps are available, separate processes for training, evaluation, calibration, and timeseries production run in parallel and interact through put and get requests in MDSs (Section~\ref{sec:timeseries}).
This means that re-training and re-calibration happen continuously, with a new data set defined and fed to classifiers as soon as they finish processing their previous sets.
When each task begins processing a new data set, it polls the relevant MDS to obtain all the required data products without waiting for the asynchronous processes to complete.
For example, several evaluation strides may use the same set of models because the training jobs generally take longer to complete than evaluation.
Nonetheless, as soon as a training job completes, the evaluation jobs will automatically retrieve the new model.

Training jobs usually take the longest to complete, with runtimes of $\mathcal{O}(\mathrm{hours})$, and it is possible, then, that trained models will respond relatively slowly to detector non-stationarity.
However, if the model becomes out-of-date, the much faster evaluation and calibration jobs, which complete in $\mathcal{O}(\mathrm{sec})$, will detect the decreased performance and update iDQ's probabilistic statements accordingly.
For example, if a model suddenly cannot distinguish between $G$ and $C$ samples, the calibration jobs will update the conditioned likelihoods to $p(r|G)\sim p(r|C)$, and therefore return the prior odds.

Asynchronous, parallel processing lowers the latency required to produce timeseries at the cost additional complexity in data discovery.
Because data may occasionally be dropped before it reaches iDQ, each process analyzes data in small strides and internally manages timeout logic, skipping data if it takes too long to arrive.
In production, we generally find iDQ achieves duty cycles above \result{$99\%$}, thereby essentially guaranteeing results will be reliably provided to GW searches in low-latency.
In fact, iDQ's information was distributed at the same time as the calibrated GW strain during the third observing run.

While many analyses will utilize the batch workflow, we expect streaming processes to be the most relevant in the coming years as iDQ's predictions are incorporated further into low-latency GW searches.
Indeed, the examples in Figs.~\ref{fig:example gw170817} and~\ref{fig:example whistle} were all derived from streaming analyses.
Section~\ref{sec:applications} enumerates a few other possible applications.

\section{Examples}\label{sec:examples}

\begin{figure*}
    \includegraphics[width=1.0\textwidth]{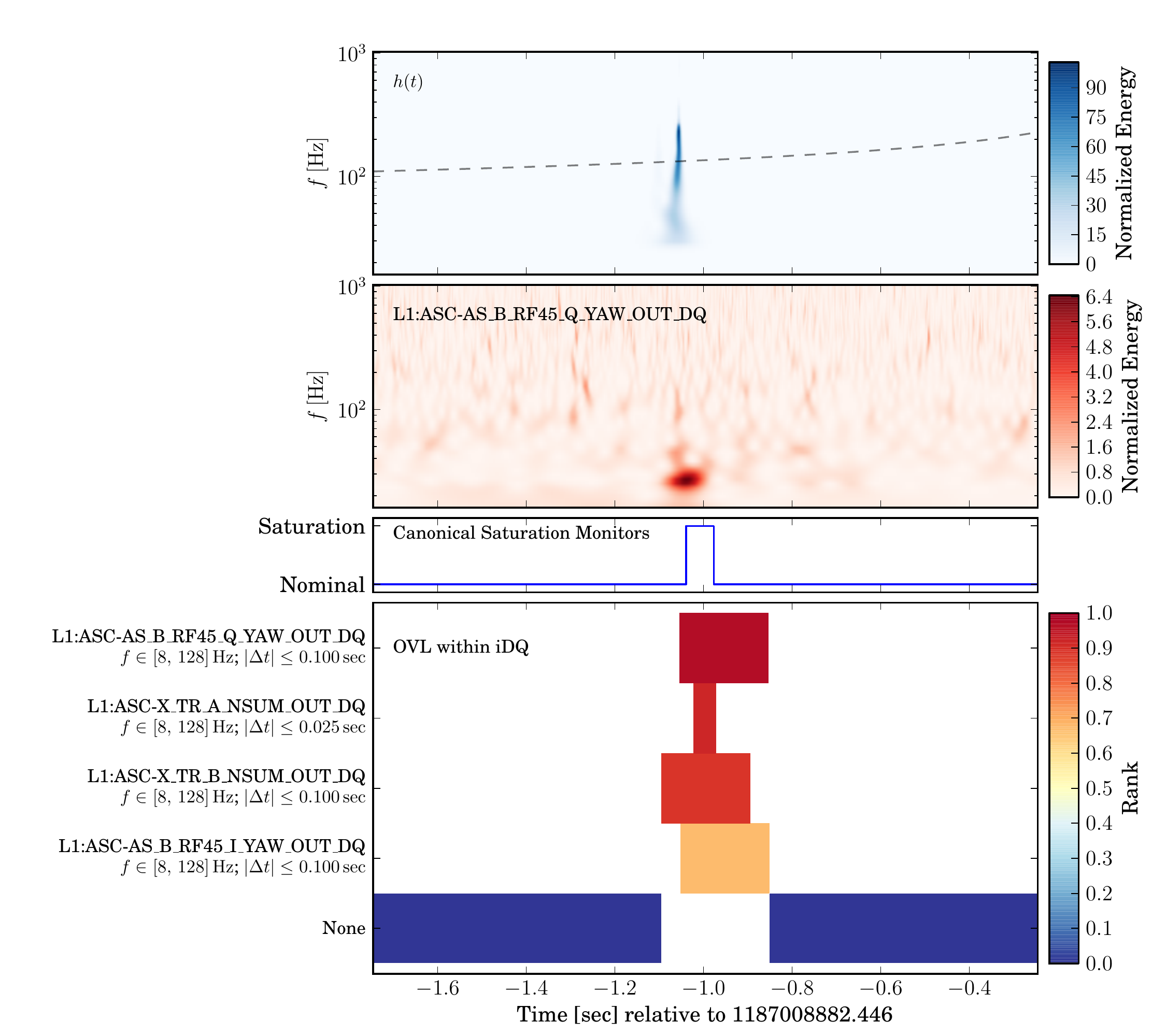}
    \caption{
        iDQ's low-latency predictions surrounding the non-Gaussian noise transient coincident with GW170817 in the LIGO Livingston detector.
        (\emph{top panel}) Time-frequency decomposition of the GW strain chosen to highlight the noise transient's short duration; GW170817's inspiral track is shown for reference (\emph{dashed line}), with times measured relative to the coalescence time.
        (\emph{2$^\mathrm{nd}$ panel}) Time-frequency representation of the top-ranked auxiliary witness active at this time, autonomously identified by iDQ as correlated with non-Gaussian noise without \textit{a priori} knowledge about the type of noise present in the detectors.
        (\emph{3$^\mathrm{rd}$ panel}) A canonical saturation monitor, which identified the time as problematic.
        (\emph{bottom panel}) OVL's feature importance, showing multiple veto configurations (Appendix~\ref{sec:OVL}) active in coincidence with the glitch.
        Color denotes OVL's rank, with rank$\rightarrow1$ indicating high confidence in the presence of a glitch.
    }
    \label{fig:example gw170817}
\end{figure*}

\begin{figure*}
    \includegraphics[width=1.0\textwidth]{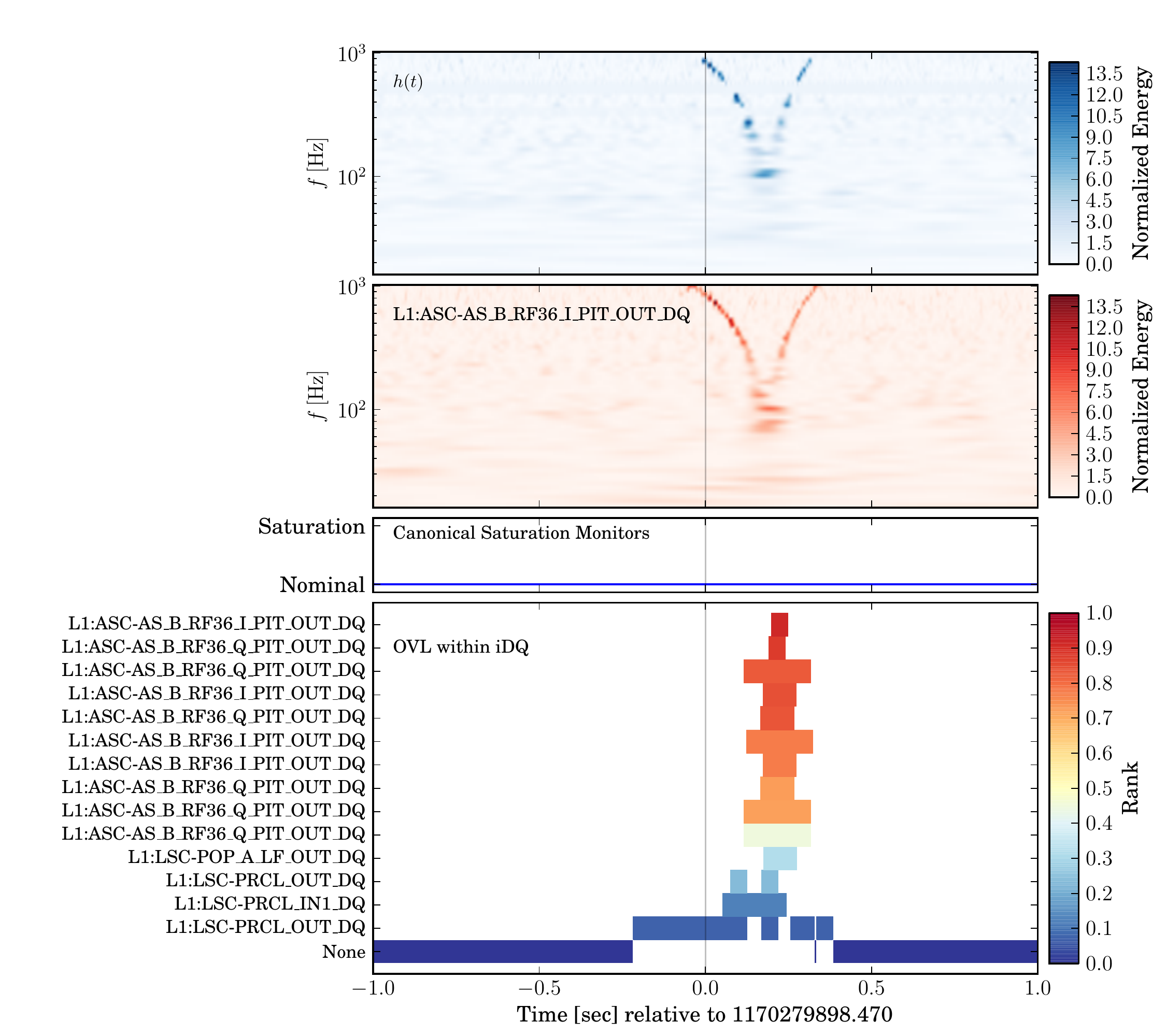}
    \caption{
        Example iDQ output surrounding a GW candidate identified by a search for unmodeled transients~\cite{Lynch2017} on 4 February 2017.
        (\emph{top panel}) A time-frequency representation of $h(t)$, showing the radio-frequency \emph{whistle} in coincidence with the candidate.
        (\emph{2$^\mathrm{nd}$ panel}) A time-frequency representation of the highest-ranked auxiliary channel active at this time, showing a similar whistle.
        (\emph{3$^\mathrm{rd}$ panel}) Canonical saturation monitors, inactive during the whistle, as expected.
        Canonical monitors for whistles do not exist in low-latency and can often miss whistles even in high latency.
        (\emph{bottom panel}) OVL's feature importance for active auxiliary channels.
        Color indicates OVL's rank, with rank$\rightarrow1$ corresponding to $p_G\rightarrow1$.
        We see that iDQ clearly identified the whistle in low-latency while canonical monitors were silent or otherwise unavailable.
    }
    \label{fig:example whistle}
\end{figure*}

We present a few examples of iDQ's behavior with real detector data from the first two observing runs.
First, and perhaps most importantly given the context, is the non-Gaussian noise artifact coincident with GW170817 in the LIGO Livingston interferometer ~\cite{Pankow:2018qpo}.
Although similar noise transients are witnessed several times per day in each IFO, their exact cause is not known.
They are, however, often associated with saturations within the interferometric control systems, and monitors exist to flag such saturations.
Figure~\ref{fig:example gw170817} shows a time-frequency projection highlighting the noise transient's short duration, as well as the behavior of the canonical monitors for saturations.
iDQ, at the same time and without prior knowledge of the existence of saturations or which auxiliary degrees of freedom correlate with noise in $h$, autonomously identified witnesses for such events and flagged the time as very likely to be a glitch in real-time.
This information was automatically made available within \result{8 sec} of the candidate being reported to the Gravitational Wave Candidate Event DataBase (GraceDB~\cite{gracedb}), thereby informing decisions in real-time about the candidate's probability of being astrophysical in origin and the resulting announcement to the broader astronomical community~\cite{GW170817GCN}.
GW170817 serves as an example of how iDQ can independently identify noise sources already known to human analysts.
Additionally, the witnesses iDQ identifies sometimes flag problematic time associated with loud noise transients that go unnoticed by more conventional monitors, including saturation-like glitches that happen to not saturate the control signals being monitored (e.g.~\cite{S190822cGCN}).

Pursuing this further, Figure~\ref{fig:example whistle} presents a radio-frequency \emph{whistle} identified as a possible GW candidate by oLIB, a search for unmodeled GW bursts~\cite{Lynch2017}, on 4 February 2017.
iDQ vetoed this event within \result{7 seconds}.
As with GW170817, iDQ autonomously developed witnesses for such noise and clearly identifies the time as glitchy.
We note that canonical monitors for saturations did not flag this event, as expected, and iDQ's low-latency predictions were the only data quality products available at the time that could reject this candidate as noise without relying on the human inspection of the signal morphology in $h(t)$.
Rejecting candidates from unmodeled transient searches based on $h(t)$ morphology itself is risky, and therefore iDQ's auxiliary witnesses provide much greater confidence the transient was of terrestrial origin.
At the time of writing, no low-latency monitors exist for such radio-frequency whistles besides iDQ.

Figures~\ref{fig:example gw170817} and~\ref{fig:example whistle} show two examples of noise transients that are typically witnessed well by the auxiliary degrees of freedom used within iDQ.
Both were identified in low-latency the OVL algorithm (\cite{Essick2013}; Appendix~\ref{sec:OVL}) running within iDQ's framework.
We typically find that, in agreement with Ref.~\cite{Biswas2013}, OVL performs as well as, if not better than, more complex algorithms, and we primarily uses its predictions to identify noise in low-latency.
In general, different classifiers may witness different noise sources, and combining the ranks from multiple classifiers (creating a \emph{boosted classifier}) is an active area of research.

\begin{figure*}
    \includegraphics[height=0.4\textwidth, clip=True, trim=0.65cm 0.40cm 0.30cm 0.50cm]{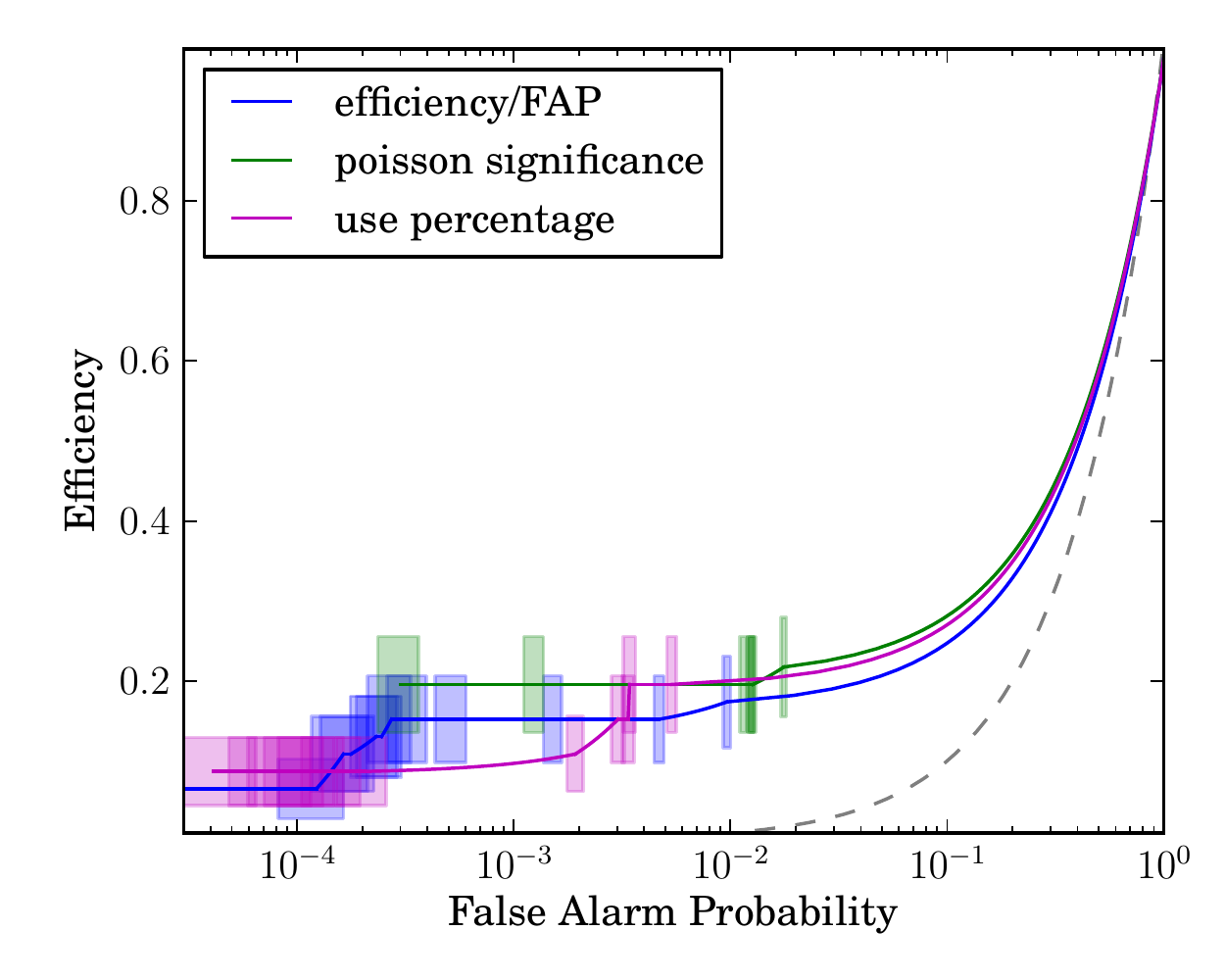}
    \includegraphics[height=0.4\textwidth, clip=true, trim=1.85cm 0.40cm 0.30cm 0.50cm]{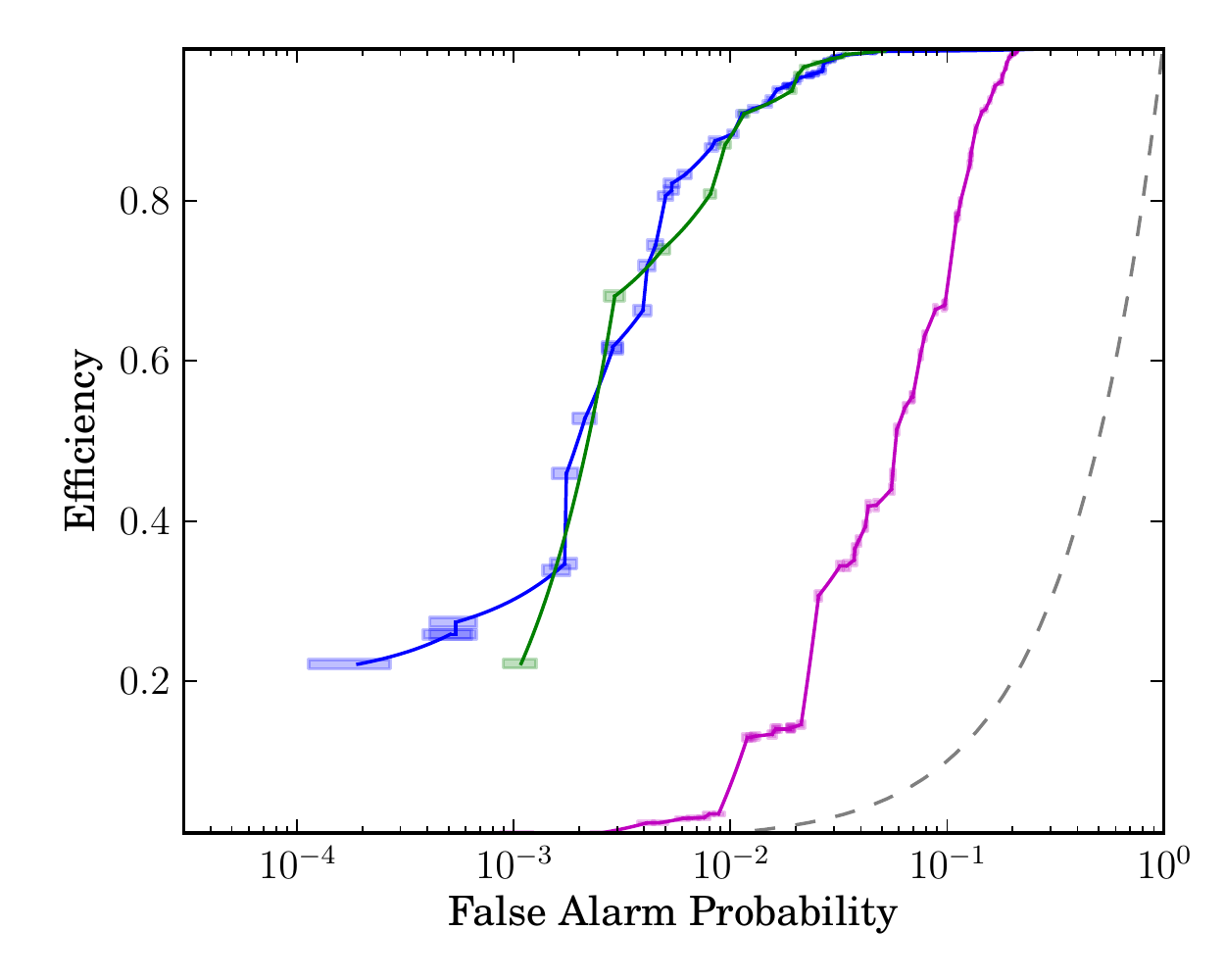}
    \caption{
        Example receiver operating characteristic (ROC) curves generated within iDQ relating the glitch detection efficiency to the false alarm probability.
        (\emph{left}) ROC curves from a batch analysis of LIGO Hanford data from December 18--20, 2015 ($\sim$2 days, 101 glitches).
        (\emph{right}) A batch analysis of LIGO Hanford data from December 22, 2015 (1 hour, 5504 glitches).
        Shaded boxes approximate 1-$\sigma$ uncertainty from the counting experiment used to measure the ROC curves at the points where they are measured, with linear interpolation between.
        Grey dashed lines correspond to uninformative classifier, and colored lines correspond to different ranking metrics within OVL (Appendix~\ref{sec:OVL veto performance metrics}).
        Each of these analyses used iDQ's internal acausal cross-validation scheme with two bins and two segments per bin, and these ROC curves therefore account for generalization error.
        These two examples represent approximate limits on the performance seen within iDQ, and primarily differ by the type of noise present within the detector.
        Nonstationary changes in non-Gaussian noise can be sudden and severe, necessitating automatic re-training and re-calibration.
    }
    \label{fig:example roc}
\end{figure*}

Figure~\ref{fig:example roc} presents a few ROC curves showing iDQ's typical performance, also demonstrating iDQ's ability to run multiple classifiers simultaneously over the same data.
We focus on OVL with 3 different choices for the ranking metric (Appendix~\ref{sec:OVL veto performance metrics}).
OVL, running within iDQ's framework, typically identifies \result{$\sim10\%$--$40\%$} of non-Gaussian noise artifacts (KleineWelle~\cite{kleine-welle} triggers between 32 and 2048 Hz with $\rho\gtrsim8$) at the cost of \result{$\lesssim0.1\%$--$1\%$} false alarm probability.
This depends on the mixture of noise sources present within the detectors, though, as a high fraction of well-witnessed noise will lead to correspondingly more impressive ROC curves.
Figure~\ref{fig:example roc} shows one such example from the LIGO Hanford detector a few days before GW151226~\cite{Abbott:2016nmj}.
An intense radio-frequency \emph{glitch-storm} produced a large number of clearly witnessed noise transients, and OVL identified $\geq90\%$ of them with $\leq1\%$ false alarm probability.
It is worth noting that, at that time, offline canonical monitors for radio-frequency noise had become less effective due to detector non-stationarity~\cite{PhysRevLett.116.241103, Abbott_2018}.
iDQ automatically detected new witnesses without human intervention and retained a high glitch-detection efficiency in low-latency.

\section{Applications within Gravitational-Wave Searches}\label{sec:applications}

The optimal incorporation of probabilistic data quality information within GW searches remains largely unaddressed in the literature.
In this section, we present a likelihood ratio test that incorporates imperfect knowledge of the presence of non-Gaussian noise within our detectors.
We first review the current state of the field and how data quality information is often incorporated into searches, demonstrating how iDQ's products could be used within existing methodologies while discussing different approaches' relative advantages and drawbacks.
We then formulate a search based on first-principles models of noise processes within detectors and the incorporation our imperfect knowledge of data quality based on both auxiliary and GW strain information.

\subsection{Current Veto and Gating Strategies}\label{sec:current applications}

Data quality products are currently applied within searches in two main ways.
Data quality \emph{vetoes} are applied after filtering, meaning after a search has produced a list of candidates.
Usually, vetoes are specified as a list of segments and any candidate that falls within these segment is rejected.
\emph{Gating}, on the other hand, attempts to remove problematic data before filtering, thereby preventing false positives from appearing in candidate lists at any point.
Importantly, both approaches assume \emph{binary} data quality information.
That is, the data is either declared clean or glitchy with complete certainty.
iDQ extends this by providing probabilistic measures of data quality.

Let us begin with vetoes applied after candidates have already been identified.
A naive approach is to perform a simple coincidence experiment between the search's candidates and data quality monitors.
If, for example, iDQ's false alarm rate timeseries dips below a threshold anywhere within a coincidence window surrounding a GW candidate, this may indicate that a non-Gaussian noise artifact is present near the candidate, which may suggest it is of terrestrial origin.
Other methods to identify problematic data based on auxiliary channels, often constructed by hand, are often used to define such veto segments.
Of course, this depends on the precise way GW candidate reference times are recorded, as low-mass compact binary coalescences can sometimes coalesce several seconds after the non-Gaussian noise that caused the false alarm.
In effect, this maps timeseries output into \emph{binary veto segments} with a window and a threshold.

iDQ's output was used in this way with modest success during the first two observing runs~\cite{O1O2Emfollow}, but vetoes suffer from several limitations.
First, searching for extreme excursions in any of iDQ's timeseries within a coincidence window naturally introduces an additional trials factor.
For example, if the coincidence window is longer than the typical separation between glitches, then the probability of obtaining $p_G\sim1$ for at least one point in that window is almost surely $1$.
This complicates the statistical interpretation of iDQ's predictions, since, for example, what iDQ reports as the false alarm probability will not generally correspond to the false alarm probability of finding a large excursion within a large window.
Furthermore, the mapping from threshold--window pairs to the effective false alarm probability will depend strongly on the quantity used, and there is no single obvious choice.
One may threshold on the false alarm probability in an attempt to bound the probability of false alarms, but one may alternatively threshold on the likelihood ratio $\Lambda^G_C$ as this may be more appropriate for likelihood ratio tests.
This could be tuned by hand, but that then negates any efforts to calibrate model predictions within iDQ, thereby defeating the purpose of a large part of the pipeline.
Nonetheless, the conceptual simplicity and ease of implementation make this attractive for practical applications.
For this reason, veto segments remain a core data quality product within GW searches.

A perhaps more sophisticated approach to vetoing based on extremized timeseries is to incorporate the extremum value within a window as part of a likelihood ratio test.
While this may remove some ambiguity about the effective false alarm probability (the likelihood ratio test will naturally account for the probability of seeing such extrema within \emph{clean} data) and remove the need to tune thresholds by hand, the ambiguities associated with the choice of window and statistic still remain.
Although properly constructed likelihood ratios should be able to simultaneously incorporate arbitrarily many window--statistic pairs, one quickly encounters the pragmatic issues with modeling high-dimensional probability distributions that led us to employ machine learning as dimensional reduction in the first place (Section~\ref{sec:supervised learning}).

Ref.~\cite{iDQ+GstLAL} explores a somewhat simpler construction, in which the iDQ $\log\Lambda^G_C$ timeseries is maximized over 1-second windows, slightly transformed and then applied directly as a multiplicative factor to a likelihood ratio detection statistic.
The transformation for iDQ's $\log\Lambda^G_C$ was empirically determined and fixed \textit{a priori}, essentially assuming the functional form for the trials factor introduced by the maximization and how that modified the likelihood of a signal being present.
While the assumed mapping may not be optimal, iDQ was found to moderately benefit current searches even with this simple approach.

Another approach is the idea of \emph{gating} in some form, in that we should remove all problematic (glitchy) times before filtering, thereby removing the need to select a specific extremization procedure to veto \textit{post hoc}.
Heuristically, the logic is that candidates are generated with large $\rho$ by glitches ringing up templates, and we can model the correlations between $\rho$ and the presence of a glitch either explicitly within a likelihood ratio test over many variates or implicitly by removing the contribution of $h(t)$ due to a glitch from the matched filter response altogether.

Such \emph{gating} schemes are now ubiquitous within the field and trivial to implement within white noise.
However, IFOs generate colored noise and the presence of loud glitches can ring whitening filters, polluting surrounding data that would otherwise be unaffected by the noise-transient.
This prompted the development of \emph{inverse-Tukey window} gating~\cite{Messick:2016aqy, Usman_2016} as well as more complicated \emph{in-painting} techniques designed to zero the filter response within a specified window after whitening~\cite{Zackay:2019kkv}.
We derive similar approaches to in-painting in Section~\ref{sec:optimal applications} from first-principles.

While gating mechanisms become more complex, we note that current approaches all rely on the same premise, that there is a predefined set of times declared glitchy that must be removed from the analysis.
These gates are typically defined by extremization processes over monitors and hard thresholds, and therefore suffer from the same ambiguities in determining appropriate settings as \textit{post hoc} veto segments, although without the ability to simultaneously consider multiple choices as would be possible in a likelihood ratio test.
Nonetheless, it is sometimes the case that filtering artifacts from gating or the ambiguity in defining gates are preferable to the original noise transient.

One could define gating conditions based on iDQ timeseries, but again we must face the selection of thresholds with no more obvious metric than guessing and checking how this affects searches' sensitivities.
One could be tempted to soften the hard thresholds with an \emph{adaptive gate}, such that the matched filter response of a timeseries $h$ with a filter $f$ would be modified to
\begin{equation}
    \rho_C(t) = \int d\tau\, h(t-\tau) f(\tau) p(C|\mathcal{M}(\vec{a}(\tau)))
\end{equation}
in effect estimating $\rho$ by probabilistically keeping the times expected to be clean based on auxiliary degrees of freedom.
Again, while heuristically appealing, it is not clear that this approach is optimal.
Indeed, it suffers from the same issues of whether to apply the adaptive gate before or after whitening the data as normal gates.

While GW searches have benefited from data quality information made available in the past (e.g.~\cite{Abbott_2018}), the issues associated with choosing or optimizing \textit{ad hoc} prescriptions for applying that information beg the question of whether there is a self-consistent framework that would provide a natural motivation for a particular approach.
We present such a framework, and additionally prescribe how the greater information available from probabilistic knowledge of data quality can be used without the need to cast that information into binary flags.

\subsection{Optimal Searches with Imperfect Knowledge of non-Gaussian Noise}
\label{sec:optimal applications}

We now formulate the problem from a first-principles model of the noise processes within our detectors.
As a reminder, we assume linear additive noise so that the detector output is given by $h=n+s+g$, where we only observe $h$ and the auxiliary state $\vec{a}$, meaning we must marginalize over the unobserved latent processes $n$, $s$, and $g$.
We begin by formulating probability distributions for detector noise in the target channel and auxiliary features conditioned on whether the IFO is in a \emph{glitchy} or \emph{clean} state.

In clean states, we assume the noise is generated by a stationary process, at least over timescale relevant to the filter, such that the autocorrelation function is given by
\begin{equation}
    \mathcal{C}_{ij}(\tau) = \left< n(t_i) n(t_j=t_i+\tau) \right> = \int df\, e^{2\pi i f \tau} S(f)
\end{equation}
where $S(f)$ is the PSD.
Adopting the Einstein summation convention, we then obtain
\begin{equation}
    p(n|C) \propto \exp\left(-\frac{1}{2}n_j \mathcal{C}^{-1}_{ij} n_i\right)
\end{equation}
by assuming Gaussianity.
Furthermore, we expect $n$ to be independent of $\vec{a}$ and declare $g=0\ \forall\ t\in C$ such that
\begin{equation}
    p(n, s, g, h, \vec{a}|C) = p(n|C)p(\vec{a}|C)\delta(g)p(s)\delta(h-(n+s+g))
\end{equation}
where the astrophysical strain induced in the detector is assumed to be independent of the instantaneous detector behavior.
We use iDQ's conditioned likelihood to model $p(\vec{a}|C)\sim p(\mathcal{M}(\vec{a})|C)$ and assume an astrophysically-motivated prior for signals $p(s)$.

In \emph{glitchy} states, we still assume $n$ is distributed as in clean times and that $n$ is independent of $(\vec{a},\,s,\,g)$.
\begin{equation}
    p(n, s, g, h, \vec{a}|G) = p(n|C)p(g|\vec{a},G)p(\vec{a}|G)p(s)\delta(h-(n+s+g))
\end{equation}
We note that the only difference is that we demand $p(g|C)=\delta(g)$ but leave $p(g|\vec{a},G)$ as a completely unknown function.
Assuming that the set of data $\{g(t)\ |\ t\in G\}$ is distributed somehow, we can construct a combined likelihood spanning predefined labeling of both $G$ and $C$ samples as
\begin{widetext}
\begin{equation}
    p(n,s,g,h,\vec{a}) = p(n|C) \delta(h-(n+s+g))p(s) \left[\prod\limits_{i\in C} p(\vec{a}_i|C)\delta(g_i)\right] \left[p(\{g_j|j\in G\}|\vec{a},G) \prod\limits_{j\in G}p(\vec{a}_j|G)\right]
\end{equation}
Marginalizing over latent processes yields
\begin{align}
    p(h,\vec{a}) & = \int \mathcal{D}n \mathcal{D}s \mathcal{D}g\, p(n,s,g,h,\vec{a}) \nonumber \\
                 & = \int \mathcal{D}s\, p(s) \prod\limits_{i\in C} p(\vec{a}_i|C) \prod\limits_{j\in G} p(\vec{a}_j|G) \int \mathcal{D}g \left(p(n=h-s-g|C) p(\{g_j|j\in G\}|\vec{a},G) \prod\limits_{i\in C}\delta(g_i)\right)
\end{align}
The result, in general, depends on how $g\in G$ is distributed, which is unknown.
We note that assuming $g$ is Gaussian-distributed with divergent variances, so that any $g$ is equally likely, incurs a large Occam factor from the normalization.
While not a problem when considering a single known permutation of which times are glitchy and which are clean, as was done in Ref.~\cite{Zackay:2019kkv}, this Occam factor can present issues when comparing different permutations, as we do below.
Instead, we note that 
\begin{align}
    \int \mathcal{D}g \left( p(n=h-s-g|C) p(\{g_j|j\in G\}|\vec{a},G) \prod\limits_{i\in C}\delta(g_i) \right) & \leq \max\limits_{n_j\in G} \left\{ p(n|C) \right\}\ \left|\  n_i = h_i-s_i \ \forall\ i\in C \right. \nonumber \\
        & = \frac{1}{\sqrt{(2\pi)^N\mathrm{det}|\mathcal{C}|}} \exp\left(-\frac{1}{2}\sum\limits_{i,j\in C} (h_i-s_i) \mathcal{C}^{-1}_{ij} (h_j-s_j) \right) \nonumber \\
        & \equiv \not{p}_C(n=h-s|\mathrm{perm})
\end{align}
where $\not{p}_C$ is the distribution for $n$ restricted to $n(t)\neq0$ iff $t\in C$, which depends on the specific permutation of which times are clean and which are glitchy.
This, then, implies
\begin{align}
    p(h,\vec{a}|\mathrm{perm}) & = \int \mathcal{D}s\, p(s)\, \prod\limits_{i\in C}p(\vec{a}_i|C)\prod\limits_{j\in G}p(\vec{a}_j|G) \int \mathcal{D}g \left( \left[\prod\limits_{i\in C}\delta(g_i)\right] p(\{g_j|j\in G\}|\vec{a},G) p(n=h-s-g|C) \right) \nonumber \\
                               & \leq \int \mathcal{D}s\, p(s)\, \prod\limits_{i\in C}p(\vec{a}_i|C)\prod\limits_{j\in G}p(\vec{a}_j|G) \not{p}_C(n=h-s|\mathrm{perm})
\end{align}
\end{widetext}
which is not a proper distribution (i.e., normalizable) because of the maximization, but instead is an upper bound on the marginal likelihood for $h$ and $\vec{a}$ given a permutation.

We, in effect, construct an inference only using times declared clean.
This is necessary because we do not know how $g$ is distributed within glitchy times.
Alternatively, machine learning models that infer $g$ directly from $\vec{a}$, or other assumptions about how $g$ is distributed (e.g., \cite{Cornish_2015}), would allow us to retain observations of $h\in G$ and marginalize over $g$ directly.
We note that maximized likelihood does not break the assumption of Gaussianity or stationarity; we simply restrict our selves to times known to be clean and treat glitchy times as if they were not observed.
In spirit, then, this is similar to gating.

However, we only have probabilistic knowledge of which times are glitchy and which are clean based on $h$ and $\vec{a}$.
We therefore must marginalize over all permutations weighed by their relative prior probabilities.
That is
\begin{multline}
    p(h,\vec{a}) \leq \sum\limits_\mathrm{perm}\left( p(\mathrm{perm}) \prod\limits_{i\in C} p(\vec{a}_i|C) \prod\limits_{j\in G} p(\vec{a}_j|G) \right.\\
        \left. \times \int \mathcal{D}s\,p(s) \not{p}_C(n=h-s|\mathrm{perm}) \right)
\end{multline}
with
\begin{equation}
    p(\mathrm{perm}) = p(C)^{N_C} p(G)^{N_G}
\end{equation}
so that the resulting likelihood is
\begin{widetext}
\begin{align}
    p(h,\vec{a}) & \leq \sum\limits_\mathrm{perm} \left( \prod\limits_{i\in C} p(\vec{a}_i|C)p(C) \prod\limits_{j\in G} p(\vec{a}_j|G)p(G) \int \mathcal{D}s\,p(s) \not{p}_C(n=h-s|\mathrm{perm})  \right) \nonumber \\
                 & = \sum\limits_\mathrm{perm} \left( p(\vec{a}|\mathrm{perm})p(\mathrm{perm}) \int \mathcal{D}s\,p(s) \not{p}_C(n=h-s|\mathrm{perm})  \right) \nonumber \\
\end{align}
\end{widetext}
and we remember that the specific sets $C$ and $G$ depend on the permutation.
Note, beyond the conditioned likelihoods from iDQ, even non-trivial prior odds alone can affect the resulting inference.
For example, this tells us that if $p(G) = p(C)/10$ and $p(\vec{a}|G)=p(\vec{a}|C)\ \forall\ t$, then one should marginalize over all possible covariance matrices that account for the prior knowledge that one glitch occurs for every 10 clean samples, on average.

Existing gating approaches assume exact data quality knowledge ($p(C|\vec{a})$ is either exactly 0 or exactly 1), thereby selecting a single permutation and a single noise model.
In Appendix~\ref{sec:optimal appendix}, we show how marginalization allowing for the presence of glitches, but without \textit{a priori} knowledge of when the glitches occur, can effectively gate glitches automatically.
This recovers and indeed expands upon current data quality approaches even without knowledge of auxiliary data.
Proper prior odds also allow us to infer how many glitches are likely to be present and where they are likely to be, all without \textit{a priori} knowledge beyond the relative frequencies of $G$ and $C$ samples.

With these marginal-maximized likelihoods in hand, we construct a likelihood ratio for the presence or absence of a signal in the data.
\begin{widetext}
\begin{equation}\label{eqn:optimal lrt}
    \Lambda^{S}_{!S} = \frac{
       \int \mathcal{D}s\, p(s) \sum\limits_\mathrm{perm} p(\vec{a}|\mathrm{perm})p(\mathrm{perm}) \not{p}_C(n=h-s|\mathrm{perm})
    }{
       \sum\limits_\mathrm{perm} p(\vec{a}|\mathrm{perm})p(\mathrm{perm}) \not{p}_C(n=h|\mathrm{perm})
    }
\end{equation}
\end{widetext}
Appendix~\ref{sec:optimal toy model} shows how the statistic behaves in the presence of glitches with either quiet or loud signals; in both cases it remains sensitive to GWs while being insensitive to the presence of glitches.

However, there is always the chance that our noise model remains insufficient to properly characterize IFO behavior, in which case other well-developed \textit{ad hoc} signal consistency tests, such as $\chi^2$ goodness-of-fit tests, could be combined with $\Lambda^S_{!S}$ within a larger likelihood ratio test, similar to how some searches currently include the matched filter $\rho$~\cite{Messick:2016aqy, Sachdev:2019vvd}, itself a proxy for the likelihood ratio in stationary Gaussian noise.
If $\Lambda^S_{!S}$ is sufficient, additional variates like \textit{ad hoc} $\chi^2$ tests will not hurt the search's sensitivity.
If it is not, they may continue to be vital.

However, Appendix~\ref{sec:signal consistency tests} shows how $\Lambda^S_{!S}$'s marginalization acts as an explicit signal consistency test without the need for additional \textit{ad hoc} $\chi^2$ tests.
In effect, the marginalization over whether any particular data sample is declared glitchy or clean constitutes a signal consistency test conditioned on the rest of the clean data in that permutation: given the other clean data, is it more likely that the observed datum in question was generated by stationary Gaussian noise or a glitch?

The computational cost of direct marginalization may be prohibitively high, though, as the combinatorics of the number of different permutations grow exponentially with the length of the signal.
Appendix~\ref{sec:computational cost} discusses a few possible approximations that may render this more tractable.

We note that many pipelines may not assume the GW signal comes from a known template family and instead search for unmodeled transients~\cite{Lynch2017, PhysRevD.93.042004, Sutton_2010, PhysRevD.83.083004}.
Such searches are often constructed by finding a maximum likelihood estimator for $s$ and the corresponding maximum likelihood, subject to loose constraints on the GW waveform morphology or polarization to avoid degeneracies from underconstrained inferences.
While we leave the explicit development of analogous searches to future work, we note that similar maximum likelihood techniques should work just as well with our marginal-maximized $\Lambda^S_{!S}$.

Similarly, one could formulate the probability of observing Gaussian noise in terms of a time-frequency decomposition of the data instead of the time-domain formulation provided here.
In this case, one could construct an analogous marginal-maximized $\Lambda^S_{!S}$ where the individual time-frequency pixels were assigned glitch or clean labels, marginalizing over all permutations.
Several parallel streams of probabilistic data quality, each targeting a specific frequency band, could be produced with parallel instances of the existing iDQ framework.
Such a formulation is perhaps closer in spirit to the time-frequency glitch model considered in Ref.~\cite{Cornish_2015} than it is to gating, although it would only exacerbate any computational issues associated with direct marginalization already present in our time-domain formulation.

\section{Discussion}\label{sec:discussion}

We present a statistical learning framework to infer the presence of non-Gaussian noise within gravitational-wave detectors: iDQ.
Using a supervised learning approach to classification, we decompose the inference into training, evaluation, calibration, and finally timeseries production.
iDQ accounts for non-stationarity in the detectors by continuously re-training classifiers to autonomously detect and exploit new witnesses for non-Gaussian noise without human intervention.
iDQ's framework can accommodate any supervised learning algorithm that operates on tabular data, and iDQ supports multiple modes of operation.
Offline, or \emph{batch} operation reproduces and expands upon previous functionality in the literature, fairly evaluating different algorithms' relative performance.
Online, or \emph{stream} operation manages multiple asynchronous processes to continuously re-train and re-calibrate statistical data quality inferences, producing robust probabilistic data quality information in real-time.

While some issues remain open, such as the best way to recover from detector non-stationarity (possibly through the use of weighted training sets) as well as the development and incorporation of novel feature sets and boosted classifiers, we show how iDQ has already proven invaluable within real GW searches.
Specifically, we show several examples in which iDQ either autonomously reproduced the behavior of canonical data quality monitors without \textit{a priori} knowledge of the type of noise in GW detectors or robustly identified noise sources otherwise unflagged by canonical monitors.
Ref.~\cite{iDQ+GstLAL} describes how iDQ has already been included within some searches.
We reiterate that iDQ has operated in low-latency throughout the entire advanced detector era and provided robust data quality information in low-latency for all detections to-date~\cite{catalog}.

We also explore current methods of incorporating data quality information within GW searches, discussing their relative merits and drawbacks, and note that all rely on absolute knowledge of data quality.
That is, most existing techniques implicitly require analysts to assume they know whether the detector is in a glitchy or clean state with absolute certainty at all times.
iDQ moves beyond this assumption.
We instead introduce a way to incorporate probabilistic data quality information that accounts for our imperfect knowledge of the presence or absence of non-Gaussian noise based on safe auxiliary channels alone.
This approach presents several attractive features, automatically incorporating optimally located gates without \textit{a priori} knowledge of where gates should be placed, as well as providing signal-consistency tests from first-principles noise models rather than \textit{ad hoc} $\chi^2$ tests.
While we remark that the computational expense of direct marginalization over imperfect data quality information may be large, we also suggest several possible solutions, leaving their full development to future work.

With increasing detection rates and improved detector sensitivity, robust data quality information will only become more important over the next few years.
Indeed, the importance of low-latency information for multi-messenger astronomy cannot be overstated.
iDQ has provided robust low-latency probabilistic data quality information throughout the advanced detector era, and will continue to do so as GWs reveal new astrophysical phenomena in the most extreme environments found anywhere in the universe.

\acknowledgments

The authors thank Ruslan Vaulin and acknowledge his vital contributions to earlier versions of iDQ.
R.~E. is supported at the University of Chicago by the Kavli Institute for Cosmological Physics through an endowment from the Kavli Foundation and its founder Fred Kavli.
P.~G. and C.~H. are supported by the National Science Foundation (NSF) through OAC-1841480,
OAC-1642391, and PHY-1454389.
The authors also gratefully acknowledge the computational resources provided by the LIGO Laboratory and supported by NSF grants PHY-0757058 and PHY-0823459.
Computations for this research were also performed on the Pennsylvania State
University’s Institute for Computational and Data Sciences Advanced
CyberInfrastructure (ICDS-ACI).

\bibliography{refs.bib}

\begin{thebibliography}{76}%
\makeatletter
\providecommand \@ifxundefined [1]{%
 \@ifx{#1\undefined}
}%
\providecommand \@ifnum [1]{%
 \ifnum #1\expandafter \@firstoftwo
 \else \expandafter \@secondoftwo
 \fi
}%
\providecommand \@ifx [1]{%
 \ifx #1\expandafter \@firstoftwo
 \else \expandafter \@secondoftwo
 \fi
}%
\providecommand \natexlab [1]{#1}%
\providecommand \enquote  [1]{``#1''}%
\providecommand \bibnamefont  [1]{#1}%
\providecommand \bibfnamefont [1]{#1}%
\providecommand \citenamefont [1]{#1}%
\providecommand \href@noop [0]{\@secondoftwo}%
\providecommand \href [0]{\begingroup \@sanitize@url \@href}%
\providecommand \@href[1]{\@@startlink{#1}\@@href}%
\providecommand \@@href[1]{\endgroup#1\@@endlink}%
\providecommand \@sanitize@url [0]{\catcode `\\12\catcode `\$12\catcode
  `\&12\catcode `\#12\catcode `\^12\catcode `\_12\catcode `\%12\relax}%
\providecommand \@@startlink[1]{}%
\providecommand \@@endlink[0]{}%
\providecommand \url  [0]{\begingroup\@sanitize@url \@url }%
\providecommand \@url [1]{\endgroup\@href {#1}{\urlprefix }}%
\providecommand \urlprefix  [0]{URL }%
\providecommand \Eprint [0]{\href }%
\providecommand \doibase [0]{http://dx.doi.org/}%
\providecommand \selectlanguage [0]{\@gobble}%
\providecommand \bibinfo  [0]{\@secondoftwo}%
\providecommand \bibfield  [0]{\@secondoftwo}%
\providecommand \translation [1]{[#1]}%
\providecommand \BibitemOpen [0]{}%
\providecommand \bibitemStop [0]{}%
\providecommand \bibitemNoStop [0]{.\EOS\space}%
\providecommand \EOS [0]{\spacefactor3000\relax}%
\providecommand \BibitemShut  [1]{\csname bibitem#1\endcsname}%
\let\auto@bib@innerbib\@empty
\bibitem [{\citenamefont {{J. Aasi \textit{et al.} (LIGO Scientific
  Collaboration)}}(2015)}]{LIGO}%
  \BibitemOpen
  \bibfield  {author} {\bibinfo {author} {\bibnamefont {{J. Aasi \textit{et
  al.} (LIGO Scientific Collaboration)}}},\ }\href {\doibase
  10.1088/0264-9381/32/7/074001} {\bibfield  {journal} {\bibinfo  {journal}
  {Class. Quantum Grav.}\ }\textbf {\bibinfo {volume} {32}},\ \bibinfo {eid}
  {074001} (\bibinfo {year} {2015})},\ \Eprint {http://arxiv.org/abs/1411.4547}
  {arXiv:1411.4547 [gr-qc]} \BibitemShut {NoStop}%
\bibitem [{\citenamefont {{F. Acernese \textit{et al.} (Virgo
  Collaboration)}}(2015)}]{Virgo}%
  \BibitemOpen
  \bibfield  {author} {\bibinfo {author} {\bibnamefont {{F. Acernese \textit{et
  al.} (Virgo Collaboration)}}},\ }\href {\doibase
  10.1088/0264-9381/32/2/024001} {\bibfield  {journal} {\bibinfo  {journal}
  {Class. Quantum Grav.}\ }\textbf {\bibinfo {volume} {32}},\ \bibinfo {eid}
  {024001} (\bibinfo {year} {2015})},\ \Eprint {http://arxiv.org/abs/1408.3978}
  {arXiv:1408.3978 [gr-qc]} \BibitemShut {NoStop}%
\bibitem [{\citenamefont {Matichard}\ \emph {et~al.}(2015)\citenamefont
  {Matichard} \emph {et~al.}}]{Matichard:2015eva}%
  \BibitemOpen
  \bibfield  {author} {\bibinfo {author} {\bibfnamefont {F.}~\bibnamefont
  {Matichard}} \emph {et~al.},\ }\href {\doibase
  10.1088/0264-9381/32/18/185003} {\bibfield  {journal} {\bibinfo  {journal}
  {Class. Quant. Grav.}\ }\textbf {\bibinfo {volume} {32}},\ \bibinfo {pages}
  {185003} (\bibinfo {year} {2015})},\ \Eprint
  {http://arxiv.org/abs/1502.06300} {arXiv:1502.06300 [physics.ins-det]}
  \BibitemShut {NoStop}%
\bibitem [{\citenamefont {Graef~Rollins}(2016)}]{Rollins:2016hlk}%
  \BibitemOpen
  \bibfield  {author} {\bibinfo {author} {\bibfnamefont {J.}~\bibnamefont
  {Graef~Rollins}},\ }\href@noop {} {\  (\bibinfo {year} {2016})},\ \Eprint
  {http://arxiv.org/abs/1604.01456} {arXiv:1604.01456 [astro-ph.IM]}
  \BibitemShut {NoStop}%
\bibitem [{\citenamefont {Viets}\ \emph {et~al.}(2018)\citenamefont {Viets}
  \emph {et~al.}}]{Viets:2017yvy}%
  \BibitemOpen
  \bibfield  {author} {\bibinfo {author} {\bibfnamefont {A.}~\bibnamefont
  {Viets}} \emph {et~al.},\ }\href {\doibase 10.1088/1361-6382/aab658}
  {\bibfield  {journal} {\bibinfo  {journal} {Class. Quant. Grav.}\ }\textbf
  {\bibinfo {volume} {35}},\ \bibinfo {pages} {095015} (\bibinfo {year}
  {2018})},\ \Eprint {http://arxiv.org/abs/1710.09973} {arXiv:1710.09973
  [astro-ph.IM]} \BibitemShut {NoStop}%
\bibitem [{\citenamefont {Abbott}\ \emph
  {et~al.}(2016{\natexlab{a}})\citenamefont {Abbott} \emph
  {et~al.}}]{GW150914}%
  \BibitemOpen
  \bibfield  {author} {\bibinfo {author} {\bibfnamefont {B.~P.}\ \bibnamefont
  {Abbott}} \emph {et~al.} (\bibinfo {collaboration} {LIGO Scientific
  Collaboration and Virgo Collaboration}),\ }\href {\doibase
  10.1103/PhysRevLett.116.061102} {\bibfield  {journal} {\bibinfo  {journal}
  {Phys. Rev. Lett.}\ }\textbf {\bibinfo {volume} {116}},\ \bibinfo {pages}
  {061102} (\bibinfo {year} {2016}{\natexlab{a}})}\BibitemShut {NoStop}%
\bibitem [{\citenamefont {{The LIGO Scientific Collaboration}}\ \emph
  {et~al.}(2018)\citenamefont {{The LIGO Scientific Collaboration}},
  \citenamefont {{the Virgo Collaboration}}, \citenamefont {{Abbott}} \emph
  {et~al.}}]{catalog}%
  \BibitemOpen
  \bibfield  {author} {\bibinfo {author} {\bibnamefont {{The LIGO Scientific
  Collaboration}}}, \bibinfo {author} {\bibnamefont {{the Virgo
  Collaboration}}}, \bibinfo {author} {\bibfnamefont {B.~P.}\ \bibnamefont
  {{Abbott}}},  \emph {et~al.},\ }\href@noop {} {\bibfield  {journal} {\bibinfo
   {journal} {arXiv e-prints}\ ,\ \bibinfo {eid} {arXiv:1811.12907}} (\bibinfo
  {year} {2018})},\ \Eprint {http://arxiv.org/abs/1811.12907} {arXiv:1811.12907
  [astro-ph.HE]} \BibitemShut {NoStop}%
\bibitem [{\citenamefont {Abbott}\ \emph
  {et~al.}(2017{\natexlab{a}})\citenamefont {Abbott} \emph
  {et~al.}}]{GW170817}%
  \BibitemOpen
  \bibfield  {author} {\bibinfo {author} {\bibfnamefont {B.~P.}\ \bibnamefont
  {Abbott}} \emph {et~al.} (\bibinfo {collaboration} {LIGO Scientific
  Collaboration and Virgo Collaboration}),\ }\href {\doibase
  10.1103/PhysRevLett.119.161101} {\bibfield  {journal} {\bibinfo  {journal}
  {Phys. Rev. Lett.}\ }\textbf {\bibinfo {volume} {119}},\ \bibinfo {pages}
  {161101} (\bibinfo {year} {2017}{\natexlab{a}})}\BibitemShut {NoStop}%
\bibitem [{\citenamefont {Abbott}\ \emph
  {et~al.}(2020{\natexlab{a}})\citenamefont {Abbott} \emph
  {et~al.}}]{GW190425}%
  \BibitemOpen
  \bibfield  {author} {\bibinfo {author} {\bibfnamefont {B.}~\bibnamefont
  {Abbott}} \emph {et~al.} (\bibinfo {collaboration} {LIGO Scientific,
  Virgo}),\ }\href {\doibase 10.3847/2041-8213/ab75f5} {\bibfield  {journal}
  {\bibinfo  {journal} {Astrophys. J. Lett.}\ }\textbf {\bibinfo {volume}
  {892}},\ \bibinfo {pages} {L3} (\bibinfo {year} {2020}{\natexlab{a}})},\
  \Eprint {http://arxiv.org/abs/2001.01761} {arXiv:2001.01761 [astro-ph.HE]}
  \BibitemShut {NoStop}%
\bibitem [{\citenamefont {Abbott}\ \emph
  {et~al.}(2016{\natexlab{b}})\citenamefont {Abbott} \emph
  {et~al.}}]{GW170817MMA}%
  \BibitemOpen
  \bibfield  {author} {\bibinfo {author} {\bibfnamefont {B.~P.}\ \bibnamefont
  {Abbott}} \emph {et~al.},\ }\href {\doibase 10.1088/0264-9381/33/13/134001}
  {\bibfield  {journal} {\bibinfo  {journal} {Classical and Quantum Gravity}\
  }\textbf {\bibinfo {volume} {33}},\ \bibinfo {pages} {134001} (\bibinfo
  {year} {2016}{\natexlab{b}})}\BibitemShut {NoStop}%
\bibitem [{\citenamefont {Abbott}\ \emph
  {et~al.}(2017{\natexlab{b}})\citenamefont {Abbott} \emph
  {et~al.}}]{GW170817GRB}%
  \BibitemOpen
  \bibfield  {author} {\bibinfo {author} {\bibfnamefont {B.~P.}\ \bibnamefont
  {Abbott}} \emph {et~al.},\ }\href {\doibase 10.3847/2041-8213/aa920c}
  {\bibfield  {journal} {\bibinfo  {journal} {The Astrophysical Journal}\
  }\textbf {\bibinfo {volume} {848}},\ \bibinfo {pages} {L13} (\bibinfo {year}
  {2017}{\natexlab{b}})}\BibitemShut {NoStop}%
\bibitem [{\citenamefont {Coulter}\ \emph {et~al.}(2017)\citenamefont
  {Coulter}, \citenamefont {Foley}, \citenamefont {Kilpatrick}, \citenamefont
  {Drout}, \citenamefont {Piro}, \citenamefont {Shappee}, \citenamefont
  {Siebert}, \citenamefont {Simon}, \citenamefont {Ulloa}, \citenamefont
  {Kasen}, \citenamefont {Madore}, \citenamefont {Murguia-Berthier},
  \citenamefont {Pan}, \citenamefont {Prochaska}, \citenamefont {Ramirez-Ruiz},
  \citenamefont {Rest},\ and\ \citenamefont {Rojas-Bravo}}]{Coulter1556}%
  \BibitemOpen
  \bibfield  {author} {\bibinfo {author} {\bibfnamefont {D.~A.}\ \bibnamefont
  {Coulter}}, \bibinfo {author} {\bibfnamefont {R.~J.}\ \bibnamefont {Foley}},
  \bibinfo {author} {\bibfnamefont {C.~D.}\ \bibnamefont {Kilpatrick}},
  \bibinfo {author} {\bibfnamefont {M.~R.}\ \bibnamefont {Drout}}, \bibinfo
  {author} {\bibfnamefont {A.~L.}\ \bibnamefont {Piro}}, \bibinfo {author}
  {\bibfnamefont {B.~J.}\ \bibnamefont {Shappee}}, \bibinfo {author}
  {\bibfnamefont {M.~R.}\ \bibnamefont {Siebert}}, \bibinfo {author}
  {\bibfnamefont {J.~D.}\ \bibnamefont {Simon}}, \bibinfo {author}
  {\bibfnamefont {N.}~\bibnamefont {Ulloa}}, \bibinfo {author} {\bibfnamefont
  {D.}~\bibnamefont {Kasen}}, \bibinfo {author} {\bibfnamefont {B.~F.}\
  \bibnamefont {Madore}}, \bibinfo {author} {\bibfnamefont {A.}~\bibnamefont
  {Murguia-Berthier}}, \bibinfo {author} {\bibfnamefont {Y.-C.}\ \bibnamefont
  {Pan}}, \bibinfo {author} {\bibfnamefont {J.~X.}\ \bibnamefont {Prochaska}},
  \bibinfo {author} {\bibfnamefont {E.}~\bibnamefont {Ramirez-Ruiz}}, \bibinfo
  {author} {\bibfnamefont {A.}~\bibnamefont {Rest}}, \ and\ \bibinfo {author}
  {\bibfnamefont {C.}~\bibnamefont {Rojas-Bravo}},\ }\href {\doibase
  10.1126/science.aap9811} {\bibfield  {journal} {\bibinfo  {journal}
  {Science}\ }\textbf {\bibinfo {volume} {358}},\ \bibinfo {pages} {1556}
  (\bibinfo {year} {2017})},\ \Eprint
  {http://arxiv.org/abs/https://science.sciencemag.org/content/358/6370/1556.full.pdf}
  {https://science.sciencemag.org/content/358/6370/1556.full.pdf} \BibitemShut
  {NoStop}%
\bibitem [{\citenamefont {Goldstein}\ \emph {et~al.}(2017)\citenamefont
  {Goldstein}, \citenamefont {Veres}, \citenamefont {Burns}, \citenamefont
  {Briggs}, \citenamefont {Hamburg}, \citenamefont {Kocevski}, \citenamefont
  {Wilson-Hodge}, \citenamefont {Preece}, \citenamefont {Poolakkil},
  \citenamefont {Roberts}, \citenamefont {Hui}, \citenamefont {Connaughton},
  \citenamefont {Racusin}, \citenamefont {von Kienlin}, \citenamefont {Canton},
  \citenamefont {Christensen}, \citenamefont {Littenberg}, \citenamefont
  {Siellez}, \citenamefont {Blackburn}, \citenamefont {Broida}, \citenamefont
  {Bissaldi}, \citenamefont {Cleveland}, \citenamefont {Gibby}, \citenamefont
  {Giles}, \citenamefont {Kippen}, \citenamefont {McBreen}, \citenamefont
  {McEnery}, \citenamefont {Meegan}, \citenamefont {Paciesas},\ and\
  \citenamefont {Stanbro}}]{Goldstein_2017}%
  \BibitemOpen
  \bibfield  {author} {\bibinfo {author} {\bibfnamefont {A.}~\bibnamefont
  {Goldstein}}, \bibinfo {author} {\bibfnamefont {P.}~\bibnamefont {Veres}},
  \bibinfo {author} {\bibfnamefont {E.}~\bibnamefont {Burns}}, \bibinfo
  {author} {\bibfnamefont {M.~S.}\ \bibnamefont {Briggs}}, \bibinfo {author}
  {\bibfnamefont {R.}~\bibnamefont {Hamburg}}, \bibinfo {author} {\bibfnamefont
  {D.}~\bibnamefont {Kocevski}}, \bibinfo {author} {\bibfnamefont {C.~A.}\
  \bibnamefont {Wilson-Hodge}}, \bibinfo {author} {\bibfnamefont {R.~D.}\
  \bibnamefont {Preece}}, \bibinfo {author} {\bibfnamefont {S.}~\bibnamefont
  {Poolakkil}}, \bibinfo {author} {\bibfnamefont {O.~J.}\ \bibnamefont
  {Roberts}}, \bibinfo {author} {\bibfnamefont {C.~M.}\ \bibnamefont {Hui}},
  \bibinfo {author} {\bibfnamefont {V.}~\bibnamefont {Connaughton}}, \bibinfo
  {author} {\bibfnamefont {J.}~\bibnamefont {Racusin}}, \bibinfo {author}
  {\bibfnamefont {A.}~\bibnamefont {von Kienlin}}, \bibinfo {author}
  {\bibfnamefont {T.~D.}\ \bibnamefont {Canton}}, \bibinfo {author}
  {\bibfnamefont {N.}~\bibnamefont {Christensen}}, \bibinfo {author}
  {\bibfnamefont {T.}~\bibnamefont {Littenberg}}, \bibinfo {author}
  {\bibfnamefont {K.}~\bibnamefont {Siellez}}, \bibinfo {author} {\bibfnamefont
  {L.}~\bibnamefont {Blackburn}}, \bibinfo {author} {\bibfnamefont
  {J.}~\bibnamefont {Broida}}, \bibinfo {author} {\bibfnamefont
  {E.}~\bibnamefont {Bissaldi}}, \bibinfo {author} {\bibfnamefont {W.~H.}\
  \bibnamefont {Cleveland}}, \bibinfo {author} {\bibfnamefont {M.~H.}\
  \bibnamefont {Gibby}}, \bibinfo {author} {\bibfnamefont {M.~M.}\ \bibnamefont
  {Giles}}, \bibinfo {author} {\bibfnamefont {R.~M.}\ \bibnamefont {Kippen}},
  \bibinfo {author} {\bibfnamefont {S.}~\bibnamefont {McBreen}}, \bibinfo
  {author} {\bibfnamefont {J.}~\bibnamefont {McEnery}}, \bibinfo {author}
  {\bibfnamefont {C.~A.}\ \bibnamefont {Meegan}}, \bibinfo {author}
  {\bibfnamefont {W.~S.}\ \bibnamefont {Paciesas}}, \ and\ \bibinfo {author}
  {\bibfnamefont {M.}~\bibnamefont {Stanbro}},\ }\href {\doibase
  10.3847/2041-8213/aa8f41} {\bibfield  {journal} {\bibinfo  {journal} {The
  Astrophysical Journal}\ }\textbf {\bibinfo {volume} {848}},\ \bibinfo {pages}
  {L14} (\bibinfo {year} {2017})}\BibitemShut {NoStop}%
\bibitem [{\citenamefont {Abbott}\ \emph
  {et~al.}(2016{\natexlab{c}})\citenamefont {Abbott} \emph
  {et~al.}}]{Martynov:2016fzi}%
  \BibitemOpen
  \bibfield  {author} {\bibinfo {author} {\bibfnamefont {B.~P.}\ \bibnamefont
  {Abbott}} \emph {et~al.},\ }\href {\doibase 10.1103/PhysRevD.93.112004,
  10.1103/PhysRevD.97.059901} {\bibfield  {journal} {\bibinfo  {journal} {Phys.
  Rev.}\ }\textbf {\bibinfo {volume} {D93}},\ \bibinfo {pages} {112004}
  (\bibinfo {year} {2016}{\natexlab{c}})},\ \bibinfo {note} {[Addendum: Phys.
  Rev.D97,no.5,059901(2018)]},\ \Eprint {http://arxiv.org/abs/1604.00439}
  {arXiv:1604.00439 [astro-ph.IM]} \BibitemShut {NoStop}%
\bibitem [{\citenamefont {Abbott}\ \emph
  {et~al.}(2020{\natexlab{b}})\citenamefont {Abbott} \emph
  {et~al.}}]{LIGOScientific:2019hgc}%
  \BibitemOpen
  \bibfield  {author} {\bibinfo {author} {\bibfnamefont {B.~P.}\ \bibnamefont
  {Abbott}} \emph {et~al.} (\bibinfo {collaboration} {LIGO Scientific,
  Virgo}),\ }\href {\doibase 10.1088/1361-6382/ab685e} {\bibfield  {journal}
  {\bibinfo  {journal} {Class. Quant. Grav.}\ }\textbf {\bibinfo {volume}
  {37}},\ \bibinfo {pages} {055002} (\bibinfo {year} {2020}{\natexlab{b}})},\
  \Eprint {http://arxiv.org/abs/1908.11170} {arXiv:1908.11170 [gr-qc]}
  \BibitemShut {NoStop}%
\bibitem [{\citenamefont {Abbott}\ \emph
  {et~al.}(2019{\natexlab{a}})\citenamefont {Abbott} \emph
  {et~al.}}]{PhysRevD.100.024017}%
  \BibitemOpen
  \bibfield  {author} {\bibinfo {author} {\bibfnamefont {B.~P.}\ \bibnamefont
  {Abbott}} \emph {et~al.} (\bibinfo {collaboration} {LIGO Scientific
  Collaboration and Virgo Collaboration}),\ }\href {\doibase
  10.1103/PhysRevD.100.024017} {\bibfield  {journal} {\bibinfo  {journal}
  {Phys. Rev. D}\ }\textbf {\bibinfo {volume} {100}},\ \bibinfo {pages}
  {024017} (\bibinfo {year} {2019}{\natexlab{a}})}\BibitemShut {NoStop}%
\bibitem [{\citenamefont {Abbott}\ \emph {et~al.}(2018)\citenamefont {Abbott}
  \emph {et~al.}}]{Abbott_2018}%
  \BibitemOpen
  \bibfield  {author} {\bibinfo {author} {\bibfnamefont {B.~P.}\ \bibnamefont
  {Abbott}} \emph {et~al.},\ }\href {\doibase 10.1088/1361-6382/aaaafa}
  {\bibfield  {journal} {\bibinfo  {journal} {Classical and Quantum Gravity}\
  }\textbf {\bibinfo {volume} {35}},\ \bibinfo {pages} {065010} (\bibinfo
  {year} {2018})}\BibitemShut {NoStop}%
\bibitem [{\citenamefont {Powell}\ \emph {et~al.}(2015)\citenamefont {Powell},
  \citenamefont {Trifirò}, \citenamefont {Cuoco}, \citenamefont {Heng},\ and\
  \citenamefont {Cavaglià}}]{Powell:2015ona}%
  \BibitemOpen
  \bibfield  {author} {\bibinfo {author} {\bibfnamefont {J.}~\bibnamefont
  {Powell}}, \bibinfo {author} {\bibfnamefont {D.}~\bibnamefont {Trifirò}},
  \bibinfo {author} {\bibfnamefont {E.}~\bibnamefont {Cuoco}}, \bibinfo
  {author} {\bibfnamefont {I.~S.}\ \bibnamefont {Heng}}, \ and\ \bibinfo
  {author} {\bibfnamefont {M.}~\bibnamefont {Cavaglià}},\ }\href {\doibase
  10.1088/0264-9381/32/21/215012} {\bibfield  {journal} {\bibinfo  {journal}
  {Class. Quant. Grav.}\ }\textbf {\bibinfo {volume} {32}},\ \bibinfo {pages}
  {215012} (\bibinfo {year} {2015})},\ \Eprint
  {http://arxiv.org/abs/1505.01299} {arXiv:1505.01299 [astro-ph.IM]}
  \BibitemShut {NoStop}%
\bibitem [{\citenamefont {Powell}\ \emph {et~al.}(2017)\citenamefont {Powell},
  \citenamefont {Torres-Forné}, \citenamefont {Lynch}, \citenamefont
  {Trifirò}, \citenamefont {Cuoco}, \citenamefont {Cavaglià}, \citenamefont
  {Heng},\ and\ \citenamefont {Font}}]{Powell:2016rkl}%
  \BibitemOpen
  \bibfield  {author} {\bibinfo {author} {\bibfnamefont {J.}~\bibnamefont
  {Powell}}, \bibinfo {author} {\bibfnamefont {A.}~\bibnamefont
  {Torres-Forné}}, \bibinfo {author} {\bibfnamefont {R.}~\bibnamefont
  {Lynch}}, \bibinfo {author} {\bibfnamefont {D.}~\bibnamefont {Trifirò}},
  \bibinfo {author} {\bibfnamefont {E.}~\bibnamefont {Cuoco}}, \bibinfo
  {author} {\bibfnamefont {M.}~\bibnamefont {Cavaglià}}, \bibinfo {author}
  {\bibfnamefont {I.~S.}\ \bibnamefont {Heng}}, \ and\ \bibinfo {author}
  {\bibfnamefont {J.~A.}\ \bibnamefont {Font}},\ }\href {\doibase
  10.1088/1361-6382/34/3/034002} {\bibfield  {journal} {\bibinfo  {journal}
  {Class. Quant. Grav.}\ }\textbf {\bibinfo {volume} {34}},\ \bibinfo {pages}
  {034002} (\bibinfo {year} {2017})},\ \Eprint
  {http://arxiv.org/abs/1609.06262} {arXiv:1609.06262 [astro-ph.IM]}
  \BibitemShut {NoStop}%
\bibitem [{\citenamefont {Zevin}\ \emph {et~al.}(2017)\citenamefont {Zevin},
  \citenamefont {Coughlin}, \citenamefont {Bahaadini}, \citenamefont {Besler},
  \citenamefont {Rohani}, \citenamefont {Allen}, \citenamefont {Cabero},
  \citenamefont {Crowston}, \citenamefont {Katsaggelos}, \citenamefont
  {Larson}, \citenamefont {Lee}, \citenamefont {Lintott}, \citenamefont
  {Littenberg}, \citenamefont {Lundgren}, \citenamefont {{\O}sterlund},
  \citenamefont {Smith}, \citenamefont {Trouille},\ and\ \citenamefont
  {Kalogera}}]{Zevin2017}%
  \BibitemOpen
  \bibfield  {author} {\bibinfo {author} {\bibfnamefont {M.}~\bibnamefont
  {Zevin}}, \bibinfo {author} {\bibfnamefont {S.}~\bibnamefont {Coughlin}},
  \bibinfo {author} {\bibfnamefont {S.}~\bibnamefont {Bahaadini}}, \bibinfo
  {author} {\bibfnamefont {E.}~\bibnamefont {Besler}}, \bibinfo {author}
  {\bibfnamefont {N.}~\bibnamefont {Rohani}}, \bibinfo {author} {\bibfnamefont
  {S.}~\bibnamefont {Allen}}, \bibinfo {author} {\bibfnamefont
  {M.}~\bibnamefont {Cabero}}, \bibinfo {author} {\bibfnamefont
  {K.}~\bibnamefont {Crowston}}, \bibinfo {author} {\bibfnamefont {A.~K.}\
  \bibnamefont {Katsaggelos}}, \bibinfo {author} {\bibfnamefont {S.~L.}\
  \bibnamefont {Larson}}, \bibinfo {author} {\bibfnamefont {T.~K.}\
  \bibnamefont {Lee}}, \bibinfo {author} {\bibfnamefont {C.}~\bibnamefont
  {Lintott}}, \bibinfo {author} {\bibfnamefont {T.~B.}\ \bibnamefont
  {Littenberg}}, \bibinfo {author} {\bibfnamefont {A.}~\bibnamefont
  {Lundgren}}, \bibinfo {author} {\bibfnamefont {C.}~\bibnamefont
  {{\O}sterlund}}, \bibinfo {author} {\bibfnamefont {J.~R.}\ \bibnamefont
  {Smith}}, \bibinfo {author} {\bibfnamefont {L.}~\bibnamefont {Trouille}}, \
  and\ \bibinfo {author} {\bibfnamefont {V.}~\bibnamefont {Kalogera}},\ }\href
  {\doibase 10.1088/1361-6382/aa5cea} {\bibfield  {journal} {\bibinfo
  {journal} {Classical and Quantum Gravity}\ }\textbf {\bibinfo {volume}
  {34}},\ \bibinfo {pages} {064003} (\bibinfo {year} {2017})}\BibitemShut
  {NoStop}%
\bibitem [{\citenamefont {Mueller}\ \emph {et~al.}(2016)\citenamefont {Mueller}
  \emph {et~al.}}]{Mueller:2016hex}%
  \BibitemOpen
  \bibfield  {author} {\bibinfo {author} {\bibfnamefont {C.~L.}\ \bibnamefont
  {Mueller}} \emph {et~al.} (\bibinfo {collaboration} {aLIGO}),\ }\href
  {\doibase 10.1063/1.4936974} {\bibfield  {journal} {\bibinfo  {journal} {Rev.
  Sci. Instrum.}\ }\textbf {\bibinfo {volume} {87}},\ \bibinfo {pages} {014502}
  (\bibinfo {year} {2016})},\ \Eprint {http://arxiv.org/abs/1601.05442}
  {arXiv:1601.05442 [physics.ins-det]} \BibitemShut {NoStop}%
\bibitem [{\citenamefont {Staley}(2015)}]{Staley:2015nie}%
  \BibitemOpen
  \bibfield  {author} {\bibinfo {author} {\bibfnamefont {A.}~\bibnamefont
  {Staley}},\ }\emph {\bibinfo {title} {{Locking the Advanced LIGO
  Gravitational Wave Detector: with a focus on the Arm Length Stabilization
  Technique}}},\ \href {\doibase 10.7916/D8X34WQ4} {Ph.D. thesis},\ \bibinfo
  {school} {Columbia U.} (\bibinfo {year} {2015})\BibitemShut {NoStop}%
\bibitem [{\citenamefont {Effler}\ \emph {et~al.}(2015)\citenamefont {Effler},
  \citenamefont {Schofield}, \citenamefont {Frolov}, \citenamefont {González},
  \citenamefont {Kawabe}, \citenamefont {Smith}, \citenamefont {Birch},\ and\
  \citenamefont {McCarthy}}]{Effler:2014zpa}%
  \BibitemOpen
  \bibfield  {author} {\bibinfo {author} {\bibfnamefont {A.}~\bibnamefont
  {Effler}}, \bibinfo {author} {\bibfnamefont {R.~M.~S.}\ \bibnamefont
  {Schofield}}, \bibinfo {author} {\bibfnamefont {V.~V.}\ \bibnamefont
  {Frolov}}, \bibinfo {author} {\bibfnamefont {G.}~\bibnamefont {González}},
  \bibinfo {author} {\bibfnamefont {K.}~\bibnamefont {Kawabe}}, \bibinfo
  {author} {\bibfnamefont {J.~R.}\ \bibnamefont {Smith}}, \bibinfo {author}
  {\bibfnamefont {J.}~\bibnamefont {Birch}}, \ and\ \bibinfo {author}
  {\bibfnamefont {R.}~\bibnamefont {McCarthy}},\ }\href {\doibase
  10.1088/0264-9381/32/3/035017} {\bibfield  {journal} {\bibinfo  {journal}
  {Class. Quant. Grav.}\ }\textbf {\bibinfo {volume} {32}},\ \bibinfo {pages}
  {035017} (\bibinfo {year} {2015})},\ \Eprint {http://arxiv.org/abs/1409.5160}
  {arXiv:1409.5160 [astro-ph.IM]} \BibitemShut {NoStop}%
\bibitem [{\citenamefont {Essick}\ and\ \citenamefont
  {Godwin}(2018{\natexlab{a}})}]{idq-repo}%
  \BibitemOpen
  \bibfield  {author} {\bibinfo {author} {\bibfnamefont {R.}~\bibnamefont
  {Essick}}\ and\ \bibinfo {author} {\bibfnamefont {P.}~\bibnamefont
  {Godwin}},\ }\href@noop {} {\enquote {\bibinfo {title} {idq source code},}\
  }\bibinfo {howpublished} {\url{https://git.ligo.org/reed.essick/iDQ}}
  (\bibinfo {year} {2018}{\natexlab{a}})\BibitemShut {NoStop}%
\bibitem [{\citenamefont {Essick}\ and\ \citenamefont
  {Godwin}(2018{\natexlab{b}})}]{idq-docs}%
  \BibitemOpen
  \bibfield  {author} {\bibinfo {author} {\bibfnamefont {R.}~\bibnamefont
  {Essick}}\ and\ \bibinfo {author} {\bibfnamefont {P.}~\bibnamefont
  {Godwin}},\ }\href@noop {} {\enquote {\bibinfo {title} {idq documentation},}\
  }\bibinfo {howpublished} {\url{https://docs.ligo.org/reed.essick/iDQ}}
  (\bibinfo {year} {2018}{\natexlab{b}})\BibitemShut {NoStop}%
\bibitem [{\citenamefont {{Abbott}}\ \emph {et~al.}(2018)\citenamefont
  {{Abbott}} \emph {et~al.}}]{ObservingScenarios}%
  \BibitemOpen
  \bibfield  {author} {\bibinfo {author} {\bibfnamefont {B.~P.}\ \bibnamefont
  {{Abbott}}} \emph {et~al.},\ }\href {\doibase
  https://doi.org/10.1007/s41114-018-0012-9} {\bibfield  {journal} {\bibinfo
  {journal} {Living Rev Relativity}\ ,\ \bibinfo {pages} {21: 3}} (\bibinfo
  {year} {2018})}\BibitemShut {NoStop}%
\bibitem [{\citenamefont {Essick}(2017)}]{Essick2017}%
  \BibitemOpen
  \bibfield  {author} {\bibinfo {author} {\bibfnamefont {R.}~\bibnamefont
  {Essick}},\ }\emph {\bibinfo {title} {{Detectability of dynamical tidal
  effects and the detection of gravitational-wave transients with LIGO}}},\
  \href {http://hdl.handle.net/1721.1/115024} {\bibinfo {type} {Theses}},\
  \bibinfo  {school} {{http://hdl.handle.net/1721.1/115024}} (\bibinfo {year}
  {2017})\BibitemShut {NoStop}%
\bibitem [{\citenamefont {Biswas}\ \emph {et~al.}(2013)\citenamefont {Biswas},
  \citenamefont {Blackburn}, \citenamefont {Cao}, \citenamefont {Essick},
  \citenamefont {Hodge}, \citenamefont {Katsavounidis}, \citenamefont {Kim},
  \citenamefont {Kim}, \citenamefont {Le~Bigot}, \citenamefont {Lee},
  \citenamefont {Oh}, \citenamefont {Oh}, \citenamefont {Son}, \citenamefont
  {Tao}, \citenamefont {Vaulin},\ and\ \citenamefont {Wang}}]{Biswas2013}%
  \BibitemOpen
  \bibfield  {author} {\bibinfo {author} {\bibfnamefont {R.}~\bibnamefont
  {Biswas}}, \bibinfo {author} {\bibfnamefont {L.}~\bibnamefont {Blackburn}},
  \bibinfo {author} {\bibfnamefont {J.}~\bibnamefont {Cao}}, \bibinfo {author}
  {\bibfnamefont {R.}~\bibnamefont {Essick}}, \bibinfo {author} {\bibfnamefont
  {K.~A.}\ \bibnamefont {Hodge}}, \bibinfo {author} {\bibfnamefont
  {E.}~\bibnamefont {Katsavounidis}}, \bibinfo {author} {\bibfnamefont
  {K.}~\bibnamefont {Kim}}, \bibinfo {author} {\bibfnamefont {Y.-M.}\
  \bibnamefont {Kim}}, \bibinfo {author} {\bibfnamefont {E.-O.}\ \bibnamefont
  {Le~Bigot}}, \bibinfo {author} {\bibfnamefont {C.-H.}\ \bibnamefont {Lee}},
  \bibinfo {author} {\bibfnamefont {J.~J.}\ \bibnamefont {Oh}}, \bibinfo
  {author} {\bibfnamefont {S.~H.}\ \bibnamefont {Oh}}, \bibinfo {author}
  {\bibfnamefont {E.~J.}\ \bibnamefont {Son}}, \bibinfo {author} {\bibfnamefont
  {Y.}~\bibnamefont {Tao}}, \bibinfo {author} {\bibfnamefont {R.}~\bibnamefont
  {Vaulin}}, \ and\ \bibinfo {author} {\bibfnamefont {X.}~\bibnamefont
  {Wang}},\ }\href {\doibase 10.1103/PhysRevD.88.062003} {\bibfield  {journal}
  {\bibinfo  {journal} {Phys. Rev. D}\ }\textbf {\bibinfo {volume} {88}},\
  \bibinfo {pages} {062003} (\bibinfo {year} {2013})}\BibitemShut {NoStop}%
\bibitem [{\citenamefont {Essick}\ \emph {et~al.}(2013)\citenamefont {Essick},
  \citenamefont {Blackburn},\ and\ \citenamefont {Katsavounidis}}]{Essick2013}%
  \BibitemOpen
  \bibfield  {author} {\bibinfo {author} {\bibfnamefont {R.}~\bibnamefont
  {Essick}}, \bibinfo {author} {\bibfnamefont {L.}~\bibnamefont {Blackburn}}, \
  and\ \bibinfo {author} {\bibfnamefont {E.}~\bibnamefont {Katsavounidis}},\
  }\href {\doibase 10.1088/0264-9381/30/15/155010} {\bibfield  {journal}
  {\bibinfo  {journal} {Classical and Quantum Gravity}\ }\textbf {\bibinfo
  {volume} {30}},\ \bibinfo {pages} {155010} (\bibinfo {year}
  {2013})}\BibitemShut {NoStop}%
\bibitem [{\citenamefont {Smith}\ \emph {et~al.}(2011)\citenamefont {Smith},
  \citenamefont {Abbott}, \citenamefont {Hirose}, \citenamefont {Leroy},
  \citenamefont {MacLeod}, \citenamefont {McIver}, \citenamefont {Saulson},\
  and\ \citenamefont {Shawhan}}]{Smith2011}%
  \BibitemOpen
  \bibfield  {author} {\bibinfo {author} {\bibfnamefont {J.~R.}\ \bibnamefont
  {Smith}}, \bibinfo {author} {\bibfnamefont {T.}~\bibnamefont {Abbott}},
  \bibinfo {author} {\bibfnamefont {E.}~\bibnamefont {Hirose}}, \bibinfo
  {author} {\bibfnamefont {N.}~\bibnamefont {Leroy}}, \bibinfo {author}
  {\bibfnamefont {D.}~\bibnamefont {MacLeod}}, \bibinfo {author} {\bibfnamefont
  {J.}~\bibnamefont {McIver}}, \bibinfo {author} {\bibfnamefont
  {P.}~\bibnamefont {Saulson}}, \ and\ \bibinfo {author} {\bibfnamefont
  {P.}~\bibnamefont {Shawhan}},\ }\href {\doibase
  10.1088/0264-9381/28/23/235005} {\bibfield  {journal} {\bibinfo  {journal}
  {Classical and Quantum Gravity}\ }\textbf {\bibinfo {volume} {28}},\ \bibinfo
  {pages} {235005} (\bibinfo {year} {2011})}\BibitemShut {NoStop}%
\bibitem [{\citenamefont {Isogai}\ and\ \citenamefont {the Ligo Scientific
  Collaboration~a Collaboration}(2010)}]{Isogai2010}%
  \BibitemOpen
  \bibfield  {author} {\bibinfo {author} {\bibfnamefont {T.}~\bibnamefont
  {Isogai}}\ and\ \bibinfo {author} {\bibnamefont {the Ligo Scientific
  Collaboration~a Collaboration}},\ }\href {\doibase
  10.1088/1742-6596/243/1/012005} {\bibfield  {journal} {\bibinfo  {journal}
  {Journal of Physics: Conference Series}\ }\textbf {\bibinfo {volume} {243}},\
  \bibinfo {pages} {012005} (\bibinfo {year} {2010})}\BibitemShut {NoStop}%
\bibitem [{\citenamefont {{Cavaglia}}\ \emph {et~al.}(2018)\citenamefont
  {{Cavaglia}}, \citenamefont {{Staats}},\ and\ \citenamefont
  {{Gill}}}]{Cavaglia2018}%
  \BibitemOpen
  \bibfield  {author} {\bibinfo {author} {\bibfnamefont {M.}~\bibnamefont
  {{Cavaglia}}}, \bibinfo {author} {\bibfnamefont {K.}~\bibnamefont
  {{Staats}}}, \ and\ \bibinfo {author} {\bibfnamefont {T.}~\bibnamefont
  {{Gill}}},\ }\href@noop {} {\bibfield  {journal} {\bibinfo  {journal} {arXiv
  e-prints}\ ,\ \bibinfo {eid} {arXiv:1812.05225}} (\bibinfo {year} {2018})},\
  \Eprint {http://arxiv.org/abs/1812.05225} {arXiv:1812.05225
  [physics.data-an]} \BibitemShut {NoStop}%
\bibitem [{\citenamefont {Colgan}\ \emph {et~al.}(2019)\citenamefont {Colgan},
  \citenamefont {Corley}, \citenamefont {Lau}, \citenamefont {Bartos},
  \citenamefont {Wright}, \citenamefont {Marka},\ and\ \citenamefont
  {Marka}}]{Colgan:2019lyo}%
  \BibitemOpen
  \bibfield  {author} {\bibinfo {author} {\bibfnamefont {R.~E.}\ \bibnamefont
  {Colgan}}, \bibinfo {author} {\bibfnamefont {K.~R.}\ \bibnamefont {Corley}},
  \bibinfo {author} {\bibfnamefont {Y.}~\bibnamefont {Lau}}, \bibinfo {author}
  {\bibfnamefont {I.}~\bibnamefont {Bartos}}, \bibinfo {author} {\bibfnamefont
  {J.~N.}\ \bibnamefont {Wright}}, \bibinfo {author} {\bibfnamefont
  {Z.}~\bibnamefont {Marka}}, \ and\ \bibinfo {author} {\bibfnamefont
  {S.}~\bibnamefont {Marka}},\ }\href@noop {} {\  (\bibinfo {year} {2019})},\
  \Eprint {http://arxiv.org/abs/1911.11831} {arXiv:1911.11831 [astro-ph.IM]}
  \BibitemShut {NoStop}%
\bibitem [{\citenamefont {Abbott}\ \emph
  {et~al.}(2017{\natexlab{c}})\citenamefont {Abbott} \emph
  {et~al.}}]{GBM:2017lvd}%
  \BibitemOpen
  \bibfield  {author} {\bibinfo {author} {\bibfnamefont {B.}~\bibnamefont
  {Abbott}} \emph {et~al.} (\bibinfo {collaboration} {LIGO Scientific, Virgo,
  Fermi GBM, INTEGRAL, IceCube, AstroSat Cadmium Zinc Telluride Imager Team,
  IPN, Insight-Hxmt, ANTARES, Swift, AGILE Team, 1M2H Team, Dark Energy Camera
  GW-EM, DES, DLT40, GRAWITA, Fermi-LAT, ATCA, ASKAP, Las Cumbres Observatory
  Group, OzGrav, DWF (Deeper Wider Faster Program), AST3, CAASTRO, VINROUGE,
  MASTER, J-GEM, GROWTH, JAGWAR, CaltechNRAO, TTU-NRAO, NuSTAR, Pan-STARRS,
  MAXI Team, TZAC Consortium, KU, Nordic Optical Telescope, ePESSTO, GROND,
  Texas Tech University, SALT Group, TOROS, BOOTES, MWA, CALET, IKI-GW
  Follow-up, H.E.S.S., LOFAR, LWA, HAWC, Pierre Auger, ALMA, Euro VLBI Team, Pi
  of Sky, Chandra Team at McGill University, DFN, ATLAS Telescopes, High Time
  Resolution Universe Survey, RIMAS, RATIR, SKA South Africa/MeerKAT}),\ }\href
  {\doibase 10.3847/2041-8213/aa91c9} {\bibfield  {journal} {\bibinfo
  {journal} {Astrophys.\ J.}\ }\textbf {\bibinfo {volume} {848}},\ \bibinfo
  {pages} {L12} (\bibinfo {year} {2017}{\natexlab{c}})},\ \Eprint
  {http://arxiv.org/abs/1710.05833} {arXiv:1710.05833 [astro-ph.HE]}
  \BibitemShut {NoStop}%
\bibitem [{\citenamefont {Abbott}\ \emph
  {et~al.}(2019{\natexlab{b}})\citenamefont {Abbott} \emph
  {et~al.}}]{LIGOScientific:2019gag}%
  \BibitemOpen
  \bibfield  {author} {\bibinfo {author} {\bibfnamefont {B.~P.}\ \bibnamefont
  {Abbott}} \emph {et~al.} (\bibinfo {collaboration} {LIGO Scientific,
  Virgo}),\ }\href {\doibase 10.3847/1538-4357/ab0e8f} {\bibfield  {journal}
  {\bibinfo  {journal} {Astrophys. J.}\ }\textbf {\bibinfo {volume} {875}},\
  \bibinfo {pages} {161} (\bibinfo {year} {2019}{\natexlab{b}})},\ \Eprint
  {http://arxiv.org/abs/1901.03310} {arXiv:1901.03310 [astro-ph.HE]}
  \BibitemShut {NoStop}%
\bibitem [{\citenamefont {Vajente}\ \emph {et~al.}(2020)\citenamefont
  {Vajente}, \citenamefont {Huang}, \citenamefont {Isi}, \citenamefont
  {Driggers}, \citenamefont {Kissel}, \citenamefont
  {Szczepa\ifmmode~\acute{n}\else \'{n}\fi{}czyk},\ and\ \citenamefont
  {Vitale}}]{PhysRevD.101.042003}%
  \BibitemOpen
  \bibfield  {author} {\bibinfo {author} {\bibfnamefont {G.}~\bibnamefont
  {Vajente}}, \bibinfo {author} {\bibfnamefont {Y.}~\bibnamefont {Huang}},
  \bibinfo {author} {\bibfnamefont {M.}~\bibnamefont {Isi}}, \bibinfo {author}
  {\bibfnamefont {J.~C.}\ \bibnamefont {Driggers}}, \bibinfo {author}
  {\bibfnamefont {J.~S.}\ \bibnamefont {Kissel}}, \bibinfo {author}
  {\bibfnamefont {M.~J.}\ \bibnamefont {Szczepa\ifmmode~\acute{n}\else
  \'{n}\fi{}czyk}}, \ and\ \bibinfo {author} {\bibfnamefont {S.}~\bibnamefont
  {Vitale}},\ }\href {\doibase 10.1103/PhysRevD.101.042003} {\bibfield
  {journal} {\bibinfo  {journal} {Phys. Rev. D}\ }\textbf {\bibinfo {volume}
  {101}},\ \bibinfo {pages} {042003} (\bibinfo {year} {2020})}\BibitemShut
  {NoStop}%
\bibitem [{\citenamefont {Ormiston}\ \emph {et~al.}(2020)\citenamefont
  {Ormiston}, \citenamefont {Nguyen}, \citenamefont {Coughlin}, \citenamefont
  {Adhikari},\ and\ \citenamefont {Katsavounidis}}]{Ormiston:2020ele}%
  \BibitemOpen
  \bibfield  {author} {\bibinfo {author} {\bibfnamefont {R.}~\bibnamefont
  {Ormiston}}, \bibinfo {author} {\bibfnamefont {T.}~\bibnamefont {Nguyen}},
  \bibinfo {author} {\bibfnamefont {M.}~\bibnamefont {Coughlin}}, \bibinfo
  {author} {\bibfnamefont {R.~X.}\ \bibnamefont {Adhikari}}, \ and\ \bibinfo
  {author} {\bibfnamefont {E.}~\bibnamefont {Katsavounidis}},\ }\href@noop {}
  {\  (\bibinfo {year} {2020})},\ \Eprint {http://arxiv.org/abs/2005.06534}
  {arXiv:2005.06534 [astro-ph.IM]} \BibitemShut {NoStop}%
\bibitem [{\citenamefont {Abbott}\ \emph
  {et~al.}(2016{\natexlab{d}})\citenamefont {Abbott} \emph
  {et~al.}}]{TheLIGOScientific:2016zmo}%
  \BibitemOpen
  \bibfield  {author} {\bibinfo {author} {\bibfnamefont {B.~P.}\ \bibnamefont
  {Abbott}} \emph {et~al.} (\bibinfo {collaboration} {LIGO Scientific,
  Virgo}),\ }\href {\doibase 10.1088/0264-9381/33/13/134001} {\bibfield
  {journal} {\bibinfo  {journal} {Class. Quant. Grav.}\ }\textbf {\bibinfo
  {volume} {33}},\ \bibinfo {pages} {134001} (\bibinfo {year}
  {2016}{\natexlab{d}})},\ \Eprint {http://arxiv.org/abs/1602.03844}
  {arXiv:1602.03844 [gr-qc]} \BibitemShut {NoStop}%
\bibitem [{Note1()}]{Note1}%
  \BibitemOpen
  \bibinfo {note} {We note that elevated detection rates expected with advanced
  detectors at design sensitivity and other planned detectors may cause this
  assumption to break down, necessitating further curation of training sets
  within our supervised learning framework.}\BibitemShut {Stop}%
\bibitem [{Note2()}]{Note2}%
  \BibitemOpen
  \bibinfo {note} {Note that a few key detections were essentially made with
  data from a single interferometer (e.g., GW170817 was initially detected as a
  a single-interferometer trigger at LIGO Hanford) and therefore requiring
  coincidences between detectors may not remove all astrophysical
  signals.}\BibitemShut {Stop}%
\bibitem [{Note3()}]{Note3}%
  \BibitemOpen
  \bibinfo {note} {The assumption of independent noise in each detector may
  break down in certain cases, like correlated magnetic noise due to Schumann
  resonances~\cite {Schumann1, Schumann2}.}\BibitemShut {Stop}%
\bibitem [{\citenamefont {Chatterji}\ \emph {et~al.}(2004)\citenamefont
  {Chatterji}, \citenamefont {Blackburn}, \citenamefont {Martin},\ and\
  \citenamefont {Katsavounidis}}]{Chatterji2004}%
  \BibitemOpen
  \bibfield  {author} {\bibinfo {author} {\bibfnamefont {S.}~\bibnamefont
  {Chatterji}}, \bibinfo {author} {\bibfnamefont {L.}~\bibnamefont
  {Blackburn}}, \bibinfo {author} {\bibfnamefont {G.}~\bibnamefont {Martin}}, \
  and\ \bibinfo {author} {\bibfnamefont {E.}~\bibnamefont {Katsavounidis}},\
  }\href {\doibase 10.1088/0264-9381/21/20/024} {\bibfield  {journal} {\bibinfo
   {journal} {Classical and Quantum Gravity}\ }\textbf {\bibinfo {volume}
  {21}},\ \bibinfo {pages} {S1809} (\bibinfo {year} {2004})}\BibitemShut
  {NoStop}%
\bibitem [{\citenamefont {Godwin}(2020)}]{godwin-thesis}%
  \BibitemOpen
  \bibfield  {author} {\bibinfo {author} {\bibfnamefont {P.}~\bibnamefont
  {Godwin}},\ }\href@noop {} {\enquote {\bibinfo {title} {Low-latency
  statistical data quality in the era of multi-messenger astronomy},}\ }
  (\bibinfo {year} {2020})\BibitemShut {NoStop}%
\bibitem [{\citenamefont {Essick}(2020)}]{pointypoisson}%
  \BibitemOpen
  \bibfield  {author} {\bibinfo {author} {\bibfnamefont {R.}~\bibnamefont
  {Essick}},\ }\href@noop {} {\bibfield  {journal} {\bibinfo  {journal}
  {\emph{in prep.}}\ } (\bibinfo {year} {2020})}\BibitemShut {NoStop}%
\bibitem [{\citenamefont {Abbott}\ \emph
  {et~al.}(2016{\natexlab{e}})\citenamefont {Abbott} \emph
  {et~al.}}]{GW150914DetChar}%
  \BibitemOpen
  \bibfield  {author} {\bibinfo {author} {\bibfnamefont {B.~P.}\ \bibnamefont
  {Abbott}} \emph {et~al.},\ }\href {\doibase 10.1088/0264-9381/33/13/134001}
  {\bibfield  {journal} {\bibinfo  {journal} {Classical and Quantum Gravity}\
  }\textbf {\bibinfo {volume} {33}},\ \bibinfo {pages} {134001} (\bibinfo
  {year} {2016}{\natexlab{e}})}\BibitemShut {NoStop}%
\bibitem [{\citenamefont {Neyman}\ and\ \citenamefont
  {Pearson}(1933)}]{Neyman:1933wgr}%
  \BibitemOpen
  \bibfield  {author} {\bibinfo {author} {\bibfnamefont {J.}~\bibnamefont
  {Neyman}}\ and\ \bibinfo {author} {\bibfnamefont {E.~S.}\ \bibnamefont
  {Pearson}},\ }\href {\doibase 10.1098/rsta.1933.0009} {\bibfield  {journal}
  {\bibinfo  {journal} {Phil. Trans. Roy. Soc. Lond.}\ }\textbf {\bibinfo
  {volume} {A231}},\ \bibinfo {pages} {289} (\bibinfo {year}
  {1933})}\BibitemShut {NoStop}%
\bibitem [{\citenamefont {Fisher}\ \emph {et~al.}()\citenamefont {Fisher},
  \citenamefont {Hemming}, \citenamefont {Verkindt}, \citenamefont {Bizouard},
  \citenamefont {Couvares}, \citenamefont {Robinet},\ and\ \citenamefont
  {Brown}}]{dqsegdb}%
  \BibitemOpen
  \bibfield  {author} {\bibinfo {author} {\bibfnamefont {R.}~\bibnamefont
  {Fisher}}, \bibinfo {author} {\bibfnamefont {G.}~\bibnamefont {Hemming}},
  \bibinfo {author} {\bibfnamefont {D.}~\bibnamefont {Verkindt}}, \bibinfo
  {author} {\bibfnamefont {M.-A.}\ \bibnamefont {Bizouard}}, \bibinfo {author}
  {\bibfnamefont {P.}~\bibnamefont {Couvares}}, \bibinfo {author}
  {\bibfnamefont {F.}~\bibnamefont {Robinet}}, \ and\ \bibinfo {author}
  {\bibfnamefont {D.}~\bibnamefont {Brown}},\ }\href@noop {} {\bibinfo
  {journal} {in prep.}\ }\BibitemShut {NoStop}%
\bibitem [{Note4()}]{Note4}%
  \BibitemOpen
\bibfield  {journal} {  }\bibinfo {note} {There is some evidence that our
  feature vectors are relatively sparse (i.e., it is rare to have multiple
  coincident auxiliary transients in a single channel) and the relevant
  information encoded in the feature sets is simply whether or not there were
  any non-Gaussian transients in the auxiliary channel within the specified
  window. In that case, it is perhaps not surprising that the \protect \emph
  {select-loudest} algorithm retains all the relevant information.}\BibitemShut
  {Stop}%
\bibitem [{\citenamefont {Pedregosa}\ \emph {et~al.}(2011)\citenamefont
  {Pedregosa}, \citenamefont {Varoquaux}, \citenamefont {Gramfort},
  \citenamefont {Michel}, \citenamefont {Thirion}, \citenamefont {Grisel},
  \citenamefont {Blondel}, \citenamefont {Prettenhofer}, \citenamefont {Weiss},
  \citenamefont {Dubourg}, \citenamefont {Vanderplas}, \citenamefont {Passos},
  \citenamefont {Cournapeau}, \citenamefont {Brucher}, \citenamefont {Perrot},\
  and\ \citenamefont {Duchesnay}}]{scikit-learn}%
  \BibitemOpen
  \bibfield  {author} {\bibinfo {author} {\bibfnamefont {F.}~\bibnamefont
  {Pedregosa}}, \bibinfo {author} {\bibfnamefont {G.}~\bibnamefont
  {Varoquaux}}, \bibinfo {author} {\bibfnamefont {A.}~\bibnamefont {Gramfort}},
  \bibinfo {author} {\bibfnamefont {V.}~\bibnamefont {Michel}}, \bibinfo
  {author} {\bibfnamefont {B.}~\bibnamefont {Thirion}}, \bibinfo {author}
  {\bibfnamefont {O.}~\bibnamefont {Grisel}}, \bibinfo {author} {\bibfnamefont
  {M.}~\bibnamefont {Blondel}}, \bibinfo {author} {\bibfnamefont
  {P.}~\bibnamefont {Prettenhofer}}, \bibinfo {author} {\bibfnamefont
  {R.}~\bibnamefont {Weiss}}, \bibinfo {author} {\bibfnamefont
  {V.}~\bibnamefont {Dubourg}}, \bibinfo {author} {\bibfnamefont
  {J.}~\bibnamefont {Vanderplas}}, \bibinfo {author} {\bibfnamefont
  {A.}~\bibnamefont {Passos}}, \bibinfo {author} {\bibfnamefont
  {D.}~\bibnamefont {Cournapeau}}, \bibinfo {author} {\bibfnamefont
  {M.}~\bibnamefont {Brucher}}, \bibinfo {author} {\bibfnamefont
  {M.}~\bibnamefont {Perrot}}, \ and\ \bibinfo {author} {\bibfnamefont
  {E.}~\bibnamefont {Duchesnay}},\ }\href@noop {} {\bibfield  {journal}
  {\bibinfo  {journal} {Journal of Machine Learning Research}\ }\textbf
  {\bibinfo {volume} {12}},\ \bibinfo {pages} {2825} (\bibinfo {year}
  {2011})}\BibitemShut {NoStop}%
\bibitem [{\citenamefont {Chollet}\ \emph {et~al.}(2015)\citenamefont {Chollet}
  \emph {et~al.}}]{chollet2015keras}%
  \BibitemOpen
  \bibfield  {author} {\bibinfo {author} {\bibfnamefont {F.}~\bibnamefont
  {Chollet}} \emph {et~al.},\ }\href@noop {} {\enquote {\bibinfo {title}
  {Keras},}\ }\bibinfo {howpublished} {\url{https://keras.io}} (\bibinfo {year}
  {2015})\BibitemShut {NoStop}%
\bibitem [{\citenamefont {Chen}\ and\ \citenamefont
  {Guestrin}(2016)}]{Chen:2016btl}%
  \BibitemOpen
  \bibfield  {author} {\bibinfo {author} {\bibfnamefont {T.}~\bibnamefont
  {Chen}}\ and\ \bibinfo {author} {\bibfnamefont {C.}~\bibnamefont
  {Guestrin}},\ }\href {\doibase 10.1145/2939672.2939785} {\  (\bibinfo {year}
  {2016}),\ 10.1145/2939672.2939785},\ \Eprint
  {http://arxiv.org/abs/1603.02754} {arXiv:1603.02754 [cs.LG]} \BibitemShut
  {NoStop}%
\bibitem [{\citenamefont {{Tulio Ribeiro}}\ \emph {et~al.}(2016)\citenamefont
  {{Tulio Ribeiro}}, \citenamefont {{Singh}},\ and\ \citenamefont
  {{Guestrin}}}]{2016arXiv160204938T}%
  \BibitemOpen
  \bibfield  {author} {\bibinfo {author} {\bibfnamefont {M.}~\bibnamefont
  {{Tulio Ribeiro}}}, \bibinfo {author} {\bibfnamefont {S.}~\bibnamefont
  {{Singh}}}, \ and\ \bibinfo {author} {\bibfnamefont {C.}~\bibnamefont
  {{Guestrin}}},\ }\href@noop {} {\bibfield  {journal} {\bibinfo  {journal}
  {arXiv e-prints}\ ,\ \bibinfo {eid} {arXiv:1602.04938}} (\bibinfo {year}
  {2016})},\ \Eprint {http://arxiv.org/abs/1602.04938} {arXiv:1602.04938
  [cs.LG]} \BibitemShut {NoStop}%
\bibitem [{\citenamefont {Blackburn}(2007)}]{kleine-welle}%
  \BibitemOpen
  \bibfield  {author} {\bibinfo {author} {\bibfnamefont {L.}~\bibnamefont
  {Blackburn}},\ }\href@noop {} {\enquote {\bibinfo {title} {Kleine-welle
  algorithm},}\ }\bibinfo {howpublished}
  {\url{https://dcc.ligo.org/public/0027/T060221/000/T060221-00.pdf}} (\bibinfo
  {year} {2007})\BibitemShut {NoStop}%
\bibitem [{\citenamefont {Lynch}\ \emph {et~al.}(2017)\citenamefont {Lynch},
  \citenamefont {Vitale}, \citenamefont {Essick}, \citenamefont
  {Katsavounidis},\ and\ \citenamefont {Robinet}}]{Lynch2017}%
  \BibitemOpen
  \bibfield  {author} {\bibinfo {author} {\bibfnamefont {R.}~\bibnamefont
  {Lynch}}, \bibinfo {author} {\bibfnamefont {S.}~\bibnamefont {Vitale}},
  \bibinfo {author} {\bibfnamefont {R.}~\bibnamefont {Essick}}, \bibinfo
  {author} {\bibfnamefont {E.}~\bibnamefont {Katsavounidis}}, \ and\ \bibinfo
  {author} {\bibfnamefont {F.}~\bibnamefont {Robinet}},\ }\href {\doibase
  10.1103/PhysRevD.95.104046} {\bibfield  {journal} {\bibinfo  {journal} {Phys.
  Rev. D}\ }\textbf {\bibinfo {volume} {95}},\ \bibinfo {pages} {104046}
  (\bibinfo {year} {2017})}\BibitemShut {NoStop}%
\bibitem [{\citenamefont {Pankow}\ \emph {et~al.}(2018)\citenamefont {Pankow}
  \emph {et~al.}}]{Pankow:2018qpo}%
  \BibitemOpen
  \bibfield  {author} {\bibinfo {author} {\bibfnamefont {C.}~\bibnamefont
  {Pankow}} \emph {et~al.},\ }\href {\doibase 10.1103/PhysRevD.98.084016}
  {\bibfield  {journal} {\bibinfo  {journal} {Phys. Rev.}\ }\textbf {\bibinfo
  {volume} {D98}},\ \bibinfo {pages} {084016} (\bibinfo {year} {2018})},\
  \Eprint {http://arxiv.org/abs/1808.03619} {arXiv:1808.03619 [gr-qc]}
  \BibitemShut {NoStop}%
\bibitem [{gra()}]{gracedb}%
  \BibitemOpen
  \href@noop {} {\enquote {\bibinfo {title} {Gracedb},}\ }\bibinfo
  {howpublished} {\url{https://gracedb.ligo.org/}}\BibitemShut {NoStop}%
\bibitem [{\citenamefont {{Reed Essick (on behalf of the LIGO-Virgo Scientific
  Collaborations)}}(2017)}]{GW170817GCN}%
  \BibitemOpen
  \bibfield  {author} {\bibinfo {author} {\bibnamefont {{Reed Essick (on behalf
  of the LIGO-Virgo Scientific Collaborations)}}},\ }\href@noop {} {\bibfield
  {journal} {\bibinfo  {journal} {https://gcn.gsfc.nasa.gov/gcn3/21509.gcn3}\ }
  (\bibinfo {year} {2017})}\BibitemShut {NoStop}%
\bibitem [{\citenamefont {{Brandon Piotrzkowski (on behalf of the LIGO-Virgo
  Scientific Collaborations)}}(2019)}]{S190822cGCN}%
  \BibitemOpen
  \bibfield  {author} {\bibinfo {author} {\bibnamefont {{Brandon Piotrzkowski
  (on behalf of the LIGO-Virgo Scientific Collaborations)}}},\ }\href@noop {}
  {\bibfield  {journal} {\bibinfo  {journal}
  {https://gcn.gsfc.nasa.gov/gcn3/25442.gcn3}\ } (\bibinfo {year}
  {2019})}\BibitemShut {NoStop}%
\bibitem [{\citenamefont {Abbott}\ \emph
  {et~al.}(2016{\natexlab{f}})\citenamefont {Abbott} \emph
  {et~al.}}]{Abbott:2016nmj}%
  \BibitemOpen
  \bibfield  {author} {\bibinfo {author} {\bibfnamefont {B.~P.}\ \bibnamefont
  {Abbott}} \emph {et~al.} (\bibinfo {collaboration} {LIGO Scientific,
  Virgo}),\ }\href {\doibase 10.1103/PhysRevLett.116.241103} {\bibfield
  {journal} {\bibinfo  {journal} {Phys. Rev. Lett.}\ }\textbf {\bibinfo
  {volume} {116}},\ \bibinfo {pages} {241103} (\bibinfo {year}
  {2016}{\natexlab{f}})},\ \Eprint {http://arxiv.org/abs/1606.04855}
  {arXiv:1606.04855 [gr-qc]} \BibitemShut {NoStop}%
\bibitem [{\citenamefont {Abbott}\ \emph
  {et~al.}(2016{\natexlab{g}})\citenamefont {Abbott} \emph
  {et~al.}}]{PhysRevLett.116.241103}%
  \BibitemOpen
  \bibfield  {author} {\bibinfo {author} {\bibfnamefont {B.~P.}\ \bibnamefont
  {Abbott}} \emph {et~al.} (\bibinfo {collaboration} {LIGO Scientific
  Collaboration and Virgo Collaboration}),\ }\href {\doibase
  10.1103/PhysRevLett.116.241103} {\bibfield  {journal} {\bibinfo  {journal}
  {Phys. Rev. Lett.}\ }\textbf {\bibinfo {volume} {116}},\ \bibinfo {pages}
  {241103} (\bibinfo {year} {2016}{\natexlab{g}})}\BibitemShut {NoStop}%
\bibitem [{\citenamefont {Abbott}\ \emph
  {et~al.}(2019{\natexlab{c}})\citenamefont {Abbott} \emph
  {et~al.}}]{O1O2Emfollow}%
  \BibitemOpen
  \bibfield  {author} {\bibinfo {author} {\bibfnamefont {B.~P.}\ \bibnamefont
  {Abbott}} \emph {et~al.},\ }\href {\doibase 10.3847/1538-4357/ab0e8f}
  {\bibfield  {journal} {\bibinfo  {journal} {The Astrophysical Journal}\
  }\textbf {\bibinfo {volume} {875}},\ \bibinfo {pages} {161} (\bibinfo {year}
  {2019}{\natexlab{c}})}\BibitemShut {NoStop}%
\bibitem [{\citenamefont {Godwin}\ \emph {et~al.}(2020)\citenamefont {Godwin}
  \emph {et~al.}}]{iDQ+GstLAL}%
  \BibitemOpen
  \bibfield  {author} {\bibinfo {author} {\bibfnamefont {P.}~\bibnamefont
  {Godwin}} \emph {et~al.},\ }\href@noop {} {\bibfield  {journal} {\bibinfo
  {journal} {in prep.}\ } (\bibinfo {year} {2020})}\BibitemShut {NoStop}%
\bibitem [{\citenamefont {Messick}\ \emph {et~al.}(2017)\citenamefont {Messick}
  \emph {et~al.}}]{Messick:2016aqy}%
  \BibitemOpen
  \bibfield  {author} {\bibinfo {author} {\bibfnamefont {C.}~\bibnamefont
  {Messick}} \emph {et~al.},\ }\href {\doibase 10.1103/PhysRevD.95.042001}
  {\bibfield  {journal} {\bibinfo  {journal} {Phys. Rev.}\ }\textbf {\bibinfo
  {volume} {D95}},\ \bibinfo {pages} {042001} (\bibinfo {year} {2017})},\
  \Eprint {http://arxiv.org/abs/1604.04324} {arXiv:1604.04324 [astro-ph.IM]}
  \BibitemShut {NoStop}%
\bibitem [{\citenamefont {Usman}\ \emph {et~al.}(2016)\citenamefont {Usman},
  \citenamefont {Nitz}, \citenamefont {Harry}, \citenamefont {Biwer},
  \citenamefont {Brown}, \citenamefont {Cabero}, \citenamefont {Capano},
  \citenamefont {Canton}, \citenamefont {Dent}, \citenamefont {Fairhurst},
  \citenamefont {Kehl}, \citenamefont {Keppel}, \citenamefont {Krishnan},
  \citenamefont {Lenon}, \citenamefont {Lundgren}, \citenamefont {Nielsen},
  \citenamefont {Pekowsky}, \citenamefont {Pfeiffer}, \citenamefont {Saulson},
  \citenamefont {West},\ and\ \citenamefont {Willis}}]{Usman_2016}%
  \BibitemOpen
  \bibfield  {author} {\bibinfo {author} {\bibfnamefont {S.~A.}\ \bibnamefont
  {Usman}}, \bibinfo {author} {\bibfnamefont {A.~H.}\ \bibnamefont {Nitz}},
  \bibinfo {author} {\bibfnamefont {I.~W.}\ \bibnamefont {Harry}}, \bibinfo
  {author} {\bibfnamefont {C.~M.}\ \bibnamefont {Biwer}}, \bibinfo {author}
  {\bibfnamefont {D.~A.}\ \bibnamefont {Brown}}, \bibinfo {author}
  {\bibfnamefont {M.}~\bibnamefont {Cabero}}, \bibinfo {author} {\bibfnamefont
  {C.~D.}\ \bibnamefont {Capano}}, \bibinfo {author} {\bibfnamefont {T.~D.}\
  \bibnamefont {Canton}}, \bibinfo {author} {\bibfnamefont {T.}~\bibnamefont
  {Dent}}, \bibinfo {author} {\bibfnamefont {S.}~\bibnamefont {Fairhurst}},
  \bibinfo {author} {\bibfnamefont {M.~S.}\ \bibnamefont {Kehl}}, \bibinfo
  {author} {\bibfnamefont {D.}~\bibnamefont {Keppel}}, \bibinfo {author}
  {\bibfnamefont {B.}~\bibnamefont {Krishnan}}, \bibinfo {author}
  {\bibfnamefont {A.}~\bibnamefont {Lenon}}, \bibinfo {author} {\bibfnamefont
  {A.}~\bibnamefont {Lundgren}}, \bibinfo {author} {\bibfnamefont {A.~B.}\
  \bibnamefont {Nielsen}}, \bibinfo {author} {\bibfnamefont {L.~P.}\
  \bibnamefont {Pekowsky}}, \bibinfo {author} {\bibfnamefont {H.~P.}\
  \bibnamefont {Pfeiffer}}, \bibinfo {author} {\bibfnamefont {P.~R.}\
  \bibnamefont {Saulson}}, \bibinfo {author} {\bibfnamefont {M.}~\bibnamefont
  {West}}, \ and\ \bibinfo {author} {\bibfnamefont {J.~L.}\ \bibnamefont
  {Willis}},\ }\href {\doibase 10.1088/0264-9381/33/21/215004} {\bibfield
  {journal} {\bibinfo  {journal} {Classical and Quantum Gravity}\ }\textbf
  {\bibinfo {volume} {33}},\ \bibinfo {pages} {215004} (\bibinfo {year}
  {2016})}\BibitemShut {NoStop}%
\bibitem [{\citenamefont {Zackay}\ \emph {et~al.}(2019)\citenamefont {Zackay},
  \citenamefont {Venumadhav}, \citenamefont {Roulet}, \citenamefont {Dai},\
  and\ \citenamefont {Zaldarriaga}}]{Zackay:2019kkv}%
  \BibitemOpen
  \bibfield  {author} {\bibinfo {author} {\bibfnamefont {B.}~\bibnamefont
  {Zackay}}, \bibinfo {author} {\bibfnamefont {T.}~\bibnamefont {Venumadhav}},
  \bibinfo {author} {\bibfnamefont {J.}~\bibnamefont {Roulet}}, \bibinfo
  {author} {\bibfnamefont {L.}~\bibnamefont {Dai}}, \ and\ \bibinfo {author}
  {\bibfnamefont {M.}~\bibnamefont {Zaldarriaga}},\ }\href@noop {} {\
  (\bibinfo {year} {2019})},\ \Eprint {http://arxiv.org/abs/1908.05644}
  {arXiv:1908.05644 [astro-ph.IM]} \BibitemShut {NoStop}%
\bibitem [{\citenamefont {Cornish}\ and\ \citenamefont
  {Littenberg}(2015)}]{Cornish_2015}%
  \BibitemOpen
  \bibfield  {author} {\bibinfo {author} {\bibfnamefont {N.~J.}\ \bibnamefont
  {Cornish}}\ and\ \bibinfo {author} {\bibfnamefont {T.~B.}\ \bibnamefont
  {Littenberg}},\ }\href {\doibase 10.1088/0264-9381/32/13/135012} {\bibfield
  {journal} {\bibinfo  {journal} {Classical and Quantum Gravity}\ }\textbf
  {\bibinfo {volume} {32}},\ \bibinfo {pages} {135012} (\bibinfo {year}
  {2015})}\BibitemShut {NoStop}%
\bibitem [{\citenamefont {Sachdev}\ \emph {et~al.}(2019)\citenamefont {Sachdev}
  \emph {et~al.}}]{Sachdev:2019vvd}%
  \BibitemOpen
  \bibfield  {author} {\bibinfo {author} {\bibfnamefont {S.}~\bibnamefont
  {Sachdev}} \emph {et~al.},\ }\href@noop {} {\  (\bibinfo {year} {2019})},\
  \Eprint {http://arxiv.org/abs/1901.08580} {arXiv:1901.08580 [gr-qc]}
  \BibitemShut {NoStop}%
\bibitem [{\citenamefont {Klimenko}\ \emph {et~al.}(2016)\citenamefont
  {Klimenko}, \citenamefont {Vedovato}, \citenamefont {Drago}, \citenamefont
  {Salemi}, \citenamefont {Tiwari}, \citenamefont {Prodi}, \citenamefont
  {Lazzaro}, \citenamefont {Ackley}, \citenamefont {Tiwari}, \citenamefont
  {Da~Silva},\ and\ \citenamefont {Mitselmakher}}]{PhysRevD.93.042004}%
  \BibitemOpen
  \bibfield  {author} {\bibinfo {author} {\bibfnamefont {S.}~\bibnamefont
  {Klimenko}}, \bibinfo {author} {\bibfnamefont {G.}~\bibnamefont {Vedovato}},
  \bibinfo {author} {\bibfnamefont {M.}~\bibnamefont {Drago}}, \bibinfo
  {author} {\bibfnamefont {F.}~\bibnamefont {Salemi}}, \bibinfo {author}
  {\bibfnamefont {V.}~\bibnamefont {Tiwari}}, \bibinfo {author} {\bibfnamefont
  {G.~A.}\ \bibnamefont {Prodi}}, \bibinfo {author} {\bibfnamefont
  {C.}~\bibnamefont {Lazzaro}}, \bibinfo {author} {\bibfnamefont
  {K.}~\bibnamefont {Ackley}}, \bibinfo {author} {\bibfnamefont
  {S.}~\bibnamefont {Tiwari}}, \bibinfo {author} {\bibfnamefont {C.~F.}\
  \bibnamefont {Da~Silva}}, \ and\ \bibinfo {author} {\bibfnamefont
  {G.}~\bibnamefont {Mitselmakher}},\ }\href {\doibase
  10.1103/PhysRevD.93.042004} {\bibfield  {journal} {\bibinfo  {journal} {Phys.
  Rev. D}\ }\textbf {\bibinfo {volume} {93}},\ \bibinfo {pages} {042004}
  (\bibinfo {year} {2016})}\BibitemShut {NoStop}%
\bibitem [{\citenamefont {Sutton}\ \emph {et~al.}(2010)\citenamefont {Sutton},
  \citenamefont {Jones}, \citenamefont {Chatterji}, \citenamefont {Kalmus},
  \citenamefont {Leonor}, \citenamefont {Poprocki}, \citenamefont {Rollins},
  \citenamefont {Searle}, \citenamefont {Stein}, \citenamefont {Tinto},\ and\
  \citenamefont {Was}}]{Sutton_2010}%
  \BibitemOpen
  \bibfield  {author} {\bibinfo {author} {\bibfnamefont {P.~J.}\ \bibnamefont
  {Sutton}}, \bibinfo {author} {\bibfnamefont {G.}~\bibnamefont {Jones}},
  \bibinfo {author} {\bibfnamefont {S.}~\bibnamefont {Chatterji}}, \bibinfo
  {author} {\bibfnamefont {P.}~\bibnamefont {Kalmus}}, \bibinfo {author}
  {\bibfnamefont {I.}~\bibnamefont {Leonor}}, \bibinfo {author} {\bibfnamefont
  {S.}~\bibnamefont {Poprocki}}, \bibinfo {author} {\bibfnamefont
  {J.}~\bibnamefont {Rollins}}, \bibinfo {author} {\bibfnamefont
  {A.}~\bibnamefont {Searle}}, \bibinfo {author} {\bibfnamefont
  {L.}~\bibnamefont {Stein}}, \bibinfo {author} {\bibfnamefont
  {M.}~\bibnamefont {Tinto}}, \ and\ \bibinfo {author} {\bibfnamefont
  {M.}~\bibnamefont {Was}},\ }\href {\doibase 10.1088/1367-2630/12/5/053034}
  {\bibfield  {journal} {\bibinfo  {journal} {New Journal of Physics}\ }\textbf
  {\bibinfo {volume} {12}},\ \bibinfo {pages} {053034} (\bibinfo {year}
  {2010})}\BibitemShut {NoStop}%
\bibitem [{\citenamefont {Thrane}\ \emph {et~al.}(2011)\citenamefont {Thrane},
  \citenamefont {Kandhasamy}, \citenamefont {Ott}, \citenamefont {Anderson},
  \citenamefont {Christensen}, \citenamefont {Coughlin}, \citenamefont
  {Dorsher}, \citenamefont {Giampanis}, \citenamefont {Mandic}, \citenamefont
  {Mytidis}, \citenamefont {Prestegard}, \citenamefont {Raffai},\ and\
  \citenamefont {Whiting}}]{PhysRevD.83.083004}%
  \BibitemOpen
  \bibfield  {author} {\bibinfo {author} {\bibfnamefont {E.}~\bibnamefont
  {Thrane}}, \bibinfo {author} {\bibfnamefont {S.}~\bibnamefont {Kandhasamy}},
  \bibinfo {author} {\bibfnamefont {C.~D.}\ \bibnamefont {Ott}}, \bibinfo
  {author} {\bibfnamefont {W.~G.}\ \bibnamefont {Anderson}}, \bibinfo {author}
  {\bibfnamefont {N.~L.}\ \bibnamefont {Christensen}}, \bibinfo {author}
  {\bibfnamefont {M.~W.}\ \bibnamefont {Coughlin}}, \bibinfo {author}
  {\bibfnamefont {S.}~\bibnamefont {Dorsher}}, \bibinfo {author} {\bibfnamefont
  {S.}~\bibnamefont {Giampanis}}, \bibinfo {author} {\bibfnamefont
  {V.}~\bibnamefont {Mandic}}, \bibinfo {author} {\bibfnamefont
  {A.}~\bibnamefont {Mytidis}}, \bibinfo {author} {\bibfnamefont
  {T.}~\bibnamefont {Prestegard}}, \bibinfo {author} {\bibfnamefont
  {P.}~\bibnamefont {Raffai}}, \ and\ \bibinfo {author} {\bibfnamefont
  {B.}~\bibnamefont {Whiting}},\ }\href {\doibase 10.1103/PhysRevD.83.083004}
  {\bibfield  {journal} {\bibinfo  {journal} {Phys. Rev. D}\ }\textbf {\bibinfo
  {volume} {83}},\ \bibinfo {pages} {083004} (\bibinfo {year}
  {2011})}\BibitemShut {NoStop}%
\bibitem [{\citenamefont {Schumann}(1952{\natexlab{a}})}]{Schumann1}%
  \BibitemOpen
  \bibfield  {author} {\bibinfo {author} {\bibfnamefont {W.~O.}\ \bibnamefont
  {Schumann}},\ }\href
  {https://www.degruyter.com/view/journals/zna/7/2/article-p149.xml} {\bibfield
   {journal} {\bibinfo  {journal} {Zeitschrift für Naturforschung A}\ }\textbf
  {\bibinfo {volume} {7}},\ \bibinfo {pages} {149 } (\bibinfo {year}
  {1952}{\natexlab{a}})}\BibitemShut {NoStop}%
\bibitem [{\citenamefont {Schumann}(1952{\natexlab{b}})}]{Schumann2}%
  \BibitemOpen
  \bibfield  {author} {\bibinfo {author} {\bibfnamefont {W.~O.}\ \bibnamefont
  {Schumann}},\ }\href
  {https://www.degruyter.com/view/journals/zna/7/3-4/article-p250.xml}
  {\bibfield  {journal} {\bibinfo  {journal} {Zeitschrift für Naturforschung
  A}\ }\textbf {\bibinfo {volume} {7}},\ \bibinfo {pages} {250 } (\bibinfo
  {year} {1952}{\natexlab{b}})}\BibitemShut {NoStop}%
\bibitem [{\citenamefont {Abbott}\ \emph
  {et~al.}(2019{\natexlab{d}})\citenamefont {Abbott} \emph
  {et~al.}}]{PhysRevLett.122.061104}%
  \BibitemOpen
  \bibfield  {author} {\bibinfo {author} {\bibfnamefont {B.~P.}\ \bibnamefont
  {Abbott}} \emph {et~al.} (\bibinfo {collaboration} {LIGO Scientific
  Collaboration and Virgo Collaboration}),\ }\href {\doibase
  10.1103/PhysRevLett.122.061104} {\bibfield  {journal} {\bibinfo  {journal}
  {Phys. Rev. Lett.}\ }\textbf {\bibinfo {volume} {122}},\ \bibinfo {pages}
  {061104} (\bibinfo {year} {2019}{\natexlab{d}})}\BibitemShut {NoStop}%
\bibitem [{\citenamefont {Cabero}\ \emph {et~al.}(2019)\citenamefont {Cabero},
  \citenamefont {Lundgren}, \citenamefont {Nitz}, \citenamefont {Dent},
  \citenamefont {Barker}, \citenamefont {Goetz}, \citenamefont {Kissel},
  \citenamefont {Nuttall}, \citenamefont {Schale}, \citenamefont {Schofield},\
  and\ \citenamefont {Davis}}]{Cabero_2019}%
  \BibitemOpen
  \bibfield  {author} {\bibinfo {author} {\bibfnamefont {M.}~\bibnamefont
  {Cabero}}, \bibinfo {author} {\bibfnamefont {A.}~\bibnamefont {Lundgren}},
  \bibinfo {author} {\bibfnamefont {A.~H.}\ \bibnamefont {Nitz}}, \bibinfo
  {author} {\bibfnamefont {T.}~\bibnamefont {Dent}}, \bibinfo {author}
  {\bibfnamefont {D.}~\bibnamefont {Barker}}, \bibinfo {author} {\bibfnamefont
  {E.}~\bibnamefont {Goetz}}, \bibinfo {author} {\bibfnamefont {J.~S.}\
  \bibnamefont {Kissel}}, \bibinfo {author} {\bibfnamefont {L.~K.}\
  \bibnamefont {Nuttall}}, \bibinfo {author} {\bibfnamefont {P.}~\bibnamefont
  {Schale}}, \bibinfo {author} {\bibfnamefont {R.}~\bibnamefont {Schofield}}, \
  and\ \bibinfo {author} {\bibfnamefont {D.}~\bibnamefont {Davis}},\ }\href
  {\doibase 10.1088/1361-6382/ab2e14} {\bibfield  {journal} {\bibinfo
  {journal} {Classical and Quantum Gravity}\ }\textbf {\bibinfo {volume}
  {36}},\ \bibinfo {pages} {155010} (\bibinfo {year} {2019})}\BibitemShut
  {NoStop}%
\bibitem [{\citenamefont {Hastings}(1970)}]{Hastings1970}%
  \BibitemOpen
  \bibfield  {author} {\bibinfo {author} {\bibfnamefont {W.~K.}\ \bibnamefont
  {Hastings}},\ }\href {\doibase 10.1093/biomet/57.1.97} {\bibfield  {journal}
  {\bibinfo  {journal} {Biometrika}\ }\textbf {\bibinfo {volume} {57}},\
  \bibinfo {pages} {97} (\bibinfo {year} {1970})},\ \Eprint
  {http://arxiv.org/abs/https://academic.oup.com/biomet/article-pdf/57/1/97/23940249/57-1-97.pdf}
  {https://academic.oup.com/biomet/article-pdf/57/1/97/23940249/57-1-97.pdf}
  \BibitemShut {NoStop}%
\bibitem [{\citenamefont {Ising}(1925)}]{Ising1925}%
  \BibitemOpen
  \bibfield  {author} {\bibinfo {author} {\bibfnamefont {E.}~\bibnamefont
  {Ising}},\ }\href {\doibase 10.1007/BF02980577} {\bibfield  {journal}
  {\bibinfo  {journal} {Zeitschrift für Physik}\ }\textbf {\bibinfo {volume}
  {31}},\ \bibinfo {pages} {253} (\bibinfo {year} {1925})}\BibitemShut
  {NoStop}%
\end{thebibliography}%

\appendix
\onecolumngrid

\section{Gaussian Kernel Density Estimates}
\label{sec:kde}

\newcommand{\Xmin}{\ensuremath{X_\mathrm{min}}}
\newcommand{\Xmax}{\ensuremath{X_\mathrm{max}}}
\newcommand{\logL}{\ensuremath{\log\mathcal{L}}}

We review basic features of kernel density estimates (KDEs) and describe the particular one-dimensional Gaussian kernel implemented within iDQ.
For convenience, iDQ adopts a fixed bandwidth (standard deviation) for all samples and imposes \emph{reflecting boundary conditions} in order to avoid edge effects associated with the finite range of ranks ($r \in [0,1]$).
This is done by reflecting the samples across the bounds of their range so that
\begin{equation}
    \{x_i\} \rightarrow \{x_i\}\oplus\{2\Xmin-x_i\}\oplus\{2\Xmax-x_i\},
\end{equation}
which forces the KDE's derivative to vanish at \Xmin~and \Xmax.

We define a Gaussian kernel between two points ($x$, $y$) given a bandwidth ($b$) as
\begin{equation}
  K(x,y;b) = \frac{1}{\sqrt{2\pi}b} e^{-(x-y)^2 / 2b^2}.
\end{equation}
We consider a set of observed samples ($x_i$) with associated weights ($w_i$). 
The samples are assumed to be independently and identically distributed according to $p(x)$.
iDQ assigns equal weights to each sample.
Similarly, without loss of generality, we can set $\sum_i w_i = 1$, but this is not strictly necessary.
We consider the following estimate for the probability density function $p(y)$ given $b$ within the prior bounds $x_i \in [\Xmin, \Xmax]$.
\begin{gather}
  \hat{p}(y|b, \{x_i\}) = \frac{1}{\mathcal{N}} \sum\limits_i w_i \left(K(y, x_i;b) + K(y, 2\Xmin - x_i;b) + K(y, 2\Xmax - x_i; b)\right) \\
  \mathcal{N} = \int\limits_{\Xmin}^{\Xmax} dy\, \sum\limits_i w_i \left(K(y, x_i;b) + K(y, 2\Xmin - x_i;b) + K(y, 2\Xmax - x_i; b)\right) 
\end{gather}
where the observed samples $x_i$ are explicitly reflected around the prior bounds. 
This is the basic estimator used within iDQ's continuous calibration maps (Section~\ref{sec:continuous calibration maps}) to model the conditioned likelihoods given observed evaluated sample sets.
Furthermore, iDQ estimates the corresponding survival functions
\begin{align}
    \hat{P}(y|b,\{x_i\}) & = \int\limits_{y}^{\Xmax} d\gamma\, \hat{p}(\gamma|b,\{x_i\}) \\
                         & = \frac{1}{\mathcal{N}} \sum\limits_i w_i \int\limits_y^{\Xmax} d\gamma\left(K(\gamma, x_i;b) + K(\gamma, 2\Xmin - x_i;b) + K(\gamma, 2\Xmax - x_i; b) \right)
\end{align}
with cumulative normal distributions computed during a single iteration over the sample set rather than numerical integration of $\hat{p}$.
Given a bandwidth, iDQ generates a dense grid of ranks and evaluates these estimators once for each grid point.
Calibration during timeseries production (Section~\ref{sec:timeseries}) can then be performed rapidly via linear interpolation without requiring repeated iteration over the sample set, which can be quite large: $\mathcal{O}(10^4)$.

We now consider the choice of bandwidth and how to represent the uncertainty in our KDE representation given different realizations of the observed sample set.
While all KDEs are biased estimators ($E[\hat{p}(y)] \neq p(y)$) because they smooth the true distribution according to $K(x,y;b)$, our bandwidth optimization procedure finds the best representation of $p$ possible given our kernel.
In practice, then, iDQ produces correct coverage (e.g., 50\% of samples have nominal survival functions $\leq50\%$) to within the expected statistical uncertainty for stationary distributions.

\subsection{Bandwidth Optimization}
\label{sec:kde optimization}

Common practice is to define a \emph{leave-one-out} cross-validation likelihood and use this to optimize the bandwidth.
We adopt the following likelihood
\begin{equation}
  \logL(b;\{x_i\}) = \frac{1}{\sum\limits_i w_i} \sum\limits_i w_i \log\left( \frac{1}{{\sum\limits_{j \neq i} w_j}} \sum\limits_{j \neq i} w_j K(x_i, x_j;b) \right)
\end{equation}
As shown in Ref.~\cite{Lynch2017}, maximizing this likelihood is equivalent to minimizing the Kullback-Leibler divergence between the true distribution and our estimator, approximating an integral over the the measure defined by the true distribution $p(x)$ with a Monte-Carlo integral over the observed sample set.
The astute reader will note that we do not impose reflecting boundary conditions within \logL.
We expect the impact to be minor and the computational complexity is significantly lessened.

iDQ optimizes $b$ separately for $G$ and $C$ samples through direct bisection searches within prespecified prior bounds.
While this generates reliable estimators, we also note that \logL~is often quite flat near its maximum and nearby bandwidths may produce similarly well behaved estimators.
In principle, one could marginalize over the choice of bandwidth with respect to a prior
\begin{equation}
    \hat{p}_\mathrm{marg}(y|\{x_i\}) = \int\limits_{b_\mathrm{min}}^{b_\mathrm{max}} db\, p(b) \mathcal{L}(b; \{x_i\}) \hat{p}(y;b, \{x_i\}).
\end{equation}
We expect marginalization to produce more robust estimators~\cite{PhysRevLett.122.061104}, although we find that using the $b$ that maximizes \logL~works well enough in practice and avoids the additional computational burden of direct numerical marginalization.

\subsection{Representating an Estimator's Uncertainty as a $\beta$-distribution}
\label{sec:kde beta-distribution}

Because iDQ has access to only a finite number of samples, $\hat{p}(x)$ will not perfectly reproduce $p(x)$.
iDQ models this sample uncertainty with $\beta$-distributions.
Let us consider the function
\begin{align}
    f(y;b,\{x_i\}) & = \hat{p}(y|b,\{x_i\})\frac{\sqrt{2\pi}b}{3} \nonumber \\
                   & = \frac{\sqrt{2\pi}b}{3\sum\limits_i w_i} \sum\limits_i w_i \left(K(x_i, y;b) + K(2\Xmin-x_i, y;b) + K(2\Xmax-x_i, y;b)\right) 
\end{align}
Because this statistic depends on the set of random variables $\{x_i\}$, it will also follow some distribution. 
We note that $f \in (0,1]$, and therefore a $\beta$-distribution is a good candidate for compactly representing the expected uncertainty.
Specifically, we fit a $\beta$-distribution to estimates of the mean and variance of $f$, based on the observed sample set.
\begin{align}
    E\left[f(y;b)\right] & = \int \left(\prod\limits_i dx_i p(x_i)\right) f(y;b, \{x_i\}) \nonumber \\
                         & = \int \prod\limits_i dx_i p(x_i) \frac{\sqrt{2\pi}b}{3\sum\limits_j w_j}\sum\limits_j w_j \left(K(x_j, y;b) + K(2\Xmin-x_j, y;b) + K(2\Xmax-x_j, y;b)\right) \nonumber \\
                         & = \frac{\sqrt{2\pi}b}{3\sum\limits_j w_j}\sum\limits_j w_j \int dx_j p(x_j) \left(K(x_j, y;b) + K(2\Xmin-x_j, y;b) + K(2\Xmax-x_j, y;b)\right) \int \prod\limits_{i \neq j} dx_i p(x_i) \nonumber \\
                         & = \frac{\sqrt{2\pi}b}{3} \int dx p(x) \left(K(x, y;b) + K(2\Xmin-x, y;b) + K(2\Xmax-x, y;b)\right) \nonumber \\
                         & \approx \frac{\sqrt{2\pi}b}{3\sum\limits_j w_j} \sum\limits_j w_j \left(K(x_j, y;b) + K(2\Xmin-x_j, y;b) + K(2\Xmax-x_j, y;b)\right)
\end{align}
where in the last line we approximate the integral over $p(x)$ as a weighed sum of our observed samples.
We also calculate the second moment as
\begin{align}
    E\left[f^2\right] & = \int \left(\prod\limits_i dx_i p(x_i)\right) f^2 \nonumber \\
                      & = \left(\frac{\sqrt{2\pi}b}{3\sum\limits_j w_j}\right)^2 \left[ \left(\sum\limits_j w_j^2 \right) \int dx p(x) \left(K(x, y;b) + K(2\Xmin-x, y;b) + K(2\Xmax-x, y;b)\right)^2 \right. \nonumber \\
                      & \quad\quad\quad\quad \left. + \left(\sum\limits_{j, k \neq j} w_j w_k\right) \left(\int dx p(x) \left(K(x_j, y;b) + K(2\Xmin-x_j, y;b) + K(2\Xmax-x_j, y;b)\right)\right)^2 \right] \nonumber \\
                      & = \left(\frac{\sqrt{2\pi}b}{3\sum\limits_j w_j}\right)^2 \left[ \left(\sum\limits_j w_j^2 \right) \frac{1}{\sum\limits_j w_j}\sum\limits_i w_i \left(K(x_i, y;b) + K(2\Xmin-x_i, y;b) + K(2\Xmax-x_i, y;b)\right)^2 \right. \nonumber \\
                      & \quad\quad\quad\quad \left. + \left(\left(\sum\limits_{j} w_j\right)^2 - \sum\limits_j w_j^2\right) \left(\frac{1}{\sum\limits_j w_j} \sum\limits_i \left(K(x_i, y;b) + K(2\Xmin-x_i, y;b) + K(2\Xmax-x_i, y;b)\right)\right)^2 \right]
\end{align}
and thereby obtain the variance directly via $V\left[f(y;b)\right] = E\left[f^2\right] - E\left[f\right]^2$.
In the case of equal weights with $N$ samples, this yields
\begin{multline}
    V\left[f(y;b)\right] = \frac{2\pi b^2}{N}\left[ \frac{1}{N}\sum\limits_i \left(K(x_j, y;b) + K(2\Xmin-x_j, y;b) + K(2\Xmax-x_j, y;b)\right)^2 \right. \\
                            \left. - \left(\frac{1}{N}\sum\limits_i \left(K(x_j, y;b) + K(2\Xmin-x_j, y;b) + K(2\Xmax-x_j, y;b)\right)\right)^2 \right] .
\end{multline}
We note that $V_{\{x\}}[f(y;b)] \propto N^{-1}V_x[K(y,x;b)]$, as expected for a Fisher-efficient estimator.
Now, the $\beta$-distribution defined by
\begin{equation}
    \left. \text{Beta}(x;\alpha,\beta) \propto x^{\alpha-1} (1-x)^{\beta-1} \quad \right| \quad x \in [0,1]
\end{equation}
has mean and variance
\begin{gather}
    E[x] = \frac{\alpha}{\alpha+\beta} \\
    V[x] = \frac{\alpha\beta}{(\alpha+\beta)^2(\alpha+\beta+1)}
\end{gather}
We choose $\alpha$ and $\beta$ to reproduce $E[f]$ and $V[f]$.
The distribution of our estimator $\hat{p}(y;b,\{x_i\})$ is therefore given by
\begin{equation}
    \hat{p}(y;b, \{x_i\}) \sim \left(\frac{3}{\sqrt{2\pi}b}\right)\, \text{Beta}(\alpha(b,\{x_i\}), \beta(b, \{x_i\}))
\end{equation}
where the fit parameters are a function of the bandwidth and observed samples.
We note that some estimates of the mean and variance have no corresponding solution in terms of $\alpha$ and $\beta$, typically corresponding to expectation values very close to 0 or 1.
We therefore impose a minimum expectation value of $\sim 10^{-6}$ for numerical stability.

iDQ estimates the best-fit $\alpha$ and $\beta$ parameters at every point on the dense grid over ranks used to compute $\hat{p}$, interpolating between neighboring grid points as needed.
This provides rapid estimates of sampling uncertainty at the same time as point estimates within timeseries production.
A similar procedure is used to model uncertainty in the cumulative distribution $\hat{P}(x)$.
Sample uncertainty for the likelihood ratio $\Lambda^G_C$ (Eqn.~\ref{eqn:likelihood}) is obtained via Monte-Carlo sampling from the $\beta$-distributions representing $\hat{p}(r|G)$ and $\hat{p}(r|C)$.
Again, these uncertainty measures are computed once for each grid point and interpolated as needed.

\section{Updates to the Ordered Veto List (OVL) Algorithm}
\label{sec:OVL}

The Ordered Veto List (OVL,~\cite{Essick2013}) algorithm has been updated since originally published.
These changes provide greater flexibility within the algorithm as well as parallelization and other computational optimizations.
We refer readers to Ref.~\cite{Essick2013} for an introduction to OVL and focus only on the updates in what follows.

\subsection{Veto Performance Metrics}
\label{sec:OVL veto performance metrics}

First, we note that OVL is very similar in concept to both hVeto~\cite{Smith2011} and UPV~\cite{Isogai2010}, in that this class of algorithm develops a hierarchically applied list of \emph{veto configurations}, each consisting of a single auxiliary channel, a significance threshold for triggers in that channel, and a symmetric time window used to construct veto segments around noise in the auxiliary channel.
By direct optimization over many channel--threshold--window tuples, the algorithms identify a preferred order in which to apply the veto conditions.
As discussed in Ref.~\cite{Essick2013}, the ordering depends on the metric used to rank channel performance, and, indeed, this is the main difference between the original OVL implementation, hVeto, and UPV.
Specifically, OVL originally used the ratio of the marginal efficiency to the marginal deadtime.
That is, the fraction of remaining noise transients removed to the fraction of remaining time removed by the veto configuration.
This behaves similarly to a likelihood ratio test and optimizes ROC curves, subject to algorithmic constraints.
hVeto uses a measure of the Poisson significance of removing coincident noise transients that, as discussed in Ref.~\cite{Essick2013}, favors veto conditions that remove many noise transients given a fixed efficiency--deadtime ratio.
UPV orders configurations by the ratio of the number of target transients removed to the number of auxiliary transients present, thereby preferring more deterministic couplings.
However, there are counterexamples (e.g., Fig.~\ref{fig:example roc}).

Because each metric has its own merits, the updated OVL algorithm now allows users to specify the metric used to rank configurations, thereby reproducing the behavior of the original OVL, hVeto, and UPV within a single framework.
We note that OVL uses exact segment logic when constructing segments, unlike hVeto~\cite{Smith2011}, and defines the use percentage slightly differently than UPV.
OVL computes the use percentage as the ratio of target transients removed to the effective number of auxiliary transients present, defined as the quotient of the vetoed time associated with an entire veto configuration to the window used to construct the veto segments.
The effective number of auxiliary transients, then, clusters nearby auxiliary noise to avoid overcounting if many neighboring auxiliary disturbances within a single channel produce nearly identical veto segments.
Anecdotally, we find that ranking by either the efficiency-to-deadtime ratio or the use percentage routinely produce better ROC curves than ranking by the Poisson significance, in agreement with Ref.~\cite{Essick2013}.

\subsection{Training}
\label{sec:OVL training}

OVL's training scheme has also been slightly updated, although it still remains largely as described in Ref.~\cite{Essick2013}.
Specifically, OVL still trains via a nested iteration, evaluating veto configurations' performance hierarchically with a given ordering, pruning ineffective configurations, and re-ordering the list within each epoch.
Pruning is done to avoid over-training and is accomplished by setting minima on various veto configuration performance metrics, like the efficiency--deadtime ratio or Poisson significance.
The precise impact of these minima has not been quantified, but typically analysts select values to balance the loss in efficiency associated with restricting the \emph{vetolist} to only the most exceptionally well-ranked veto configurations and the generalization error introduced by over-fitting, as efficient vetoes can sometimes correspond to statistically rare accidental coincidences within the training set that do not generalize well.
For this reason, it is thought that pruning based on the Poisson significance has the largest impact on over-training.

OVL learns when it re-orders its list, as this places higher ranked veto conditions first.
Because of the hierarchical nature of OVL, we take care to re-order the veto configurations to preserve as much information as possible.
Specifically, within each re-ordering, we first sort the list to place high-threshold, small-window configurations first.
All else being equal, these should produce better veto configurations.
Only then do we order the configurations by their metrics, so that configurations with the same score are ordered to prefer high thresholds and short windows.
Pragmatically, we find this makes a small but noticeable difference in the final ordering produced by the algorithm.

\subsection{Ranks}
\label{sec:OVL ranks}

Because iDQ requires classifiers to generate ranks within the unit interval and the metrics used within OVL typically span the positive real line, we map the metric scores ($m$) into ranks ($r$) according to
\begin{equation}\label{eqn:ovl rank}
    r = \frac{m}{\xi + m},
\end{equation} where the scale ($\xi$) is fixed for each metric separately to account for their very different dynamic ranges seen with typical interferometric data.
We note that this mapping is not unique, and other functional forms would accomplish the same task.
However, Eqn.~\ref{eqn:ovl rank} distributes the ranks more uniformly over the unit interval than some other mappings, and this can help iDQ's calibration build accurate representations of the resulting conditional likelihoods.

\subsection{Feature Importance}
\label{sec:OVL feature importance}

Finally, we would be remiss if we did not discuss OVL's notion of feature importance, one of the most attractive aspects of the algorithm besides its robust performance.
Because OVL only considers a single auxiliary channel at a time and applies them in a specific order, it is straightforward to determine which auxiliary features (channel, threshold, and window) are responsible for OVL's predicted rank at any time.
In this way, OVL reports which veto configurations were active as a function of time while simultaneously reporting their relative importances as their ranks.
Indeed, feature importance as a function of time is shown in Figures~\ref{fig:example gw170817} and~\ref{fig:example whistle}.

\begin{figure}
    \includegraphics[width=1.0\textwidth, clip=True, trim=1.25cm 1.50cm 0.75cm 1.75cm]{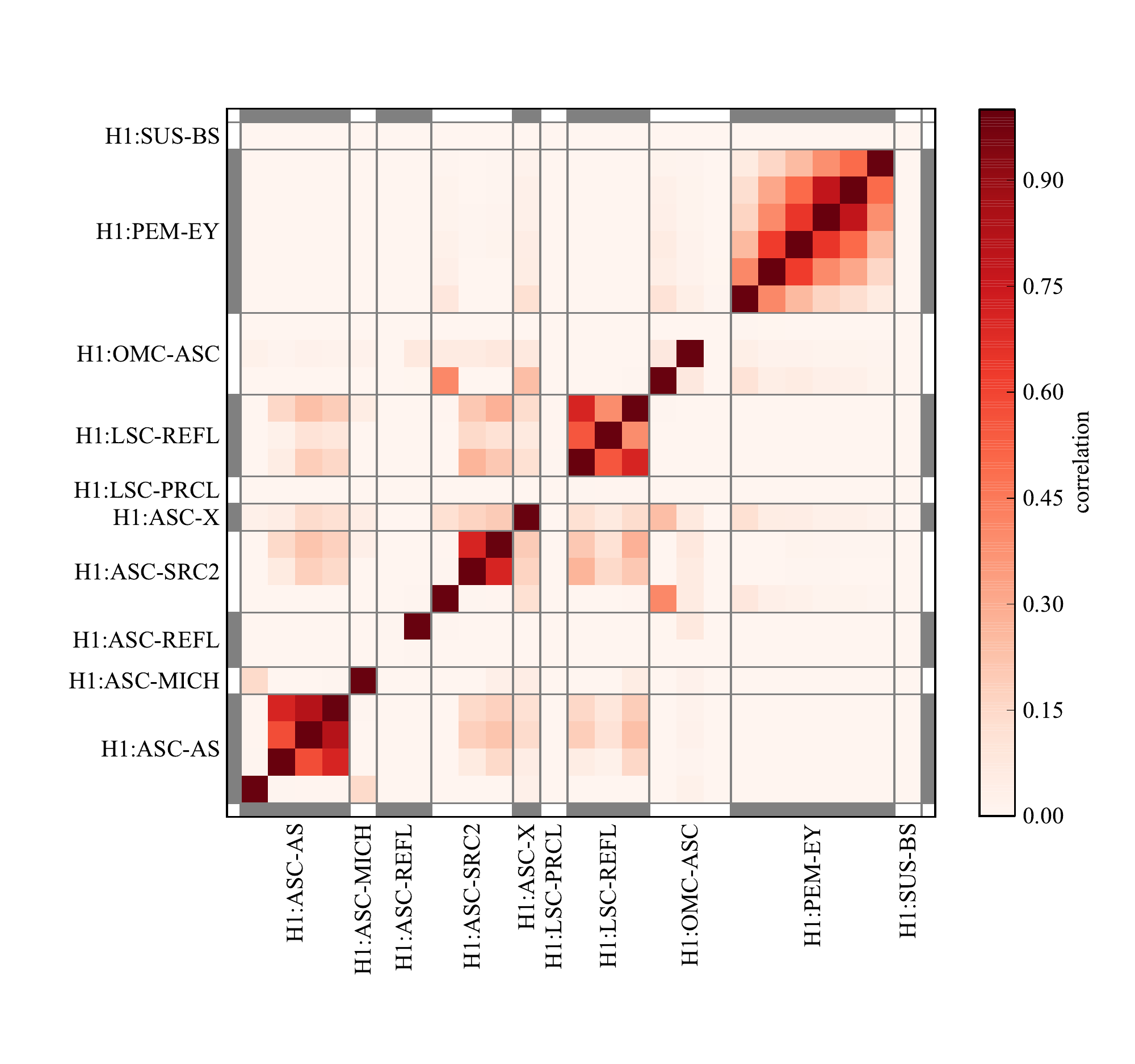}
    \caption{
        Example OVL correlation matrix showing the inter-relatedness of various veto conditions.
        Data is taken from the LIGO Hanford observatory on 16-18 August 2017 
        and the veto configurations included are based on the OVL vetolist used in low-latency during GW170817.
        For brevity, we group veto configurations by their channel subsystems and order them alphabetically.
        Diagonal elements with zero correlation correspond to veto conditions that were not active at any time in the $2\times10^5$ seconds analyzed.
    }
    \label{fig:example OVL redundancy}
\end{figure}

OVL also measures correlations between veto conditions.
Specifically, it can report the intersection of segments created by each veto configuration as a symmetric matrix.
Diagonal elements correspond to the time contained within of each set of veto segments separately.
Nearly redundant veto configurations, then, will produce intersection times close to the times of each configuration separately.
We note that OVL's training algorithm, by design, will remove redundant configurations and therefore reduce the amount of overlap in the list by applying configurations hierarchically, removing vetoed transients and time before proceeding to the next configuration.
This, combined with pruning, will tend to select a single witness configuration out of sets of highly correlated configurations.
Nonetheless, such ``covariance matrices'' between veto configurations may prove useful when diagnosing the source of specific noise transients identified by the selected auxiliary witnesses.
Figure~\ref{fig:example OVL redundancy} shows examples of correlations between the veto configurations used surrounding GW170817.

\section{Synthetic Feature Generation}
\label{sec:mock classifier data}

In addition to supporting multiple possible feature sources, each supplying distinct sets of features, iDQ can also generate synthetic data on-the-fly.
This is done for testing purposes and to benchmark algorithms.
Briefly, iDQ simulates an arbitrary number of stationary Poisson processes representing sources of noise.
Each of these synthetic processes is described by a separate rate as well as distributions over frequency and $\rho$.
Synthetic processes are then witnessed by user-specified sets of channels, and each witness records values scattered around the true value within the process (e.g., central times recorded in witness channels are Gaussian-distributed around the central times produced by the process, with separate standard deviations for each witness).
In this way, the synthetic processes entangle the features witnessed by several channels, and noise in a single channel can be modeled by a separate process witnessed only by that channel.
Each channel may witness multiple streams, generating arbitrarily complex correlations within the feature set.
This implementation realizes the probabilistic graphical model depicted in Fig.~\ref{fig:probabilistic graph}.

\section{Futher Discussion of Optimal Searches}
\label{sec:optimal appendix}

We now explore some attractive features and limiting cases of $\Lambda^S_{!S}$ (Eqn.~\ref{eqn:optimal lrt}) in order to build further intuition for how marginalization over probabilistic data quality information benefits searches.
We note that if $p(C|\vec{a})$ is \emph{binary}, that is we assume perfect knowledge of which data is clean and which is glitchy, only a single permutation retains non-trivial probability.
Current gating schemes, then, are equivalent to assuming perfect knowledge of data quality within the detectors at all times, at best an exaggeration of the current state of the field since the sources of many non-Gaussian noise transients remain unknown (e.g., \cite{Cabero_2019, GW150914DetChar}).

We also note that, assuming trivial conditioned likelihoods from iDQ ($p(r|C)=p(r|G)\ \forall\ r$), the weight assigned to each permutation is $p(\mathrm{perm})=p(C)^{N_C} p(G)^{N_G} = p(C)^{N_C} (1-p(C))^{N-N_C}$, which is just the binomial distribution with $N$ trials, $N_C$ successes, and a probability of success given by $p(C)$.
This has the appealing interpretation of marginalizing over the number and placement of glitches given knowledge about their relative frequency but nothing else.
Indeed, this is the most basic piece of data quality information that could be incorporated and, as we will see, it could already significantly improve search backgrounds.

\subsection{Toy Model}
\label{sec:optimal toy model}

Let us consider a toy model of stationary white noise in three observed data.
We assume constant prior odds for $G$ vs. $C$, but otherwise assume $\vec{a}$ is uninformative.
The marginal-maximized likelihood then becomes
\begin{align}
    p(h,\vec{a}) & = \frac{1}{(2\pi)^{3/2}\sigma^3} \left[ p(C)^3 \exp\left( -\frac{|h_1-s_1|^2 + |h_2-s_2|^2 + |h_3-s_3|^2}{2\sigma^2} \right) \nonumber \right. \\
                 & \quad\quad\quad\quad + p(C)^2 p(G) \left(\exp\left( -\frac{|h_1-s_1|^2 + |h_2-s_2|^2}{2\sigma^2} \right) \right. \nonumber \\
                 & \quad\quad\quad\quad\quad\quad\quad\quad\quad\quad\quad\quad\quad \left. + \exp\left( -\frac{|h_1-s_1|^2 + |h_3-s_3|^2}{2\sigma^2} \right) + \exp\left( -\frac{|h_2-s_2|^2 + |h_3-s_3|^2}{2\sigma^2} \right)\right) \nonumber \\
                 & \quad\quad\quad\quad + p(C) p(G)^2 \left(\exp\left( -\frac{|h_1-s_1|^2}{2\sigma^2} \right) + \exp\left( -\frac{|h_2-s_2|^2}{2\sigma^2} \right) + \exp\left( -\frac{|h_3-s_3|^2}{2\sigma^2} \right)\right) \nonumber \\
                 & \quad\quad\quad\quad \left. + p(G)^3 \right]
\end{align}
Now, let us further assume $t_1,\ t_3 \in C$ and $t_2\in G$ for concreteness such that
\begin{gather}
    h_1-s_1 = n_1 \sim \sigma \\
    h_2-s_2 = n_2+g_2 \gg \sigma \\
    h_3-s_3 = n_3 \sim \sigma
\end{gather}
In this case, we obtain
\begin{align}
    p(h,\vec{a}) & \approx \frac{1}{(2\pi)^{3/2}\sigma^3} \left[ p(C)^3 \exp\left( -\frac{|g_2|^2}{2\sigma^2} \right) + p(C)^2 p(G) \left(\exp\left(-\frac{|n_1|^2+|n_3|^2}{2\sigma^2}\right) + 2\exp\left(-\frac{|g_2|^2}{2\sigma^2}\right)\right) \nonumber \right. \\
                 & \quad\quad\quad\quad \left. + p(C) p(G)^2 \left(\exp\left( -\frac{|n_1|^2}{2\sigma^2} \right) + \exp\left( -\frac{|g_2|^2}{2\sigma^2} \right) + \exp\left( -\frac{|n_3|^2}{2\sigma^2} \right)\right) + p(G)^3 \right] \\
                 & \approx \frac{1}{(2\pi)^{3/2}\sigma^3} \left[ p(C)^3 e^{-|g_2|^2/2\sigma^2} + p(C)^2 p(G) e^{-(|n_1|^2+|n_3|^2)/2\sigma^2} + p(C) p(G)^2 \left(e^{-|n_1|^2/2\sigma^2} + e^{-|n_3|^2/2\sigma^2}\right) + p(G)^3 \right]
\end{align}
If we now assume glitches are relatively rare \textit{a priori} ($p(C)/p(G) \gg e^{-1/2}$), then the second term dominates over the third and fourth terms.
Because the glitch is loud, we additionally have $e^{-|g_2|^2/2\sigma^2} \ll p(G)/p(C)$ and the second term also dominates the first term, yielding
\begin{equation}
    p(h,\vec{a}) \approx \frac{p(C)^2 p(G)}{(2\pi)^{3/2}\sigma^3} \exp\left(-\frac{|h_1-s_1|^2+|h_3-s_3|^2}{2\sigma^2}\right)
\end{equation}
which is equivalent to the what we would obtain if we knew the correct sample to gate \emph{a priori}, even though we did not, up to a normalization constant.
Specifically, the marginalization automatically detects the correct placement for gates based on the relative frequency of $G$ and $C$ samples and the data's consistency with Gaussian noise without any other \textit{a priori} knowledge.
We note that, if the signal is quiet ($h-s \sim h \sim \sigma$), then we obtain
\begin{align}
    \log \Lambda^S_{!S} & \approx -\frac{|h_1-s_1|^2+|h_3-s_3|^2}{2\sigma^2} + \frac{|h_1|^2 + |h_3|^2}{2\sigma^2}
\end{align}
exactly as expected for stationary white Gaussian noise with the glitch gated with precise \textit{a priori} knowledge of the glitch's location.
If instead the signal is loud, the inference is more complicated as the noise-only model may confuse what is really a loud signal with a loud glitch, although we have
\begin{equation}
    \Lambda^S_{!S} \approx \left(\frac{p(C)}{p(G)}\right)^2 \exp\left(-\frac{|h_1-s_1|^2 + |h_3-s_3|^2)}{2\sigma^2}\right) \sim \left(\frac{p(C)}{p(G)}\right)^2 \gg 1
\end{equation}
which is still large due to the prior odds and therefore still strongly in favor of a signal.
While the full solution with more samples and colored noise is more challenging technically, is follows the same basic principles.

We again note that the benefits of marginalization seen within this toy model assume uninformative auxiliary information and simply accounts for the possibility that glitches exist within the detectors and our imperfect knowledge of data quality.
Informative supervised learning models based on $\vec{a}$ with correct calibration, such as those provided by iDQ, can only further improve the inference.
In the case of our toy model, this would simply add additional weight to the correct permutation that gated the second sample.

\subsection{Signal Consistency Tests}
\label{sec:signal consistency tests}

We remark that signal consistency tests, like $\chi^2$ goodness-of-fit, require the data to be consistent over several smaller, independent trials, looking at the distribution of $\sum_t \rho(t)^2$ rather than $(\sum_t \rho(t))^2$.
We note that such $\chi^2$ statistics are \textit{ad hoc} and not uniquely defined.
Therefore, there is no particular reason we should expect to derive the form of any such statistic from first-principles considerations.
Nonetheless, we show that marginalization naturally defines a signal consistency requirement similar in spirit to, but different in detail from, existing $\chi^2$ tests.

Let us further consider the model comparisons implicit with the marginalization over permutations.
As an example, let us assume that the $G$ or $C$ assignments are known perfectly for all samples except one: $h_k$.
The explicit marginalization over this single unknown sample is then
\begin{align}
    p(h,\vec{a}) & = p(\vec{a}_k|C)p(C) \left(\frac{\exp\left(-\frac{1}{2}\sum\limits_{i,j\in C+k} (h_i-s_i) \mathcal{C}_{ij} (h_j-s_j)\right)}{\sqrt{(2\pi)^{N}\mathrm{det}|\mathcal{C}|}} \right) + p(\vec{a}_k|G)p(G) \left(\frac{\exp\left(-\frac{1}{2}\sum\limits_{i,j\in C} (h_i-s_i) \mathcal{C}_{ij} (h_j-s_j)\right)}{\sqrt{(2\pi)^{N}\mathrm{det}|\mathcal{C}|}}\right) \nonumber \\
                 & = p(\vec{a}_k|C)p(C) \not{p}_C(n=h-s) \left( \exp\left(-\sum\limits_{i\in C} (h_i-s_i)\mathcal{C}^{-1}_{ik} (h_k-s_k) - \frac{1}{2}(h_k-s_k)^2\mathcal{C}^{-1}_{kk} \right) + \frac{p(G|\vec{a}_k)}{p(C|\vec{a}_k)}\right)
\end{align}
We note that
\begin{equation}
    \exp\left(-\sum\limits_{i\in C} (h_i-s_i)\mathcal{C}^{-1}_{ik} (h_k-s_k) - \frac{1}{2}(h_k-s_k)^2\mathcal{C}^{-1}_{kk} \right) = \sqrt{2\pi \frac{\mathrm{det}|\mathcal{C}|_{C+k}}{\mathrm{det}|\mathcal{C}|_C}} p(h_k-s_k|\{h_i-s_i \ \forall \ i \in C\})
\end{equation}
which is the likelihood of observing ($h_k-s_k$) as part of the stationary Gaussian noise process conditioned on the observations of the rest of the Gaussian noise process known to be clean ($h_i-s_i\ \forall\ i \in C$) multiplied by the prior volume allowed by the additional degrees of freedom.
This is a signal consistency test that checks whether the $k^\mathrm{th}$ sample agrees with the signal seen in the other $N_C$ samples.
Marginalization compares this consistency test against the posterior odds that the $k^\mathrm{th}$ sample was a glitch based on $\vec{a}_k$, effectively placing a lower bound on the probability of seeing \emph{any} $h_k-s_k$.
Within likelihood ratio tests, this prevents the noise-only model from becoming vanishingly small in the presence of loud glitches, thereby preventing the likelihood ratio from diverging and rendering the search much less sensitive to glitches.
Indeed, this is exactly the behavior seen in our toy model.

\subsection{Computational Cost of Marginalization}
\label{sec:computational cost}

We note that the marginalization over all permutations proposed within $\Lambda^S_{!S}$ is combinatorially expensive.
Efficiently implementing such a calculation is an open problem, but we discuss a few possible solutions below.

First, one could Monte-Carlo integrate instead of performing the entire sum.
However, Monte-Carlo integrals have variances that scale as
\begin{equation}
    \mathrm{Var}\left[\frac{1}{N}\sum_i f_i\right] \sim \frac{1}{N} \mathrm{Var}\left[f\right]
\end{equation}
The integral's variance, then, could be quite large in the presence of loud glitches as the variance between different permutations would be large.
Such integrals may require many samples to converge.

Alternatively, one could sample from the sum in a scheme similar to the Metropolis-Hasting algorithm~\cite{Hastings1970}.
Jump proposals would consist of flipping the label of one sample from $G$ to $C$ or vice versa, much like an Ising spin model~\cite{Ising1925}.
However, Markov-Chain Monte-Carlo estimates of marginal likelihoods are not without their own computational challenges and may not scale well in practice.

Regardless of the sampling procedure, we note that the number of possible permutations could be exponentially reduced by labeling small contiguous segments as $G$ or $C$ instead of labeling every time sample separately.
This would greatly reduce the computational cost, perhaps to something tractable, but introduces issues of how to select the window size.
Based on our considerations of model comparisons occurring within the marginalization, windows of comparable size to the stationary Gaussian noise's autocorrelation times may be appropriate, as this is the relevant timescale over which $h_k|h_{i\in C}$ becomes less in formed by $h_{i\in C}$ and therefore less stringent of a test.

We would need to compute the probability that there was a glitch present at any time within each segment.
This coarse-graining should be straightforward, though, as
\begin{equation}
    p(G\in\mathrm{window}|\vec{a}) = 1 - \prod\limits_{t_i\in\mathrm{window}} p(C|\vec{a}(t_i))
\end{equation}

A related approach would be to round the probabilities up or down based on some threshold.
That is, if $p(C|\vec{a})$ is above some threshold, we only consider permutations where that sample is labeled $C$.
Similarly, if $p(C|\vec{a})$ is below another threshold, we only consider permutations where that sample is labeled $G$, marginalizing over only the samples in the middle.
In effect, this would define
\begin{equation}
    p_\mathrm{eff}(C|\vec{a}) = \left\{\begin{matrix} 1 & p(C|\vec{a}) \geq p_\mathrm{max} \\ p(C|\vec{a}) & p_\mathrm{min} < p(C|\vec{a}) < p_\mathrm{max} \\ 0 & p(C|\vec{a}) \leq p_\mathrm{min} \end{matrix}\right.
\end{equation}
There is no single obvious choice of thresholds, however, so care would be needed.

We note that this is similar to the \emph{auto-gating} implemented within some existing searches~\cite{Messick:2016aqy, Sachdev:2019vvd}, although the thresholds are placed on $\rho$ and are not currently determined by $p(G|\vec{a})/p(C|\vec{a})$.
What's more, they ignore the conditioning on other data already declared clean.
Indeed, this could be considered a conservative choice as the threshold to declare a sample clean based on the conditioned likelihood $h_k|h_{i\in C}$ should only be more stringent than a threshold derived without the observations of $h_{i\in C}$.

Such coarse-graining is likely to result in information loss and therefore less sensitive searches.
However, the impact may be small enough and the computational speed-ups large enough to make this tractable, thereby improving GW search sensitivity compared to current approaches that do not marginalize over imperfect data quality information.
Ref.~\cite{iDQ+GstLAL} implements one such coarse-graining procedure, although they do not attempt to marginalize over probabilistic data quality and instead directly modify their likelihood ratio with a multiplicative factor that depends on iDQ's output.
Even this simple approach already shows modest improvements in search sensitivity.

\end{document}